\documentclass[12pt, preprint]{aastex}
\usepackage{amsmath}

\begin{document}

\title{On the secular behavior of dust particles in an eccentric protoplanetary disk with an embedded massive gas giant planet}

\author{He-Feng Hsieh and Pin-Gao Gu}
\affil{Institute of Astronomy and Astrophysics, Academia Sinica,
    Taipei 10617, Taiwan} \email{hfhsieh@asiaa.sinica.edu.tw \quad gu@asiaa.sinica.edu.tw}

\begin{abstract}
We investigate the dust velocity and spatial distribution in an eccentric
protoplanetary disk under the secular gravitational perturbation of an embedded
planet of about 5 Jupiter masses. We first employ the FARGO code to obtain the
two-dimensional density and velocity profiles of the eccentric gas disk exterior to
the gap opened up by the embedded planet in the quasi-steady state. We then apply
the secular perturbation theory and incorporate the gas drag to estimate the dust
velocity and density on the secular timescale. The dust-to-gas ratio of the
unperturbed disk is simply assumed to be 0.01. In our fiducial disk model with the
planet at 5 AU, we find that 0.01 cm- to 1 m-sized dust particles are well coupled
to the gas. Consequently, the particles behave similarly to the gas and exhibit
asymmetric dynamics as a result of eccentric orbits. The dust surface density is
enhanced around the apocenter of the disk. However, for the case of a low-density
gaseous disk (termed ``transition disk" henceforth in this work) harboring the
planet at 100 AU, the azimuthal distributions of dust of various sizes can deviate
significantly. Overall, the asymmetric structure exhibits a phase correlation
between the gas velocity fields and dust density distribution. Therefore, our study
potentially provides a reality check as to whether an asymmetric disk gap detected
at sub-millimeter and centimeter wavelengths is a signpost of a massive gas giant
planet.

\end{abstract}

\section{INTRODUCTION}
Past theoretical models have demonstrated that the tidal interactions between a
protoplanetary disk and an embedded giant planet
clean out the region around the planet's orbit and open up a gap
\citep[e.g.][]{LP93,Ward97}. Therefore, the presence of a cavity in a
protoplanetary disk as revealed by dust continuum emissions
has at times been postulated as a potential signpost of an embedded gas giant
planet \citep[e.g.][and references therein]{Andrews11}. Non-planetary explanations
have also been proposed, e.g. grain growth, fast radial drift of dust, and
photoevaporation \citep[e.g.][]{WC11, TCL05}. Evidently, a thorough understanding
of the dust dynamics in a gaseous disk is essential for interpretation of such
observational results.

The aerodynamic drag is a well-known interaction between gas and dust in
protoplanetary disks. The presence of a radial pressure gradient causes the
azimuthal gas velocity to deviate from the circular Keplerian motion, resulting in
the super-Keplerian or sub-Keplerian motion. On the other hand, dust particles are
unaffected by the pressure within the disk. The resulting discrepancy in the
velocities gives rise to the drag force and hence a drift of the particles toward
the pressure maxima \citep[e.g.][]{HB03}.
This dust trapping mechanism has been previously applied to a circular disk with an
embedded gas giant planet. For instance, the outer edge of the gap opened up by the
planet can act as a filter to stall the radial drift of large particles in the
outer disk \citep{Riceet06}. In addition, dust that is well-coupled to the gas can
be temporarily trapped by the spiral density waves excited by the planet as the
waves pass by \citep{PM06}.

If the planet is sufficiently massive,
the disk exterior to the planet's orbit can become moderately eccentric and start
precessing on the secular timescale, depending on the disk mass and viscosity
\citep{Papa01,KD06}. Eccentric protoplanetary disks in binary systems have also
been observed in simulations as a result of the influence of a companion star
\citep[e.g.][]{MK12}. This phenomenon is reminiscent of eccentric disks in the SU
Ursae Majoris systems, which have been long proposed to account for the superhump
light curves \citep[e.g.][]{Wh88}. The eccentricity is excited exponentially and
secularly by the tidal forcing of the companion star via the 3:1 eccentric Lindblad
resonance \citep{Lubow91}. \citet{GO06} carried out a linear analysis for the
secular perturbations with the azimuthal mode number $m=1$, and derived the
so-called eccentricity equation for a gaseous disk. The equation
describes the eccentric instability in terms of ``eccentric modes", with the growth
and precession rates determined by given disk properties such as sound-speed and
density profiles \citep[also see][]{Lubow10}. In the presence of dissipation that
damps eccentricity\footnote{\citet{GO06} introduced a bulk viscosity to
parameterize the eccentricity damping. It should be kept in mind that shear
viscosity may facilitate eccentricity growth rather than damping
\citep{Kato78,O01}. Furthermore, viscosity also affects how a disk is tidally
truncated and thus determines the density at the eccentric resonances, which in
turn factors the eccentricity excitation \citep{MK12}. Nonlinear dissipation
through shocks or tidal stresses also affect the saturation of the eccentricity
growth \citep{Paardet08}.},
the slowly precessing
disk finally settles into a steady state in which the eccentricity and longitude of
pericenter are smooth functions of the disk radius.



In such an eccentric protoplanetary disk harboring a massive planet, the orbits of
dust particles would be eccentric as well due to the gas drag. The standard
gas-dust dynamics as analyzed for circular orbits \citep[e.g.][]{W77,TL02} is
unable to yield the dust velocity and the corresponding spatial density
distribution accurately. One straightforward approach for studying the dust
behavior in eccentric disks is to utilize two-fluid (gas + dust) hydrodynamic
simulations such as SPH codes and RODEO \citep{PM06}. In fact, because several
giant planets have been discovered in close stellar binaries (binary separation $<$
20 AU), a large body of the literature and considerable attention have been devoted
to eccentric circumstellar disks in these binaries to examine the outcome of
planetesimal collisions on the formation of giant planets \citep[e.g.][and
references therein]{Paardet08}. For dust particles, \citet{Fouchet10} performed 3D
SPH simulations to investigate the dynamics of dust with sizes from 0.1 mm to 1 cm
in protoplanetary disks, including an eccentric disk driven by a planet of 5
Jupiter masses on a fixed circular orbit at 40 AU. After about 100 planetary
orbits, they found that 1 cm-sized dust had accumulated in the strong spiral
density waves just around the outer region of the gap in the disk in which the
eccentricity is still evolving.
At present, SPH simulations are still computationally expensive. For a small
perturber like a planet, a tremendous length of time is required to reach
quasi-steady states for the study of any features related to secular effects.

Therefore, for the sake of convenience, we run the FARGO code \citep{M00} to obtain
the gas profile in a two-dimensional disk with a massive planet, and then employ
the secular perturbation theory and incorporate aerodynamic drag \citep{Paardet08}
to estimate the dust velocity and density that will in turn reveal the
asymmetric structures inherent to an eccentric disk.
The purpose of this study is to focus only on the secular behaviors of dust
associated with an eccentric protoplanetary disk in the presence of a massive giant
planet. Therefore, non-secular effects, such as dust-gas dynamics in the horseshoe
orbit and whether dust particles can drift toward the peaks of spiral density waves
in an eccentric disk \citep{Fouchet10}, are not studied in this work.

The paper is organized as follows. In \S2, we first describe the fiducial disk
model for the study. We then briefly review the eccentricity equation for dust and
explain our approach for solving the problem in \S3. The results for the fiducial
models of a protoplanetary disk and for a simple transition disk are presented in
\S4. We finally summarize the results and discuss the limitations of our work in
\S5.

\section{DISK MODELS}
A geometrically thin protoplanetary disk with an embedded gas giant planet is
numerically simulated using the FARGO code, a two-dimensional hydrodynamic code
employing an efficient modification of standard transport algorithm \citep{M00}.
Therefore, the disk pressure and density are vertically integrated quantities. The
simulations are conducted in polar coordinates $(r, \phi)$ centered on the
protostar.
The quantities in FARGO are dimensionless. The semi-major axis of the planet $a_p$
is taken to be the length unit and the mass of the central protostar $M_\star$ is
adopted as the mass unit. The time unit is obtained from
$\Omega_p^{-1}=(a_p^3/GM_\star)^{1/2}$ and the gravitational constant $G$ is set to 1 in
FARGO.

In this study, we consider the following fiducial model with standard parameters
for a locally isothermal and initially axi-symmetric gas disk \citep[cf.][]{KD06}:
$M_\star = 1 M_\odot$, $a_p = 5$ AU, and a constant aspect ratio $H/r = 0.05$,
where $H$ is the vertical pressure scale height of the disk. We adopt $\nu = 1
\times 10^{-5}$ and $\Sigma(t=0) = 1\times 10^{-4}$ to be the values of the
dimensionless kinematic viscosity and initial uniform surface density,
respectively. The surface density corresponds to about 35.6 g cm$^{-2}$, amounting
to approximately a disk mass of $5.03\times 10^{-3} M_\odot$ inside $4a_p=20$ AU.
Because all the quantities are dimensionless in FARGO, we also lower the surface
density by using the larger length-scale $a_p=100$ AU to simulate a simplified
model that resembles a ``transition disk"\footnote{Transition disks are commonly
referred to as a class of protoplanetary disks containing an optically thin inner
region and an optically thick outer disk, as implied from their infrared and
(sub)millimeter spectral energy distributions \citep[e.g.][]{WC11}. In this work,
the term ``transition disks" is loosely used for an idealized disk with an overall
low surface density, which may bare a resemblance to ``anemic" or ``homologously
depleted" disks discussed in the literature \citep[e.g.][]{Andrews11}.} with an
embedded gas giant planet far away from its protostar. That is, with the same disk
mass $5.03\times 10^{-3} M_\odot$ inside $4a_p=400$ AU, the surface density of the
transition disk is about 0.09 g cm$^{-2}$. A number of gas giants have been
directly imaged at $a_p
>$ 50-100 AU around their main-sequence host stars (see The Extrasolar Planets
Encyclopedia website at http://exoplanet.eu). How and when these planets lie on the
current large orbits is still an open question. Some planet formation and migration
models \citep[e.g.,][]{Boss11,Crida09} suggest that these planets could have been
situated on the large orbits back to their T Tauri epochs when their protoplanetary
disks were still present, as is considered here for the transition disk.

To excite a noticeable disk eccentricity even for a planet on a circular orbit, we
employ the initial mass ratio $q\equiv m_p/M_\star = 5 \times 10^{-3}$
\citep{KD06}. The planet mass $m_p$ is allowed to grow by setting the inverse of
the accretion timescale to $2$ \citep{K99}\footnote{Namely, in each normalized time
step $\Delta t$, the surface density of the gas inside the Roche radius of the
planet is reduced by about a factor of $2\Delta t$, which is added to the planet
mass.}.
The softening length is 0.4
times of the Hill radius of the planet. Disk self-gravity is neglected. We force
the planet on a fixed circular/eccentric orbit with the apocenter located in the
$\phi=0$ direction; namely, we omit the gravitational force exerted by the disk on
the planet and the increments in momentum due to the mass accretion. Hence,
planetary migration is not considered. Ignoring planetary migration is justified
because we are interested in physical properties of an eccentric disk on secular
timescales. These timescales are shorter than the typical Type II planetary
migration timescale $\sim r^2/\nu$ \citep{Ward97}, which is about a few $10^5$
dynamical time in our model. In the simulation, the planet initially orbits the
central star counterclockwise from $\phi=0$.
The computational domain in the $r$ and
$\phi$ is $[0.4, 4.0] \times [0, 2\pi]$ with resolutions $(N_r, N_\phi) =
(256,768)$. At both the inner and outer boundaries, we apply the non-reflecting
boundary condition to avoid unphysical reflecting waves.

As for dust particles in the gaseous disk, we do not consider any dust size
distributions. For the dust of a given size, we simply assume that the dust-to-gas
ratio of the unperturbed disk (i.e., the disk in the absence of a planet) is 0.01.
Modelling dust dynamics in the presence of a planet is described in the next
session.


\section{SECULAR EVOLUTION OF THE DUST}
Since FARGO is a hydrodynamical code and not formulated to simulate dust particles,
our approach for dust dynamics in an eccentric gaseous disk is to consider the
secular evolution of the dust using the secular perturbation theory, which gives
rise to the eccentricity equation. The equation describes the secular evolution of
a massless particle perturbed by a companion. Given the drag force, the dust motion
and surface density can then be deduced from the gas dynamics in the quasi-steady
state of an eccentric disk in FARGO. However, to compute the secular evolution of
dust particles, we need to filter out the short-term evolution on the orbital
timescale of the planet in the gas simulation. As noted in the Introduction, the
eccentric equation has been modified to include the gas drag and applied to
planetesimals in a protoplanetary disk in binaries \citep{Paardet08}. As per
Paardekooper et al., we adopt the same eccentricity equation for solids, but with
the perturbed potential in terms of the Lapalace coefficients
\citep[e.g.][]{Tremaine98,MD99} and the gas drag appropriate for small particles
(i.e. micron to meter instead of km size). The details are described as follows.

\subsection{Eccentricity equation for the dust}
In the subsection, we briefly review the eccentricity equation for dust particles
in accordance with the the linear analysis for a gas disk by \citet{GO06}, in which
the variables such as velocities and density are expanded as power series in terms
of $\epsilon \equiv H/r$.

\citet{GO06} considered a 2D gaseous disk with the axi-symmetric basic state
allowing for a small deviation from the Keplerian circular angular velocity
$\Omega_K$ to the accuracy of the order $\epsilon^2 \equiv (H/r)^2$, due to the
small background pressure gradient and orbit-averaged planet potential. After
introducing the $m=1$ perturbations in the form of
$\mathcal{R}[x'(r,t)\exp(-i\phi)]$, where $\mathcal{R}$ denotes the real part, they
showed that the complex eccentricity is related to the perturbed velocities at the
zeroth order of $\epsilon$ (denoted by the subscript 0):
\begin{equation}
u_{g,0}' = ir\Omega_K E_g, \qquad v_{g,0}' = \frac{1}{2}r\Omega_K E_g,
\label{eq:epic_g}
\end{equation}
where $u$ and $v$ are the radial and azimuthal velocities. $E_g$ is the complex
eccentricity of the form $e_g\exp(i\varpi_g)$ with the eccentricity, $e_g(r,t)$, as
the amplitude, and the longitude of pericenter, $\varpi_g(r,t)$, as the phase
angle. The subscript $g$ represents the gas to distinguish it from the subscript
$d$ for dust to be used later in the paper. The above equation describes the
Keplerian motion of the gas with a small eccentricity $e_g$, thus giving the
following relation for the epicyclic motion
\begin{equation}
u_{g,0}'=2iv_{g,0}'.\label{eq:epic_g_2}
\end{equation}
In order to examine the secular evolution of an eccentric disk resulting from the
weak companion potential and small radial pressure gradient, \citet{GO06} proceeded
to the next order (i.e. at $O(\epsilon^2)$, denoted by the subscript 2) and
obtained the evolutionary equation for the complex eccentricity $E_g$.

Likewise, the eccentricity equation for dust particles can be obtained from the
linearized equations with zero pressure. In the basic state, the dust particles in
a disk are assumed to be steady, non-self-gravitating, and axi-symmetrically
distributed. As such, the basic state of the dust motion reads\footnote{The same
expansion in terms of $\epsilon$ applied to dust equations implies that the dust
forms the same disk as the gaseous disk with the same $H/r$ under the assumption of
the constant dust-to-gas ratio in a 2D disk.}
\begin{equation}\label{bg_d_1}
r\Omega_{d,0} = \frac{\partial \Phi_0}{\partial r} = r\Omega_K,
\end{equation}
\begin{equation}\label{bg_d_2}
2r\Omega_{d,0}\Omega_{d,2} = \frac{\partial \Phi_2}{\partial r} + {1\over t_s}
(u_{d,2}-u_{g,2}),
\end{equation}
\begin{equation}\label{bg_d_3}
u_{d,2}\Omega_{d,0} = - \frac{2r}{t_s}(\Omega_{d,2} - \Omega_{g,2}),
\end{equation}
In the above equations, $\Phi_0$ is the primary potential from the central
protostar, $\Omega_{d,2}$ is the dust angular velocity departure from $\Omega_K$
due to the orbit-averaged planet potential $\Phi_2$ and the radial gas drag, and
$\Omega_{g,2}=(1/\Sigma_g)(dp_g/dr)/(2r\Omega_K)$ is the gas angular velocity
departure from $\Omega_K$ due to the radial pressure gradient $dp_g/dr$
\citep{GO06}. $u_{d,2}$ is the radial drift velocity arising from the gas drag
characterized by the stopping time in the Epstein regime \citep{W77}
\begin{equation}
t_s=\frac{\rho_d a_d}{\rho_g \bar{v}_{th}} \qquad {\rm if}\quad a_d<\frac{9}{4}
\lambda_g,
\end{equation}
where $a_d$ is the mean radius of the dust grains, $\rho_g$ is the gas mass
density, $\rho_d$ is the density of a dust particle, $\bar v_{th}$ is the mean
thermal velocity of the gas,
and $\lambda_g$ is the mean free path of the gas.
The stopping time $t_s$ measures the timescale for the coupling between the dust
and gas in the disk. Note that in the other regimes where $\lambda_g$ is much
smaller than $a_d$, the gas drag is fluid-like.
Therefore both the dust and gas velocities need to be known in advance to determine
the stopping time \citep{W77}. Because the dust velocities in eccentric orbits are
what we are solving for and thus not known beforehand, in this work, we therefore
focus only on the dust in the Epstein regime,
which can be verified in advance. As a matter of the fact, the Epstein drag is
adequate for the dust sizes and disk parameters considered in this study.

Combining Equations (\ref{bg_d_1})-(\ref{bg_d_3}), we obtain the radial drift
velocity in the basic state \citep[cf.][]{TL02}
\begin{equation}
u_{d,2}={\tau_s^{-1}u_{g,2}-\eta r\Omega_K - (1/\Omega_K)(d\Phi_2/dr) \over \tau_s
+ \tau_s^{-1}},\label{eq:u_d2}
\end{equation}
where $\tau_s=t_s \Omega_K$ is the dimensionless stopping time and
$\eta=-(c_s^2/r^2\Omega_K^2)(d\ln p_g/d\ln r)$ with the isothermal sound speed
$c_s^2=p_g/\Sigma_g$. For the rest of the work, we neglect $u_{g,2}$, which can be
related to the slow accretion flow \citep{O01}.

For the $m=1$ secular perturbations, the dust perturbed velocity at the zeroth
order can be expressed in terms of the dust eccentricity $E_d=e_d \exp(i\varpi_d)$
as has been done for the gas component:
\begin{equation}\label{dust_form}
u_{d,0}' = ir\Omega_K E_d \quad\quad v_{d,0}' = \frac{1}{2}r\Omega_K E_d,
\end{equation}
which describe the epicyclic motion with the relation
\begin{equation}\label{dust_relation}
u_{d,0}' = 2iv_{d,0}'.
\end{equation}
The eccentricity equation for dust particles can then be derived at $O(\epsilon^2)$
\citep[cf.][]{GO06}, as can be found in \citet{Paardet08}:
\begin{equation}
2r\Omega_K \frac{\partial E_d}{\partial (\epsilon^2 t)} =
-i\frac{E_d}{r}\frac{\partial }{\partial r}\left(r^2 \frac{\partial
\Phi_2}{\partial r}\right) + {i\over r^2} {\partial \over \partial r} \left( r^2
\Phi'_2 \right)- \frac{2r\Omega_K}{t_s}(E_d-E_g),\label{eq:E_d}
\end{equation}
where the terms associated with resonant interactions and the radial drift velocity
$u_{d,2}$ are ignored because we focus only on secular evolutions of eccentric
orbits. The effect of the radial drift will be estimated separately by Equation
(\ref{eq:u_d2}). In the above equation, $\Phi'_2$ is the $m=1$ component of the
planet's potential due to the planetary eccentricity $E_p$ ($=e_p
\exp(i\varpi_p)$).
Equation (\ref{eq:E_d}) presents the complex form of the eccentricity equation in
terms of $h\equiv e\sin \varpi$ and $k \equiv e\cos \varpi$ derived from the
Lagrange's planetary equation in the secular perturbation theory
\citep[e.g.,][]{Tremaine98,MD99,MS00}. Hence, we can express $\Phi_2$ and $\Phi'_2$
in terms of the Lapalace coefficients,
and simplify Equation (\ref{eq:E_d}) to
\begin{equation}
\frac{\partial E_d}{\partial (\epsilon^2 t)} = i
\frac{E_d}{t_{prec,d}}+i\frac{E_p}{t_{f}} -\frac{E_d-E_g}{t_s}, \label{eq:E_d_t}
\end{equation}
where $t_{prec,d}$ and $t_f$ are given by \citep[e.g.][]{Tremaine98,MD99}
\begin{eqnarray}
t_{prec,d}&=& \frac{4}{q}\left( \frac{r}{a_p} \right) \left(
\frac{1}{b_{3/2}^{(1)}\left( \frac{a_p}{r} \right)} \right) \frac{1}{\Omega_K},\\
t_{f}&=& \frac{4}{q}\left( \frac{r}{a_p} \right) \left(
\frac{1}{b_{3/2}^{(2)}\left( \frac{a_p}{r} \right)} \right) \frac{1}{\Omega_K},\\
\end{eqnarray}
and the Laplace coefficients $b_{3/2}^{(1)}$ and $b_{3/2}^{(2)}$ can be computed
numerically by Carlson's algorithm \citep{Presset92}. $t_{prec,d}$ is the
precession timescale of a free particle on an eccentric orbit, whereas $t_{f}$ is
related to the forced eccentricity in the secular perturbation theory (see below).

We consider the steady-state situation for dust particles. The eccentricity
equation (\ref{eq:E_d_t}) is then reduced to \citep[cf.][]{Paardet08}
\begin{equation}
E_d=E_f {1\over \sqrt{1+1/\tau_{s,sec}^2}} \exp [i\arctan (-(1/\tau_{s,sec})] + E_g
{1\over \sqrt{1+\tau_{s,sec}^2}} \exp [ i \arctan (\tau_{s,sec})].\label{EccEq}
\end{equation}
where
\begin{eqnarray}
\tau_{s,sec} &\equiv &\frac{t_s}{t_{prec,d}},\label{eq:A} \\
E_f &\equiv &\frac{t_{prec,d}}{t_f} E_p.\label{eq:E_f}
\end{eqnarray}
In the above equations, $E_f$ is the forced eccentricity from the secular
perturbation theory, with the amplitude $e_f<e_p$. In addition, we define the
dimensionless ``secular" stopping time $\tau_{s,sec}$, the ratio of the stopping
time to the precession timescale of a free particle. In analogy to the definition
of $\tau_s$, $\tau_{s,sec}$ measures the degree of the coupling between a dust and
the gas in the secular evolution.



In the case of the planet on a circular orbit, $E_f=0$ and Equation (\ref{EccEq})
implies that $\varpi_d$ lags $\varpi_g$ by the phase angle $=\arctan
(\tau_{s,sec})$. When $\tau_{s,sec}\ll 1$, $E_d \approx E_g$ and therefore
particles are well-coupled to the gas on the secular timescale.
In this case, the dust orbit is almost identical to the gas eccentric orbit.
When $\tau_{s,sec}>1$, the particles are weakly coupled to the gas, and the drag
force wields no or little effects. Thus, the dust particles are in near-Keplerian
circular motion, provided that the planetary eccentricity is zero. Nevertheless,
all dust orbits with different eccentricities and pericenters precess with the gas
at the same rate in a steady state.

On the other hand, for eccentric planet orbits ($e_p \neq 0$), $E_f$ in Equation
(\ref{EccEq}) may affect $E_d$ significantly depending on $\tau_{s,sec}$. In the
regime where $\tau_{s,sec} \ll 1$, the dust is strongly coupled to the gas and thus
$E_d \approx E_g$. In the other regime where $\tau_{s,sec} > 1$, the gas drag does
not dominate the dust motion but plays a role secondary to the planet's potential
in providing the free eccentricity to dust particles in the steady state
\citep{Paardet08}.
When $\tau_{s,sec}|E_f|> |E_g|$ for a given dust size, Equation (\ref{EccEq})
describes pericenter alignment of the dust with a libration rate equal to the
precession rate of the gas disk $\dot \varpi_g$. If $\tau_{s,sec}$ is moderately
larger than 1 as well, the alignment is close to the planet's pericenter.

\subsection{Calculations of dust velocity and density to the zeroth order}
Once $E_d$ is known from Equation (\ref{EccEq}), the secular solution of the dust
velocity as a function of $r$ and $\phi$ can be obtained as follows
\begin{eqnarray}
u_d &=& \mathcal{R}(u'_{d,0}e^{-i\phi}) + O(\epsilon^2) \nonumber \\
        &=& r\Omega_K[\mathcal{R}(E_d) \sin\phi  - \mathcal{I}(E_d) \cos\phi ] +
        O(\epsilon^2),\\
v_d &=& r\Omega_K + \mathcal{R}(v'_{d,0}e^{-i\phi})  + O(\epsilon^2) \nonumber \\
          & =& r\Omega_K + \frac{1}{2} r\Omega_K[\mathcal{R}(E_d) \cos\phi  + \mathcal{I}(E_d) \sin\phi ] + O(\epsilon^2),
\end{eqnarray}
where $\mathcal{R}$ and $\mathcal{I}$ denote the real and
imaginary parts, respectively.

To compute $E_d$ in Equation (\ref{EccEq}), we need the complex gas eccentricity
$E_g=e_g\exp(i\varpi_g)$. The values of $e_g$ and $\varpi_g$ can be derived from
the velocity fields in the simulation.
Time averaging is first performed to reveal the secular $m=1$ results associated
with the eccentric disk. We then calculate the eccentricity from the eccentricity
vector and compute the longitude of pericenter from $\phi = f + \varpi$, where
$\phi$ is the true longitude and $f$ is the true anomaly. In the FARGO code, the
direction of reference points to $\phi=0$
and for the case of $e_p \neq 0$,
$\varpi_p=\pi$ (i.e. $E_p=-e_p$).
To evaluate the stopping time, we employ the mean free path
\citep{Riceet06}
\begin{equation}
\lambda_g = \frac{4 \times 10^{-9}}{\rho_g}\text{ cm},
\end{equation}
where $\rho_g=\Sigma_g/(\sqrt{2}\pi H)$, $\Sigma_g$ is the gas surface density, and
$\rho_d = 1.25 \text{ g cm$^{-3}$}$ \citep{TL02}.

The non-axisymmetric dust spatial distribution associated with the secular
evolution in an eccentric disk is of great interest because it can be observed by
mapping the dust continuum emissions from a disk. The perturbed dust surface
density to the zeroth order is estimated from the $m=1$ linearized continuity
equation \citep{O01}:
\begin{equation}
\Sigma_{d,0}'=r\frac{\partial}{\partial r}(\Sigma_{d,0}E_d).\label{dust_den_pert}
\end{equation}
Consequently, the total dust surface density in the eccentric disk is given by
\begin{equation}
\Sigma_d = \Sigma_{d,0} +  \mathcal{R}(\Sigma_{d,0}' e^{-i\phi})+ O(\epsilon^2).
\label{eq:Sigma_d}
\end{equation}
To obtain the zeroth-order basic state of dust density $\Sigma_{d,0}$ in the above
equation, we first take the azimuthal average of the gas density in the simulation
as the zeroth-order basic state, and assume the dust-to-gas ratio to be $0.01$
everywhere. Since we omit the size distribution of the particles, the computed dust
density does not represent the true magnitude.

\section{RESULTS}
We present the results for two different planetary eccentricities, $e_p=0$ and 0.1,
for a disk of standard parameters (a protoplanetary disk) and for a low-density
disk (a transition disk). The general gas dynamics in the dimensionless units for
our fiducial model is shown and described first. The physical units are then
considered to present the results for the dust velocities and the resulting dust
surface densities in both the protoplanetary and transition disks.

\subsection{General gas dynamics}
\begin{figure}
\centering
\begin{tabular}{cc}
\scalebox{0.5}{\includegraphics[bb=0.72in 5in 7.75in 10in]{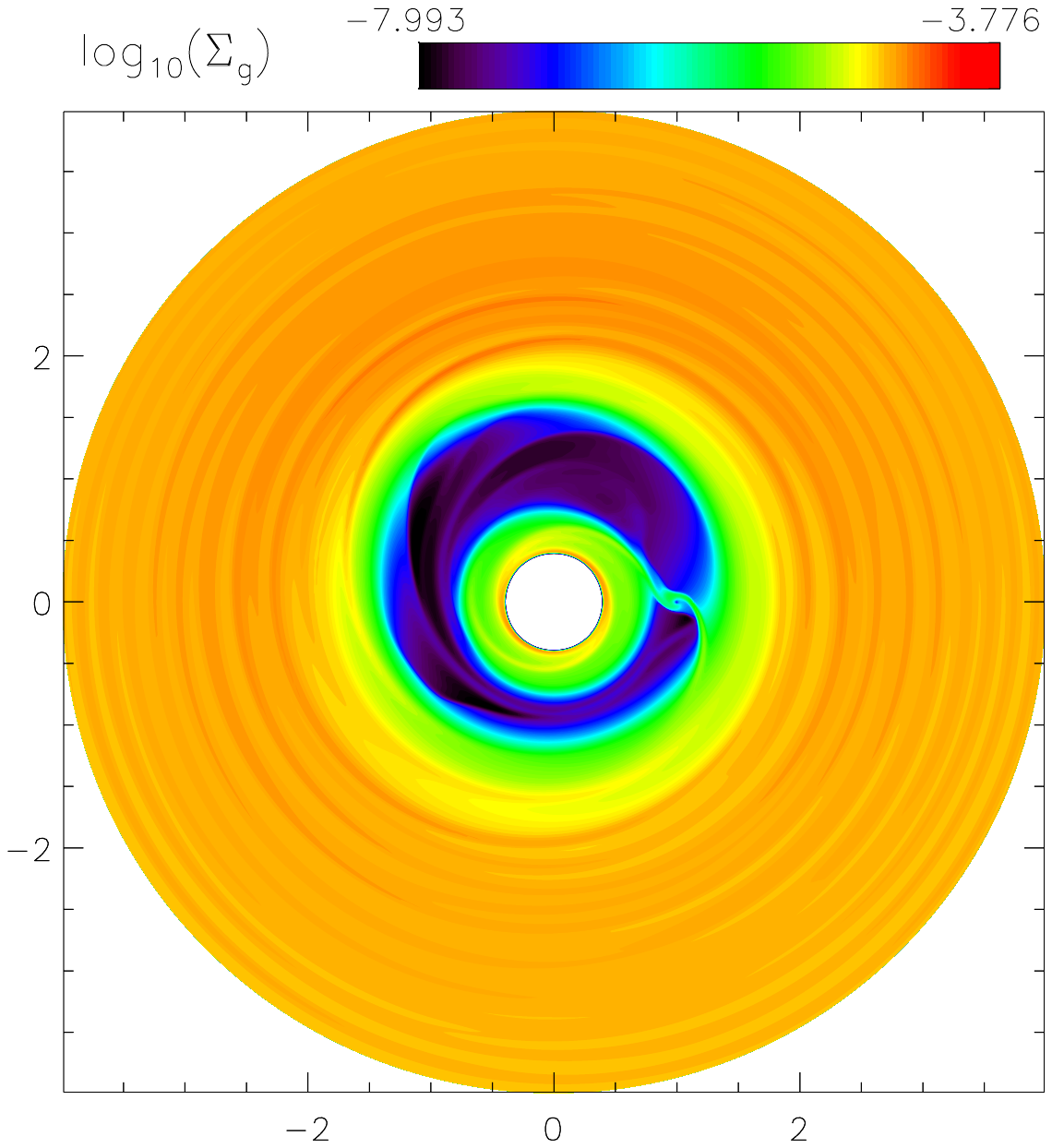}} &
\scalebox{0.5}{\includegraphics[bb=0.72in 5in 7.75in 10in]{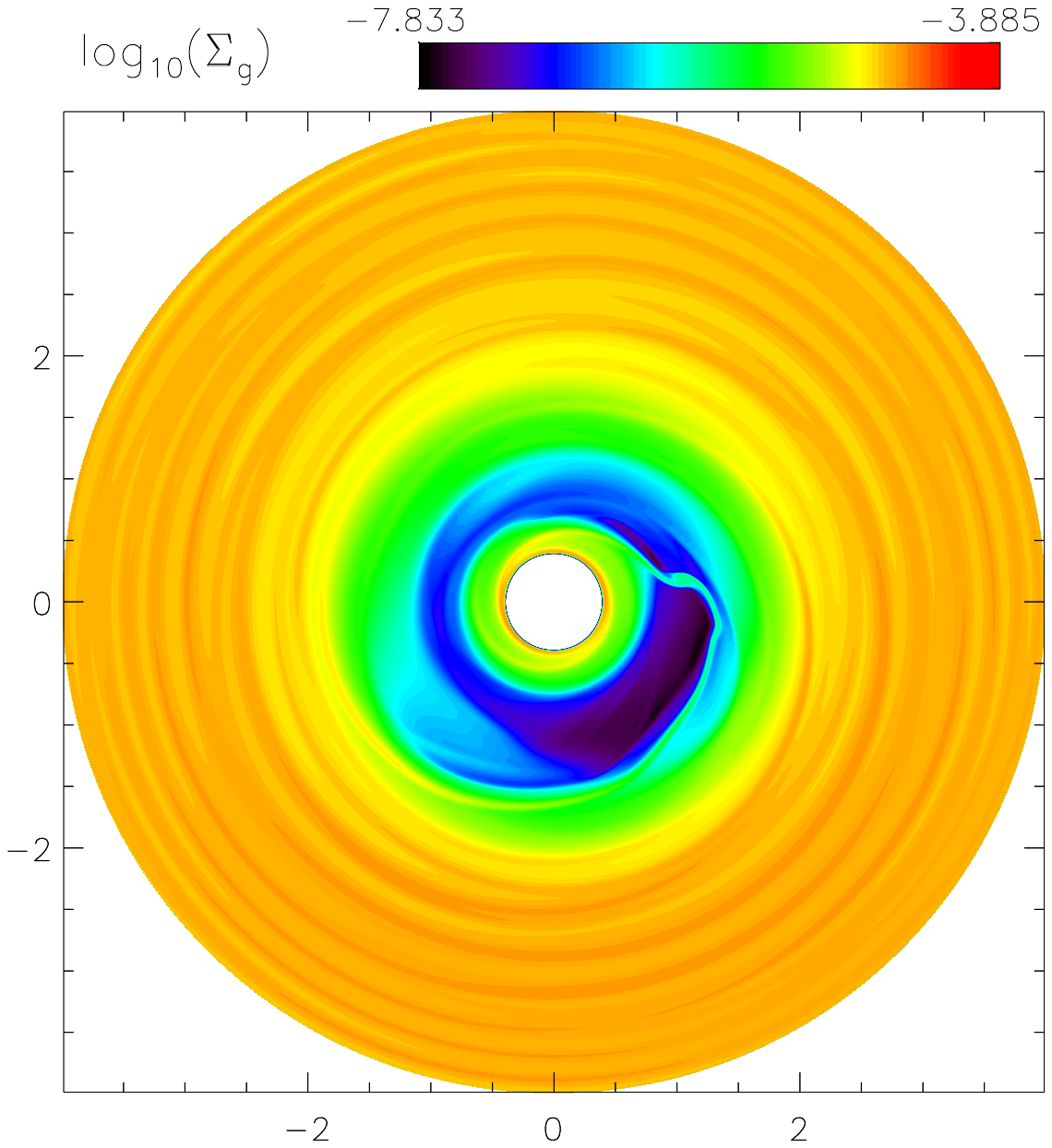}} \\
\scalebox{0.5}{\includegraphics[bb=0.72in 5in 7.75in 10in]{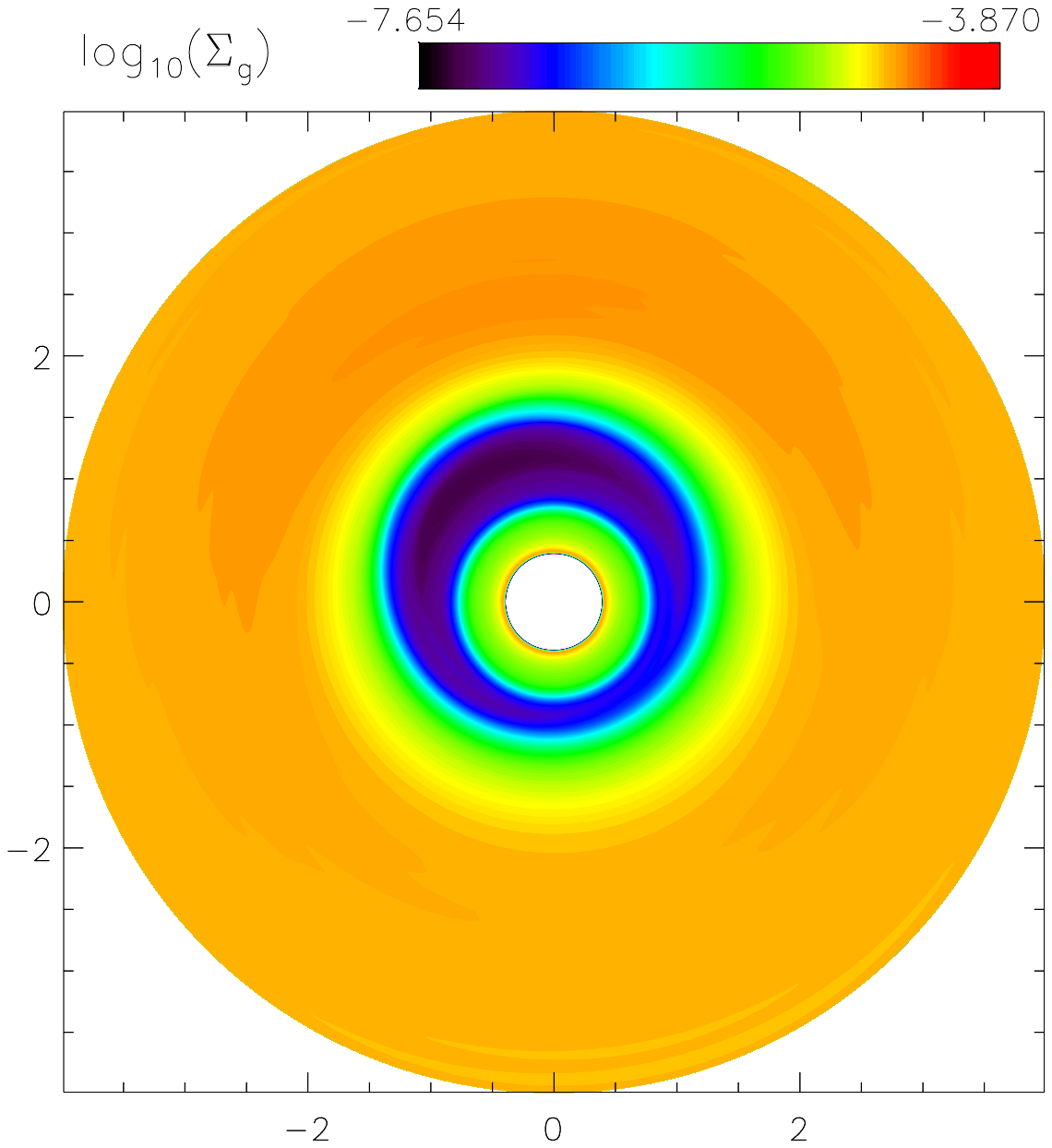}} &
 \scalebox{0.5}{\includegraphics[bb=0.72in 5in 7.75in 10in]{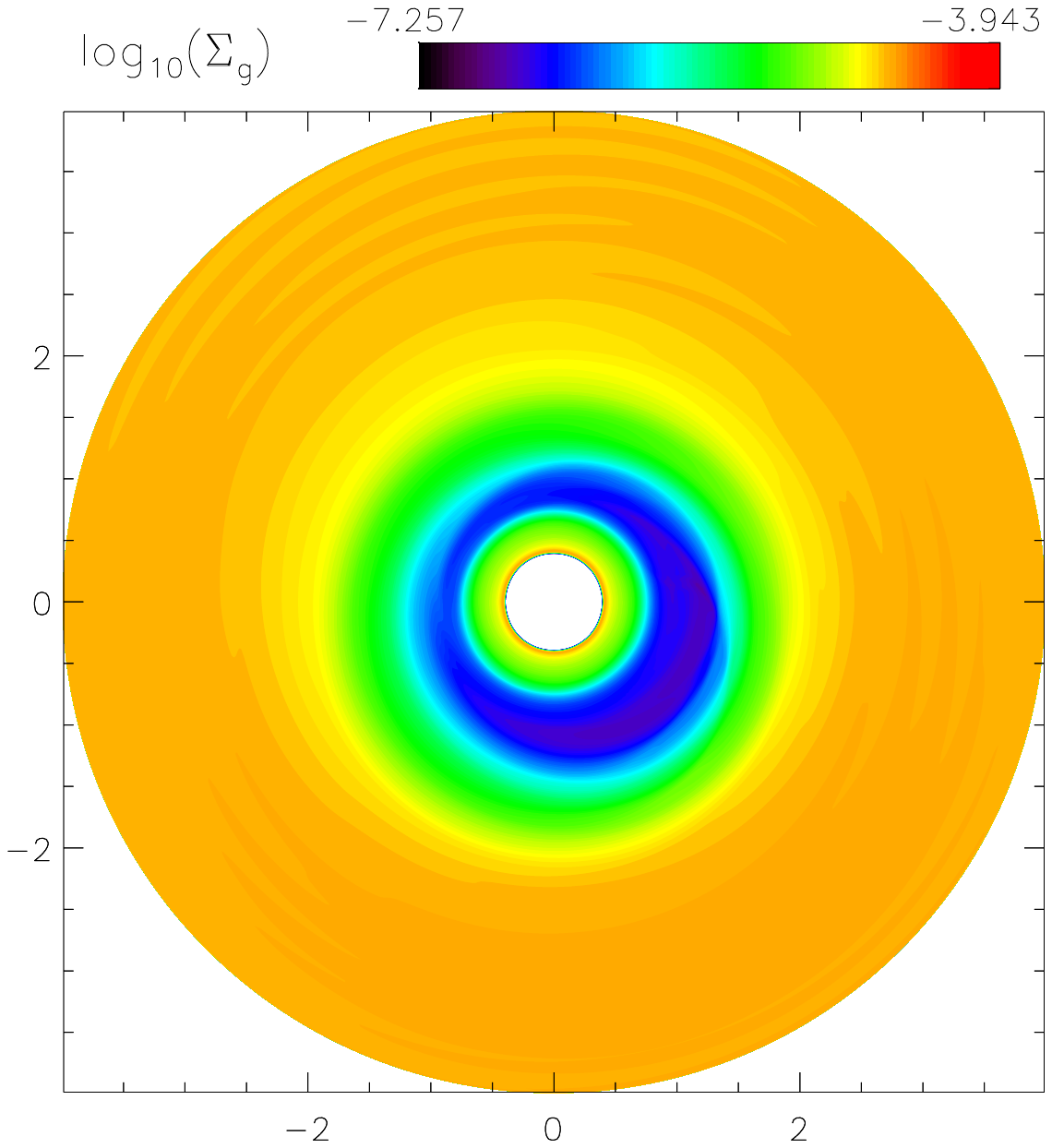}}
\end{tabular}
  \caption{The quasi-steady maps of gas surface densities in the dimensionless units for $e_p=0$
  (left panels) and $e_p=0.1$ (right panels). The top panels show the snap shots at $t=3000$ and the gas surface densities in the bottom panels are the time-averaged
  values of those shown in the top panels to erase most of the short-period spiral features.}
  \label{fig:density_g}
\end{figure}

We present the results for the cases of $e_p=0$ and $e_p=0.1$ at $t=3000$ orbits to
illustrate the typical gas dynamics in a quasi-steady state. At this time, the
planetary mass increases by about 6.8\% in the $e_p=0$ case and by about 16\% in
the $e_p=0.1$ case. The gas surface density are shown in the top panels of Figure
\ref{fig:density_g}, where the planet lies at $(r,\phi)=(1,0)$ for $e_p=0$ and at
(1.1,0) for $e_p=0.1$. We plot the densities color-coded on the logarithmic scale
to clearly display the structure of the gap. The gas disk exterior to the planetary
orbit apparently becomes eccentric, especially in the gap region. Furthermore, the
exterior disk precesses slowly at the rate of $\dot \varpi_g \approx
-(1/399)\Omega_p$ for $e_p=0$ and $-(1/152.8)\Omega_p$ for $e_p=0.1$. The absolute
values of $\dot \varpi_g$ are much smaller than $\Omega_p$ indicative of the
secular behavior of the eccentric mode with $m=1$. The negative value implies
regression, i.e. retrograde precession. According to the linear theory by
\citet{GO06} and the simulation by \citet{KPO08}, the disk regression in our
simulation is probably due to the fact that the disk global pressure
gradient\footnote{The vertically integrated pressure of the 2-D disk is
proportional to $\epsilon^2 \Sigma_{g,0}/r$. In our model with prescribed constant
values of $\epsilon$ and $\Sigma_{g,0}$, the global pressure gradient is simply
proportional to $\epsilon^2 \Sigma_{g,0} /r^2$.} plays a more important role than
the planetary potential in contributing to $\dot \varpi_g$.
In contrast, the disk interior to the planetary orbit remains almost circular.

Besides the eccentric mode, the planetary potential also induces the short-period
features manifested by tightly wound spiral density waves in the top panels of
Figure \ref{fig:density_g}. The interior disk is dominated by the $m=2$ waves, as
has been typically seen in the simulations for a circular disk \citep[e.g.][]{K99},
whereas the exterior disk is dominated by the $m=1$ waves that are distorted by the
eccentric shape of the disk. These short-period features are associated with the
predominant modes with pattern speeds equal to $\Omega_p$ and $\Omega_p/2$
\citep{Lubow91,KD06}. As a result, the gas motions associated with the density
waves are periodic and would be cancelled out when averaged over one orbital cycle
of the planet to reveal secular features.
Because of the disk regression,
the time averaging is performed in the non-rotating frame for each grid point over
the period of $0.9975 \times 2\pi/\Omega_p$ for $e_p=0$ and $0.9935 \times
2\pi/\Omega_p$ for $e_p=0.1$, instead of one planetary orbital period
$2\pi/\Omega_p$.\footnote{Owing to the non-uniform angular motion along an
eccentric orbit, the time averaging in the $e_p=0.1$ case is less perfect than that
in the $e_p=0$ case. Nevertheless, the difference between the average over one
planetary period and over the smaller period taking into consideration the disk
regression is minuscule because the disk regression is slow.}

The resulting time-averaged gas density is presented in the bottom panels of Figure
\ref{fig:density_g}. At this point, the short-period features due to the tightly
wound spiral density waves are almost removed, albeit not completely. It is likely
due to the quasi-steady results. The color-coded density maps suggest that the gap
of the disk with the planet on an eccentric orbit appears wider but shallower than
that with the planet on a circular orbit. We plot the azimuthal average of the gas
densities in Figure \ref{fig:DenAve_vs_ep}, which confirms the gap shape in
relation to the planetary eccentricity. Note that the disk starts to become tidally
truncated roughly at the 3:1 eccentric outer and inner Lindblad resonances at
$r\approx 2.08$ and at $r \approx 0.48$, respectively. Hereafter in the paper, the
gap is referred to the region from $r \approx 0.5$ to 2.

The time-averaged density distribution of the exterior disk is predominated by the
$m=1$ azimuthal distribution, a feature associated with the eccentric mode.
However, the time-averaged gas flows inside the gap from $r\approx 0.5$ to 2 is
more complicated than the eccentric motions.
We examine whether Equation~(\ref{eq:epic_g_2}) is satisfied by an eccentric disk
at 3 different disk radii ($r=$ 1.5, 2, and 3) for the time-averaged flow. This is
illustrated in Figure \ref{check} for the $e_p=0$ (left panels) and $e_p=0.1$
(right panels) cases. For $r=1.5$, the black curves show that
Equation~(\ref{eq:epic_g_2}) is poorly satisfied,
likely a result of the irregular horseshoe orbit associated with an eccentric disk
and the non-linear effects due to the strong planetary potential. Secular effects
should be predominant away from the co-orbital region within the gap. Indeed,
farther away from the planet, all the curves become much smoother as shown for
$r=2$ and 3, indicating that the eccentric mode has become predominant. In view of
the complicated gas motion inside the gap, for the rest of study, we focus only on
the secular results outside $r=2$ where the azimuthal average of gas density
$\Sigma_{g,0}
> 9\times 10^{-5}$ for $e_p=0$ and $> 5\times 10^{-5}$ for $e_p=0.1$ as indicated
in Figure \ref{fig:DenAve_vs_ep}.

The departures of the gas velocity from the Keplerian circular motion (i.e. the
perturbed velocity) for the time-averaged disk are displayed in
Figure~\ref{fig:v_gas}. Both the perturbed radial velocity $u'_g=u_g$ and the
perturbed azimuthal velocity $v'_g=v_g-r\Omega_K$ exhibit $m=1$ azimuthal profiles,
as to be expected from the bottom panels of Figure \ref{fig:density_g}. Moreover,
Figure~\ref{fig:v_gas} shows that the phase difference between $u_g$ and
$v_g-r\Omega_K$ is $90^\circ$, in agreement with the epicyclic motion for an
eccentric disk described by Equation (\ref{eq:epic_g_2}). Namely, $u_g\approx
u'_{g,0}$ and $v_g-r\Omega_K \approx v'_{g,0}$ as has been verified in Figure
\ref{check}. According to the color coding, the perturbed velocities are larger
around the outskirts of the gap region (i.e. $r\gtrsim 2$) and become progressively
smaller toward the edge of the simulated disk, implying that the disk eccentricity
decreases with the radius. This is confirmed by the azimuthally averaged
eccentricity profile of the gas disk as shown in the solid lines in
Figure~\ref{fig:ecc_ep=0}, which is similar to the eccentricity profile shown in
\citet{KD06}. Note that the eccentricity profile is calculated and plotted all the
way into the gap and inner disk region in order to facilitate comparison with the
results from \citet{KD06}. The fast decay of the eccentricity profile with radius
beyond the planet's orbit seems to be in agreement with that of a confined
eccentric mode in the linear secular theory by \citet{GO06}, even though an
eccentric disk is not a good description of a horseshoe flow within the gap (see
Figure \ref{check}). In the exterior disk, the magnitudes of the epicyclic
velocities and the corresponding disk eccentricities are similar between the
$e_p=0$ and 0.1 cases.

\subsection{Protoplanetary Disks}
Now we turn to the specific case for the protoplanetary disk with $a_p=5$ AU.
The values of the gas velocities in the eccentric disk can be obtained by
multiplying the velocities in Figure \ref{fig:v_gas} by $1.33\times 10^6$ cm
s$^{-1}$.

Figure \ref{vrd_8000_5} illustrates the perturbed velocities for the particles with
the size of $1\times10^{-2}$ cm and $100$ cm in the $e_p=0$ case. The figure shows
the $m=1$ distribution due to the eccentric exterior disk. The results for the two
particle sizes are indistinguishable because $\tau_{s,sec}\ll 1$ in
Equation~(\ref{EccEq}). More specifically, $\tau_{s,sec}$ is $1.67 \times
10^{-(6-7)}$ for the 0.01 cm-sized dust particles and
$1.67\times 10^{-(2-3)}$ for the 100 cm-sized dust particles.
Hence the particles smaller than one meter are well coupled to the gas by the
strong gas drag.
It is further validated in Figure \ref{fig:vd-vg}, which shows small fractional
differences ($\ll 1$) between the dust velocities in Figure \ref{vrd_8000_5} and
the gas velocities in the left panels of Figure~\ref{fig:v_gas}.\footnote{The
fractional velocity differences for the meter-sized particles become larger than
0.1 in the narrow black regions (i.e. out of the top scale) in the right panels of
Figure~\ref{fig:vd-vg}. It is because in these regions $u_g$ or $v'_g$ is close to
zero (see Figure~\ref{fig:v_gas}) such that the slight phase difference in the
velocity for the larger particle (i.e. meter-sized) leads to the larger fractional
differences. Hence, the narrow black regions do not indicate weak coupling between
the gas and particles.} Similar results for strong dust-gas coupling are also found
for the $e_p=0.1$ case; the perturbed dust velocities (not shown) are almost
identical to those displayed in the right panels of Figures~\ref{fig:v_gas} and the
values of $\tau_{s,sec}$ are on the same order as those for the $e_p=0$ case.
Figure~\ref{fig:ecc_ep=0} shows that in both the $e_p=0$ and $e_p=0.1$ cases, the
eccentricity profile of the gas disk and that of the well-coupled dust coincide and
decrease outwards from $e=0.08$ at $r=2$. Consequently, as in the case for the gas
disk, the $m=1$ velocity fields for dust particles are more prominent in the region
around the outer edge of the gap where the radial velocity associated with the
eccentric orbit is up to about 744.8 m/s. The forced eccentricity $e_f=|E_f|$ is
also plotted in the right panel of Figure \ref{fig:ecc_ep=0} for comparison with
the case where $e_p$ is non-zero. It is apparent that in the region that our study
focuses on (i.e. $r\geq 2$), the dust particles with size smaller than 1 m are so
well-coupled to the gas that their eccentricity profiles are clearly distinct from
$e_f$, i.e. $E_d\approx E_g$ for $\tau_{s,sec} \ll 1$ from Equation (\ref{EccEq}).
Consequently, all particles smaller than 1 m regress with the gas disk at the same
rate in the steady state.


The time-averaged surface densities for the 0.01 cm- and 100 cm-sized particles are
shown in Figure \ref{den_8000_5}. The $e_p=0$ case is plotted in the left panels
and the $e_p=0.1$ case is presented in the right panels.
By comparison with the bottom panels of Figure \ref{fig:density_g}, we can see that
the $m=1$ distributions of the dust and gas density look similar. In fact, a more
careful comparison indicates that the dust density distributions for all the cases
are almost identical to the gas density distribution, as one would expect for
well-coupled dust particles. The dust distributions are asymmetric in the exterior
eccentric disk.
The location of the azimuthally averaged longitude of pericenter $\varpi_{d,ave}$
is illustrated on the disk by the plus signs, which are almost $180^{\circ}$ apart
from the dust density peak as shown in the figure. It can be easily understood by
inspecting Equations~(\ref{dust_den_pert}) \& (\ref{eq:Sigma_d}). The perturbed
dust density $\Sigma'_{d,0}$ is approximately equal to $\Sigma_{d,0} E_d (d\ln e_d/
d\ln r$), where $d\varpi_d /dr$ has been neglected, as can be validated from Figure
\ref{den_8000_5}. Since $d\ln e_d/ d\ln r <0$ from Figure \ref{fig:ecc_ep=0},
$\Sigma'_{d,0} \propto -E_d.$\footnote{It is worthy noting that since $E_d \propto
-iu'_{d,0}$, it explains why the dust pericenters $\varpi_{d,ave}$ trace the same
radial contour as $u'_{d,0}=0$ in the first quarter of the velocity map as shown in
Figure \ref{vrd_8000_5}.} Hence the phase difference between $\Sigma'_{d,0}$ and
$E_d$ is $180^{\circ}$, implying that the density peak lies more or less at the
location of the apocenter of the disk. The result can be found quite intuitive by inspecting
the divergence of the dust velocity fields shown in Figure \ref{vrd_8000_5}.
As the dust particles move toward
the apocenter (pericenter), the divergence of the velocity fields is negative (positive)
and thus the streamlines
are being compressed (rarefied), therefore
finally reaching the maximum (minimum) surface density at the apocenter (pericenter).



We see from Figure \ref{den_8000_5} that after time averaging, there still exist
residual density fluctuations associated with short-period density waves, as has
already been noted for Figure \ref{fig:density_g} in the paper. Nevertheless, the
residual density is small enough to reveal the density profiles corresponding to
the secular $m=1$ eccentric disk. It is illustrated in Figure \ref{fig:Den_1D},
which shows the background and perturbed density profiles along the polar direction
where the time-averaged dust density $\Sigma'_{d,0}$ is at its maximum.
Corresponding to $\phi=110^\circ$ and $300^\circ$ respectively for the $e_p=0$ and
$0.1$ cases in Figure \ref{den_8000_5}, Figure~\ref{fig:Den_1D} shows that although
the residual waves are present, they are small enough for the identification of the
secular value of $\Sigma'_{d,0}$ from order-of-magnitude estimates.
In terms of the dimensionless units, the background dust surface density
$\Sigma_{d,0}$ is about $10^{-6}$. Figure \ref{fig:Den_1D} indicates that the
excess/deficit of the dust density $\Sigma'_{d,0}$ associated with the $m=1$
eccentric disk is about $10^{-7}$ (i.e. $\sim 0.1\Sigma_{d,0}$).
Other than the secular density perturbations of dust due to the eccentric disk, we
also estimate the local density perturbations of dust due to the tightly wound gas
density waves for comparison. The perturbed gas density of the spiral density waves
$\Sigma'_{g,wave}$ is about $0.1 \Sigma_{g,0}$ from the hydrodynamical simulation.
Given the dust-to-gas ratio 0.01, the perturbed dust density associated with the
density waves is then given by $0.01\Sigma'_{g,wave} \sim 0.001 \Sigma_{g,0} = 0.1
\Sigma_{d,0}$, which is about the same order as the secular density perturbations
of dust $\Sigma'_{d,0}$.


The above 2D dust analysis does not take into account the background radial drift
with the speed described by $u_{d,2}$ in Equation (\ref{eq:u_d2}). The importance
of the radial drift can be evaluated by comparing the radial drift timescale
$t_{drift}\sim r/u_{d,2}$ to the precession time $t_{prec}=2\pi(399/\Omega_p)$ for
$e_p=0$ and $2\pi (152.8/\Omega_p)$ for $e_p=0.1$. Using Equation (\ref{eq:u_d2})
with $u_{g,2}=0$ and $d\ln p_g/d\ln r=-5/4$ in our isothermal model,
we find that $t_{prec}/t_{drift}$ is about $5\times 10^{-(3-4)}$ for 0.01 cm-sized
dust particles and about 0.3-0.5 for 100 cm-sized particles. Meter-sized particles
drift inwards relatively fast at a rate almost comparable to the precession rate
because the dimensionless stopping time $\tau_s \sim 1$ for meter-sized particles
\citep{W77} in contrast to $\tau_s\ll 1$ for 0.01 cm-sized dust particles. The
values of $t_{prec}/t_{drift}$ imply that the particles smaller than 1 m slowly
drift inwards to the adjacent eccentric orbit after a number of orbital regressions
in the protoplanetary disk.

\subsection{Transition Disks}
To study the dust behavior in a ``transition" disk defined simply by a disk with a
low gas density, we apply $a_p = 100$ AU as the new length unit to the
dimensionless simulation results presented in \S4.1. The initial gas density is
then reduced to about $8.89\times 10^{-2}$\ g cm$^{-2}$. The resulting disk accretion rate $\sim
\nu \Sigma$ is diminished by a factor of 90 or so.
The drag force is weakened accordingly due to the low gas density. Hence, the
spatial distribution of the dust particles with different sizes would be
significantly different.

The departures from the Keplerian circular velocity for particles of 3 different
sizes (0.01, 1, and 100 cm) are shown in Figure \ref{fig:vrd_8000_100_ep=0} for
$e_p=0$ and in Figure \ref{fig:vrd_4000_100_ep=0.1} for $e_p=0.1$. In comparison to
the results from the gas velocity map in Figure \ref{fig:v_gas}, particles smaller
than 1 cm are still tightly coupled to the gas but meter-sized particles become
weakly coupled in the transition disk. As a result, the radial velocity of the gas
and that of the well-coupled dust can reach about 167-171 m/s around the outer edge
of the gap, while the radial velocity of meter-sized particles exhibit different
magnitudes; namely, 40-53 m/s for $e_p=0$ and 145-152 m/s for $e_p=0.1$ at the
outer edge of the gap. The larger velocity for the meter-sized particles in the
$e_p=0.1$ case is due to excitation by the planet eccentricity.
There exists a pronounced azimuthal phase difference in the velocities between the
small and large particles.

Figure \ref{fig:den_100} shows the dust density distributions for the 3 different
particle sizes in the transition disk. The dust density distribution is similar to
that for the gas in the cases of the 0.01 cm-sized and 1 cm-sized dust particles.
In both the $e_p=0$ and $e_p=0.1$ cases, the value of $\tau_{s,sec}$ in the exterior
disk is about
$10^{-(3-4)}$ for the 0.01 cm-sized particles, which is 100 times smaller than that
for the 1 cm-sized particles and 4 orders of magnitude smaller than that for 100
cm-sized particles. Meter-sized particles are still in the Epstein regime but
become weakly coupled to the gas due to the low gas density (i.e.
$\tau_{s,sec}>1$).
As a result, the spatial distribution appears to be distinctly different between
small dust grains and meter-sized particles. The location of the pericenter at each
annulus is denoted by the plus sign in Figure \ref{fig:den_100}. As in the case for
the standard disk discussed in \S4.2, the apocenters of the transition disk are
almost aligned. Further, the density excess (deficit) resides around the apocenter
(pericenter) of the dust.
In the $e_p=0$ case, the corresponding phase difference between the 1 cm- and 100
cm-sized particles is about 80$^\circ$, as shown in the left panels of Figure
\ref{fig:den_100}, in agreement with $\arctan (\tau_{s,sec})\approx
\arctan (10)$ as described in \S3.1. Despite the phase difference, particles of
different sizes regress with the gas disk at the same rate in the steady state. In
the $e_p=0.1$ case, the pericenters of meter-sized particles are always close to
the pericenter of the planet even though Figure \ref{fig:den_100} only shows a snap
shot.

The azimuthal averages of dust and gas eccentricities are shown in Figure
\ref{ecc_8000_100} for comparison. In the exterior disk for $e_p=0$ (left panel),
the eccentricity of the weakly coupled particles marked in red (1 m) is smaller
than that of well-coupled grains marked in blue (0.01 cm) and green (1 cm). In
contrast, in the case of $e_p=0.1$ (right panel), the eccentricity of meter-sized
particles become larger than that of the smaller dust in the region
that our study focuses on (i.e. $r\geq 2$); the $e_d$ of the meter-sized particles
is close to $e_f$, while the well-coupled dust (i.e. sizes smaller than 1 cm) still
lies on the same eccentric orbits of the gas and thus regress with the gas disk.
Because $\tau_{s,sec} \geq 1$ for meter-sized particles and Figure
\ref{ecc_8000_100} shows that $e_f
> e_g $, we have $\tau_{s,sec}e_f > e_g$. Thus, the orbits of the large particles librate instead
of regressing with the gas. The libration rate equals the regression rate, as noted
in \S3.1. Additionally, at $r\gtrsim 2$, the value of $\tau_{s.sec}$ of meter-sized
particles is the largest. It explains why the pericenters of meter-sized particles
at such distances are close to the planetary pericenter.

As the dust particles become less coupled to the gas, the drag force wields less
influence on the orbital motions of them. Therefore, the eccentric mode for dust
excited by aerodynamic drag becomes less pronounced.
It follows that in the absence of $e_p$, the weakly coupled particles are in
near-Keplerian circular motions and thus their spatial distribution is close to
axi-symmetric as manifested in the lower color contrast in the lower left panel of Figure
\ref{fig:den_100}. This leads to the lower density contrast for the meter-sized
particles along the azimuthal direction of the disk. In the dimensionless unit, the
density contrast for meter-sized particles is only about 5$\times 10^{-8}$ ($\sim
0.05\Sigma_{d,0}$), which is only a fraction of the density contrast $10^{-7}$ ($\sim
0.1\Sigma_{d,0}$) for smaller particles.
In the case of $e_p=0.1$,
the density contrast for the asymmetric distribution of meter-sized particles shown
in the bottom right panel of Figure \ref{fig:den_100} is also about $5\times
10^{-8}$ ($\sim 0.05\Sigma_{d,0}$), despite the fact that their orbits are actually
more eccentrically forced by the planetary eccentricity. This is because the eccentric
gradient $de_d/dr$ is smaller, as shown in the right panel of Figure
\ref{ecc_8000_100}, leading to less compressive streamlines near the apocenter for
meter-sized particles.

Finally we estimate the effect of the radial drift of the dust particles on their
secular eccentric motion in the transition disk. We find that $t_{prec}/t_{drift}$
is on the order of $10^{-(0-2)}$ for the 0.01 cm-sized dust particles and $10^{-3}$
for the 100-cm particles. The ratio $t_{prec}/t_{drift} \sim 1$ for the 0.01
cm-sized dust only occurs near the location of $r=2$ where $\tau_s \sim 1$ and thus
the radial drift is fast enough \citep{W77} to be comparable to the precession
timescale. In contrast, the meter-sized dust drifts relatively slowly in the
transition disk because their $\tau_s \gg 1$. Overall, in most of the region of the
exterior disk, the ratio $t_{pres}/t_{drift}$ is small. This means that as in the
case of the protoplanetary disk, particles smaller than 1 m drift slowly inwards to
the adjacent eccentric orbit after a number of precessions in the transition disk.

\section{SUMMARY AND DISCUSSIONS}
We investigate the secular behavior of dust particles in a 2D eccentric
protoplanetary disk harboring a massive planet. One fiducial model is considered
with the following parameters: $M_\star=1\ M_\odot$, $M_p/M_\star\approx 5\times
10^{-3}$, $H/r=0.05$, $a_p=5$ AU, $\Sigma_g(t=0)=10^{-4}\times M_*/a_p^2=35.6$
g/cm$^2$, $\nu=10^{-5}\times a_p^2\Omega_p$, and the dust-to-gas ratio of the
unperturbed disk is 0.01 everywhere.

The disk is assumed to be non-self-gravitating. We apply
the fiducial model to the planet on a fixed circular and on a fixed eccentric orbit
with $e_p=0.1$.
We employ the FARGO code to extract the gas velocity and density from the 2-D
hydrodynamical simulation, and then apply the eccentricity equation with the gas drag
in the Epstein regime to study the dust dynamics. The simulations are performed for
a sufficient length of time for the gas disk to attain quasi-steady states. To reveal the secular
features of the disk, we average the gas properties over almost one planetary orbit
in order to eliminate most of the short-period features associated with the spiral
density waves driven by the orbital motion of the planet. Because the disk interior
to the planet remains nearly circular in the model and the gas motions inside the
gap deviate considerably from simple eccentric motions, we focus our study on the
eccentric disk almost exterior to the gap, i.e. $2\leq r/a_p \leq 4$. The gap of
the gas disk opened up by the planet on the eccentric orbit is slightly wider but a
little shallower than that by the planet on the circular orbit. The gas disk
regresses at the rate of about $(1/399)\Omega_p$ in the $e_p=0$ case and about
$(1/152.8)\Omega_p$ in the $e_p=0.1$ case.

The most important parameter for determining the gas-dust coupling on the secular
timescale in an eccentric disk is $\tau_{s,sec}$, the secular stopping time defined
as the ratio of the stopping time to the precession timescale for a free particle.
In our fiducial model for both the $e_p=0$ and $e_p=0.1$ cases, the spatial density
distribution and velocity fields of dust particles of size smaller than 1 m are
almost identical to those of the gas component due to the strong coupling through
the gas drag (i.e. $\tau_{s,sec}\ll 1$). The perturbed velocities, defined as the
velocity departures from the Keplerian circular orbit, are non-axisymmetric and
exhibit the $m=1$ azimuthal profile with the phase difference of 90 degrees between
the perturbed radial and azimuthal velocities as a consequence of the eccentric
disk. The density also exhibits the $m=1$ distribution, with a $180^\circ$ phase
difference relative to the perturbed azimuthal velocity. As a result, the density
excess (deficit) lies around the apocenter (periceter) of the disk, with a
magnitude of about 10\% of the background dust density. Because the eccentricity
decreases outwards from $e\approx 0.8$ at the outer edge of the gap, the $m=1$
structure is more prominent around the outer edge of the gap region. The perturbed
velocity in this region is about 744.8 m/s.

We also apply the fiducial model to the transition disk defined in our work loosely
by a protoplanetary disk with the overall low gas density equal to $8.89\times
10^{-2}$ g cm$^{-2}$. This is achieved by simply using the dimensionless results
from the same simulation but with the planet placed at $a_p=100$ AU as the new
length unit. In this low-density disk,  meter-sized particles become weakly coupled
to the gas (i.e. $\tau_{s,sec} >1$), while smaller particles still move closely
with the gas on eccentric orbits (i.e. $\tau_{s,sec}\ll 1$). The velocity
departures from the Keplerian circular motion for small dust particles can be as
high as 167-171 m/s, and for meter-sized particles only reaches about 40-53 m/s in
the $e_p=0$ case but can be excited to about 145-152 m/s in the $e_p=0.1$ case.
Consequently in the $e_p=0$ case, meter-sized particles move in near-Keplerian
circular orbits, which, in the steady state, regress at the same rate as the orbits
of smaller dust particles but lag behind by about $80^\circ$. In the case of
$e_p=0.1$, the orbits of the meter-sized particles are more eccentrically excited
by the planetary eccentricity, with the pericenters roughly aligned with the
pericenter of the planet's orbit. The orbits of the meter-sized particles do not
regress with the gas disk but librate around the location close to the pericenter
of the planet. The $m=1$ density contrast resulting from the eccentric orbits is
about 5\% for meter-sized particles and about 10\% for the smaller particles.

The presence of gas drag also causes radial drift of the dust. In our fiducial
model for the protoplanetary and transition disks, the radial drift timescale is
smaller than the precession timescale of dust particles smaller than 1 m in most
regions of the eccentric disk.
Therefore, the particles under consideration do not migrate inwards noticeably
during the time for a few disk regressions. However, we anticipate that on the even
longer timescale the dust eventually drifts to the outer edge of the gap, as in the
case of a circular disk. Thus, the gap can act as a filter to stall the radial
drift of large particles as proposed by \citet{Riceet06}. We should also note that
in our model for the transition disk, the slow radial drift of particles results
from $\tau_s \gg 1$; namely, the gas becomes so tenuous that the particles are
almost decoupled from the gas in one orbit. When taking into account any long-term
disk evolution, it is conceivable that before the disk gradually loses mass and
evolves to the transition disk, $\tau_s$ of dust particles of certain sizes is
about 1. As a result, the eccentric disk almost loses dust particles of particular
sizes before it becomes a transition disk. This happens to some extend in the
simulation by \citet{Fouchet10}, where cm-sized particles are almost depleted in
most part of the disk by radial drift as the disk evolves. By comparison, our model
focuses on a particular stage with particles of all sizes present in the disk from
$r=2$-4$a_p$, implying that we have assumed that dust particles are continuously
replenished from the outer boundary. The evidence of gas replenishment to a
protoplanetary disk at late stages has been suggested by the observation of the AB
Aurigae disk \citep{Tang12}.

Although the simulated gas disk in our model is not turbulent, the large kinematic
viscosity $\nu$ is normally attributed to a turbulent flow. When the particles are
well coupled to the gas such that $\tau_s \ll 1$, turbulent diffusion may be
effective in smearing out the density inhomogeneity of dust. In a 2D disk, the rate
for the turbulent diffusion is proportional to the divergence of $\Sigma_g
(\nu/{\rm Sc}) \nabla (\Sigma_d/\Sigma_g)$ \citep[e.g.][]{TL02}, where Sc is the
Schmidt number given approximately by $1+\tau_s^2$ \citep{YL07}. In our model,
where a constant and uniform dust-to-gas ratio in the basic state is assumed,
$\nabla (\Sigma_d/\Sigma_g)$ is almost zero for well coupled dust particles and
thus the effect of turbulent diffusion is negligible. For meter-sized particles
sufficiently decoupled from the gas in the transition disk, $\tau_s \gg 1$ and the
timescale for the turbulent diffusion $\sim 10^5 \tau_s^2 (r/a_p)^2/\Omega_p$,
which is much larger than the regression/libration time $399/\Omega_p$ and
$152.8/\Omega_p$.
Therefore, turbulent diffusion is not important for the secular effects of our
analysis.


In a circular disk, well-coupled dust particles can be temporarily trapped in the
peaks of spiral density waves and more permanently in the Lindblad resonances
\citep{PM06}. The eccentric disk harboring a planet of 5 Jupiter masses simulated
by \citet{Fouchet10} exhibited a great concentration of 1 cm-sized dust particles
in the strong spiral wave just outside of the gap. In this work, we limit ourselves
to the secular evolution of dust. Hence, whether these transient and resonant
effects for dust trapping can also occur in an eccentric gaseous disk is not
addressed.

\citet{MK12} simulated eccentric circumstellar disks in binary star systems and
investigated how different physical parameters affect disk evolutions. They showed
that the gas eccentricity of a realistic radiative disk is smaller than that of a
locally isothermal disk. As a result, the dust eccentricity is expected to be
smaller than that in our results, and hence the asymmetric dust distribution in an
eccentric disk would present more of a challenge for detection.


Other than the aforementioned effects that are ignored, several phenomena have also
been left out in this work.  In reality, dust particles can settle toward the
mid-plane and influence the disk gas. The efficiency of the dust settling depends
also on the stopping time, which evolves with the gas disk \citep{Fouchet10}. The
resulting thin, dense disk for dust with its relative motion to the gas leads to
the streaming instability, which clumps the dust \citep{YG05}. We are unable to
consider in our 2D model these effects involving vertical motions. In our model,
dust of various sizes are assumed to be uniformly distributed after the planet-disk
system settles into a quasi-steady state. We simply assume that the dust-to-gas
ratio is so small that dust motions are passively influenced by the gas motions
without any feedback to the gas dynamics. Another limitation of our model is that
dust growth/fragmentation have not been taken into account. Collisions among large
particles can produce smaller particles, which have the effect of smearing
 out the distinct phase
differences of the asymmetric features between well and weakly coupled particles.
Orbital crossing between strongly and weakly coupled dust particles surely enhances
dust collisions and hence facilitates dust growth/fragmentation. Without
considering these effects, the dust size distribution is thus not modelled in our
study. Nevertheless, our work focussing only on secular dynamics of dust presents a
framework for future studies to explore more complicated gas-dust dynamics in an
eccentric disk.

Having described the limitations of our analysis, our results may render a basic
picture for ALMA (Atacama Large Millimeter/submillimer Array) and EVLA (Karl G.
Jansky Very Large Array) observations. The short-period features, such as the
tightly wound density waves, probably present formidable challenges to be
observationally resolved, but the secular features discussed in this study are more
large-scale, comparable to the gap size, and thus more promising detection-wise.
There exists a phase correlation for the $m=1$ structure between the gas velocity
fields and dust density distribution in an eccentric disk with an embedded massive
planet. The velocity departure from the Keplerian circular motion in our model can
be larger than 150 m/s, which is probably high enough for ALMA detectability. In
addition, the maximum dust emissions should be aligned with the apocenter of the
eccentric gap in an eccentric disk. We show that particles with sizes of 1 m and 1
cm may be distributed differently in the azimuthal direction of an eccentric disk
with a low gas density. Although large particles may be broken down into small dust
particles, perhaps an attempt to measure the spectral energy distribution with EVLA
in different azimuthal directions covering the wavelengths from 1 to tens of cm and
to ascertain whether the dust sizes depend on the azimuthal angle of a transition
disk with an eccentric gap is feasible.

\begin{figure}\center
\scalebox{0.8}{\includegraphics[bb=0.72in 5in 7.75in 10in]{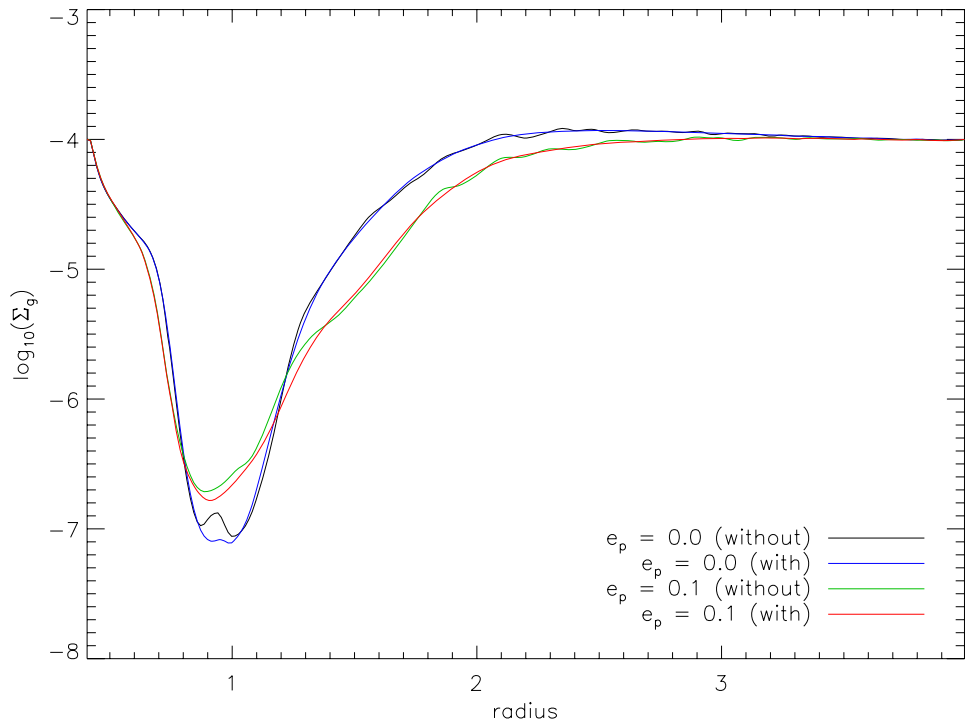}} \caption{The azimuthal average of the gas densities shown in
Figure \ref{fig:density_g}. The gap in the $e_p = 0.1$ case is wider but shallower
than that in the $e_p=0$ case. The label ``with (without)" denotes the azimuthal
average with (without) time average. } \label{fig:DenAve_vs_ep}
\end{figure}
\begin{figure}
  \centering
\begin{tabular}{cc}
  \scalebox{0.5}{\includegraphics[bb=0.72in 5in 7.75in 10in]{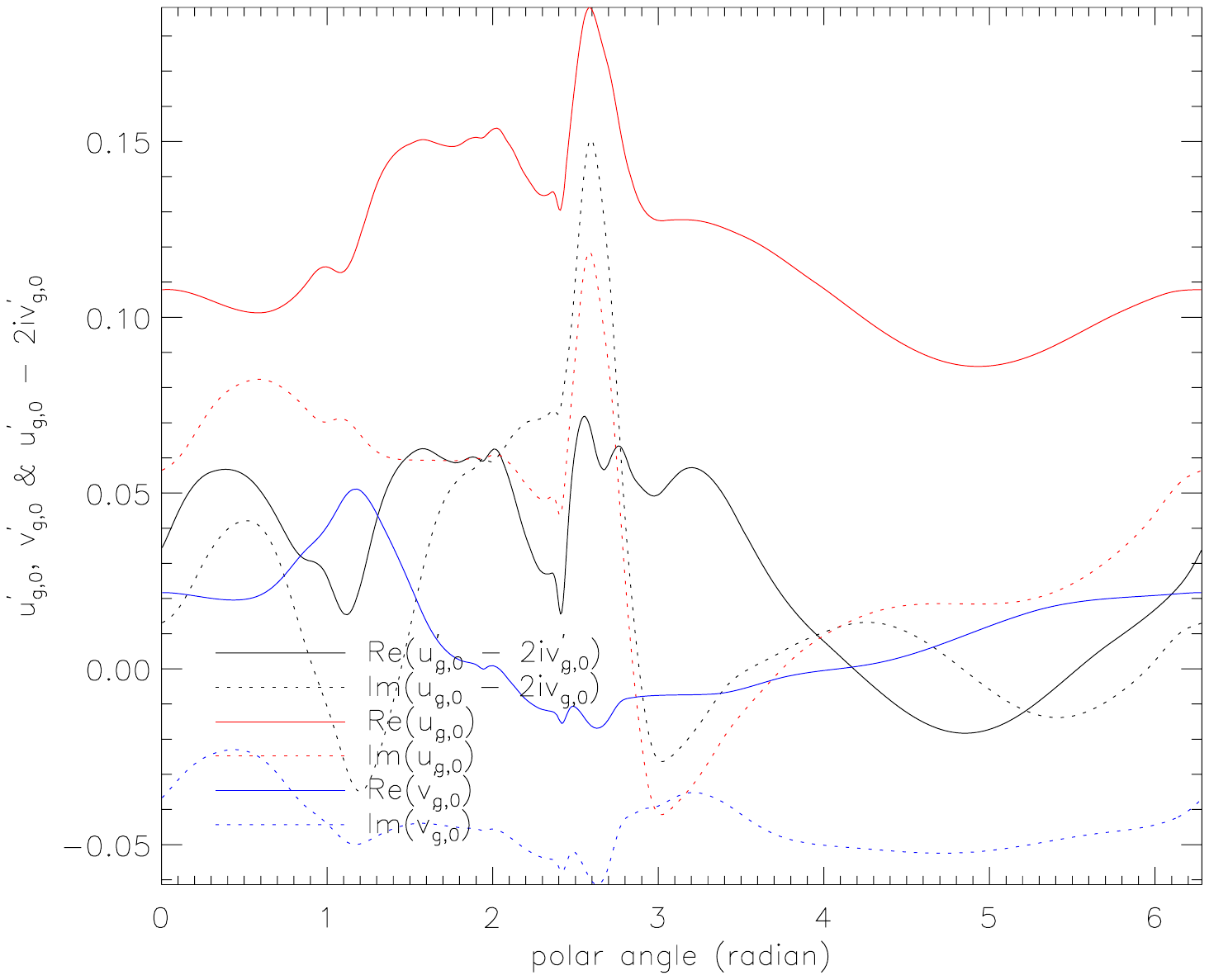}} &
 \scalebox{0.5}{\includegraphics[bb=0.72in 5in 7.75in 10in]{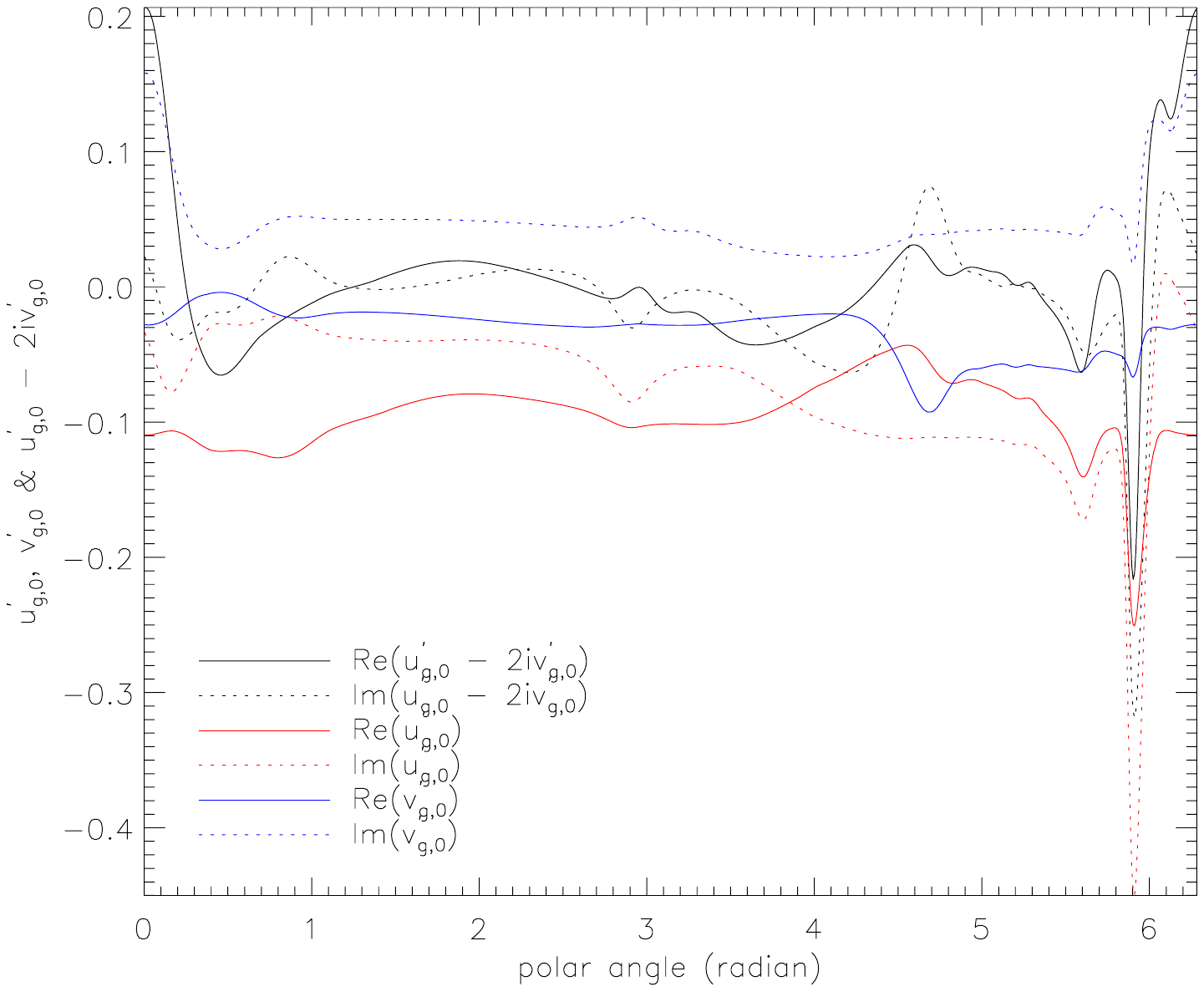}} \\
  \scalebox{0.5}{\includegraphics[bb=0.72in 5in 7.75in 10in]{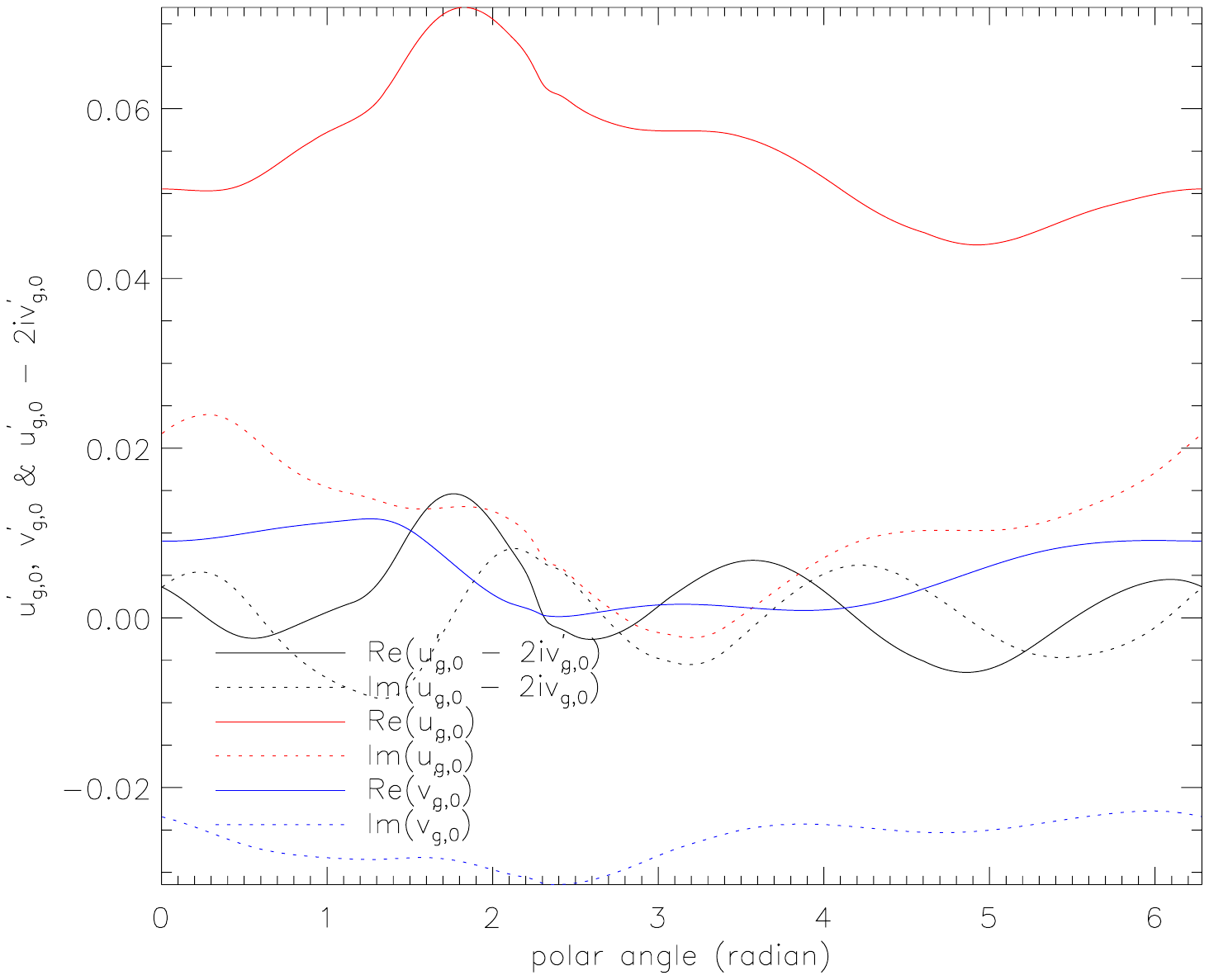}} &
   \scalebox{0.5}{\includegraphics[bb=0.72in 5in 7.75in 10in]{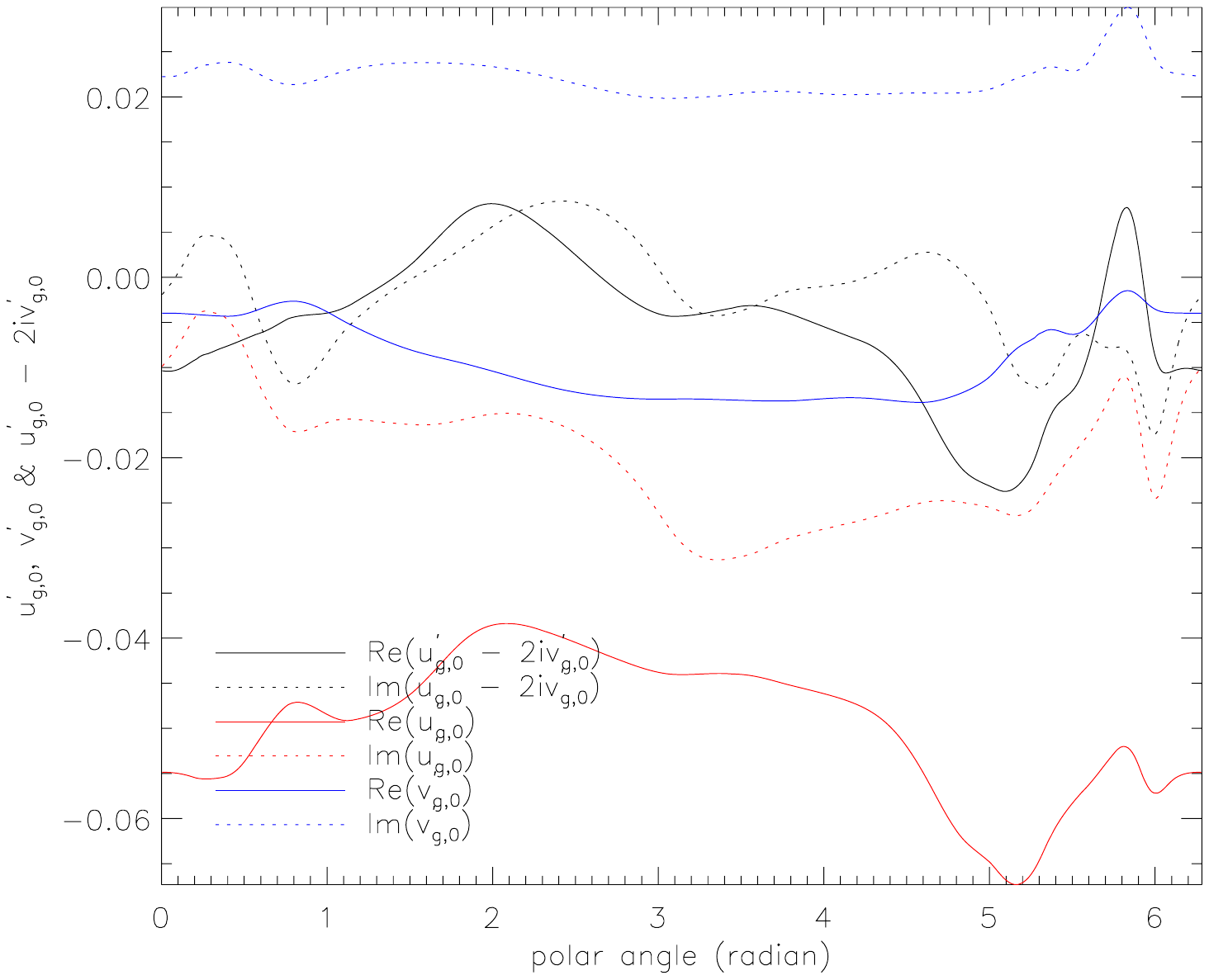}} \\
  \scalebox{0.5}{\includegraphics[bb=0.72in 5in 7.75in 10in]{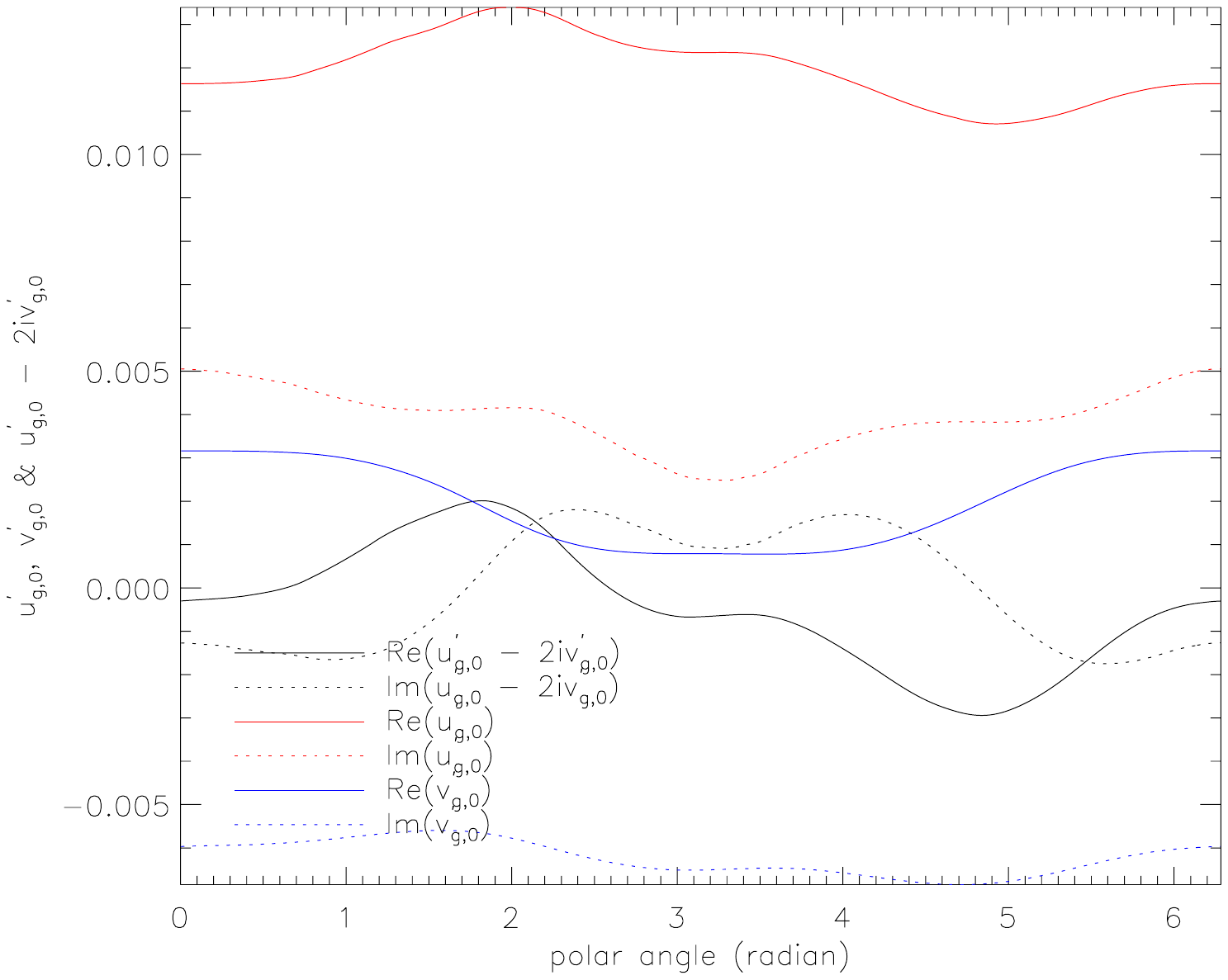}} &
   \scalebox{0.5}{\includegraphics[bb=0.72in 5in 7.75in 10in]{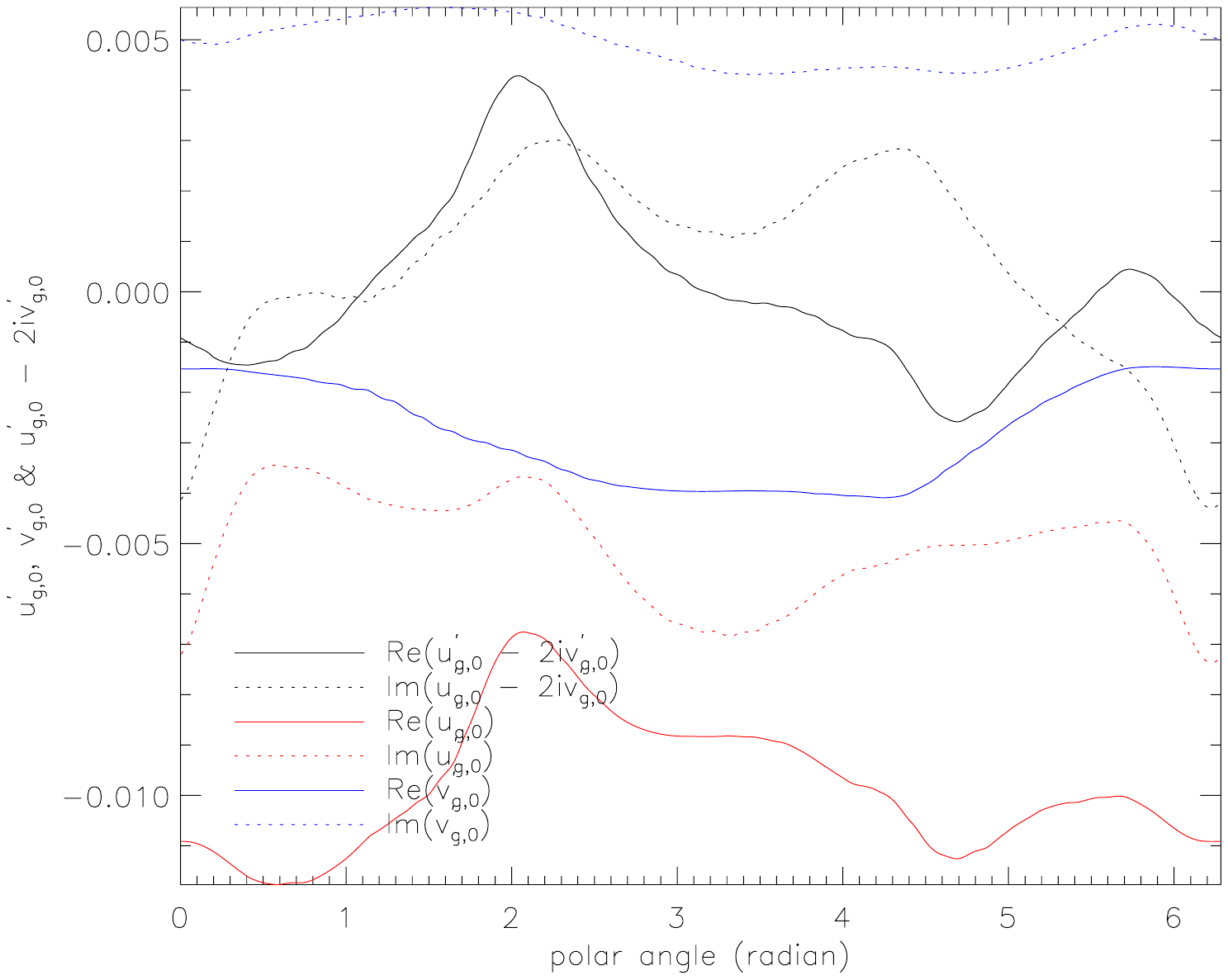}}
  \end{tabular}
  \caption{The results of verifying whether the relation $u_{g,0}' = 2iv_{g,0}'$ is satisfied
at $r = 1.5$ (top), $r = 2$ (middle) and $r = 3$ (bottom) for the cases of $e_p=0$
(left) and 0.1 (right).
           The real and imaginary parts are denoted by solid and dotted lines.
           The red lines represent the values of $u_{g,0}'$ and the blue lines are those of $v_{g,0}'$.
           The values of the relation are shown in black.}
  \label{check}
\end{figure}
\begin{figure}
\centering
\begin{tabular}{cc}
\scalebox{0.5}{\includegraphics[bb=0.72in 5in 7.75in 10in]{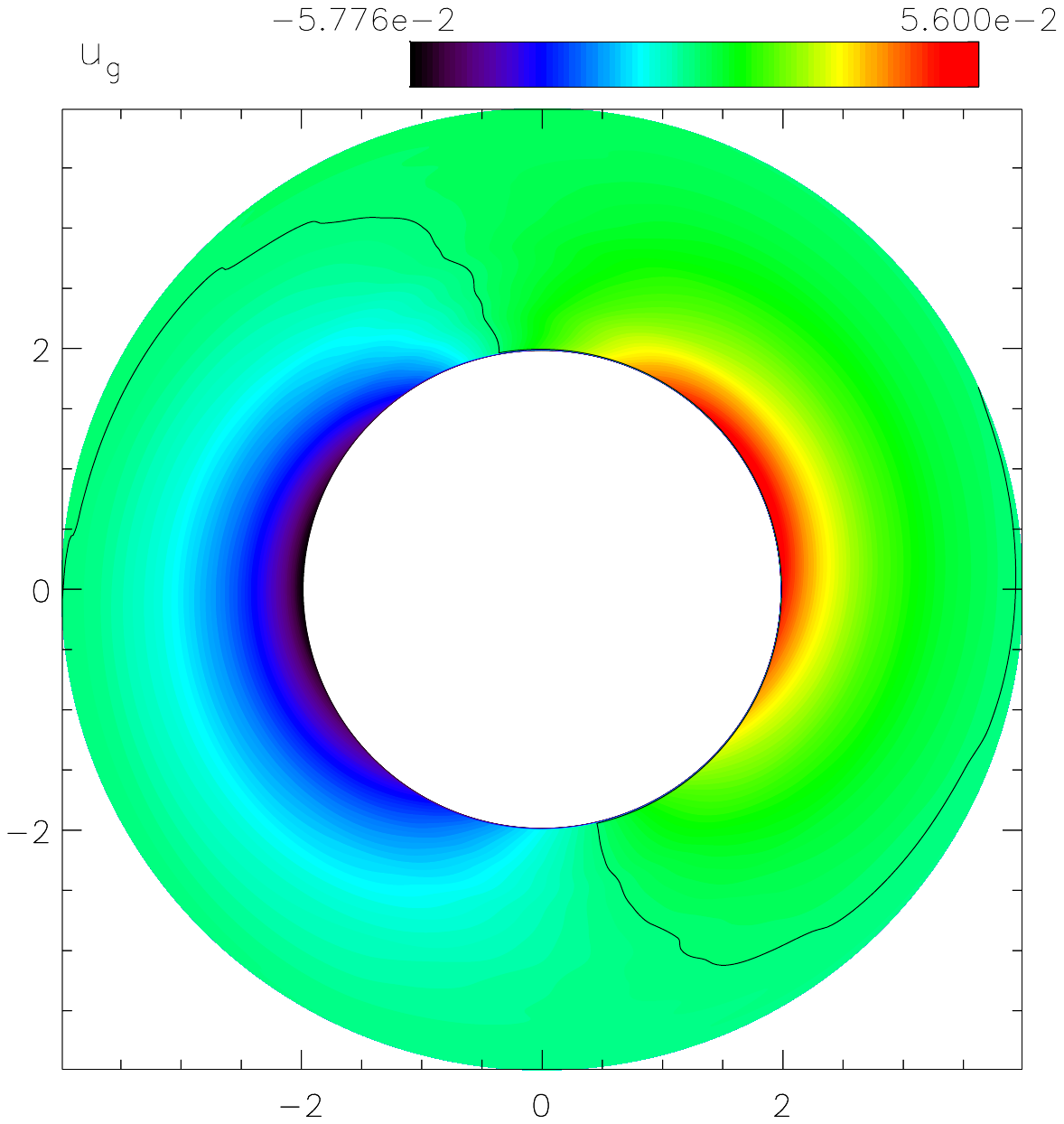}}&
\scalebox{0.5}{\includegraphics[bb=0.72in 5in 7.75in 10in]{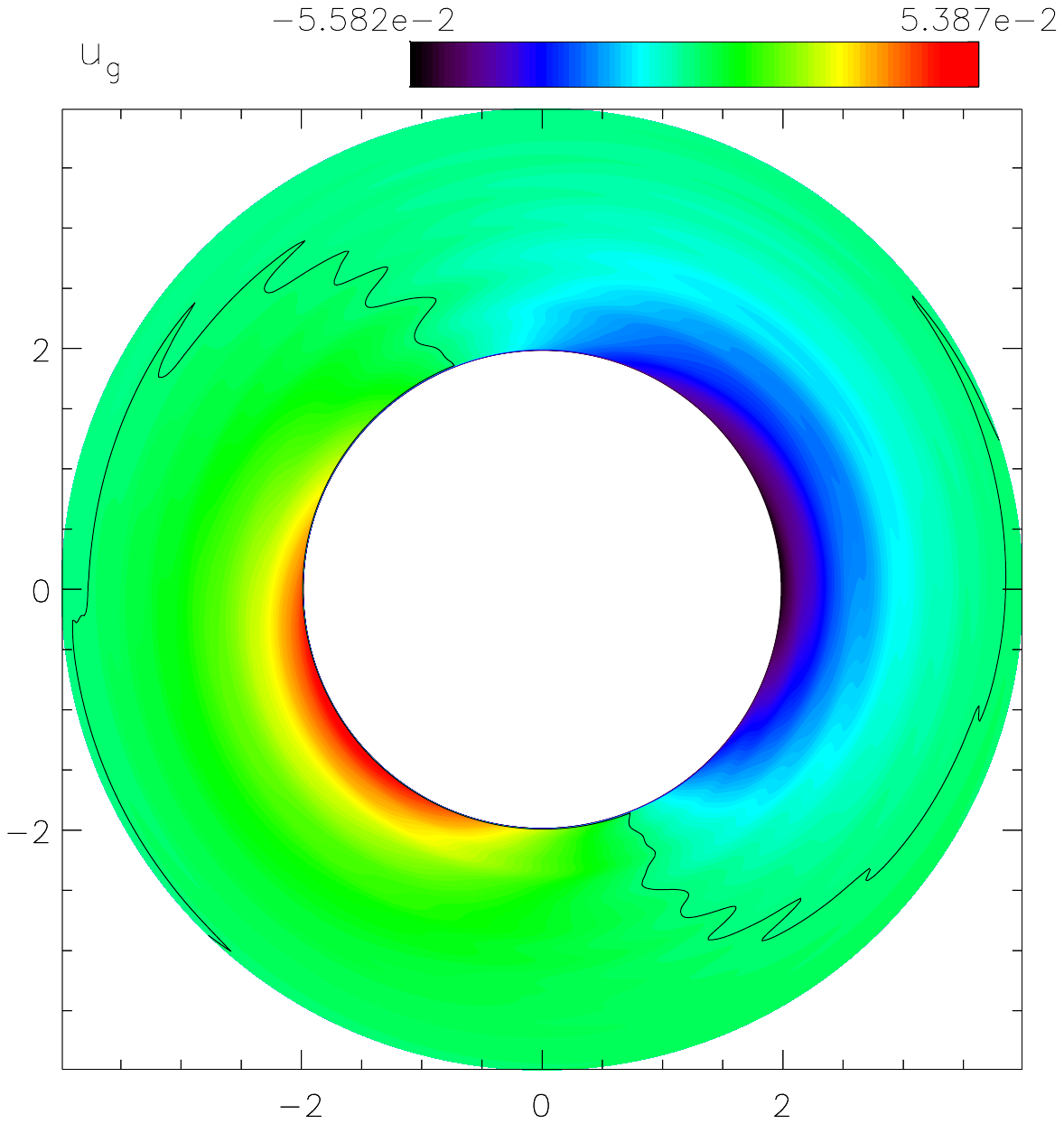}}\\
\scalebox{0.5}{\includegraphics[bb=0.72in 5in 7.75in 10in]{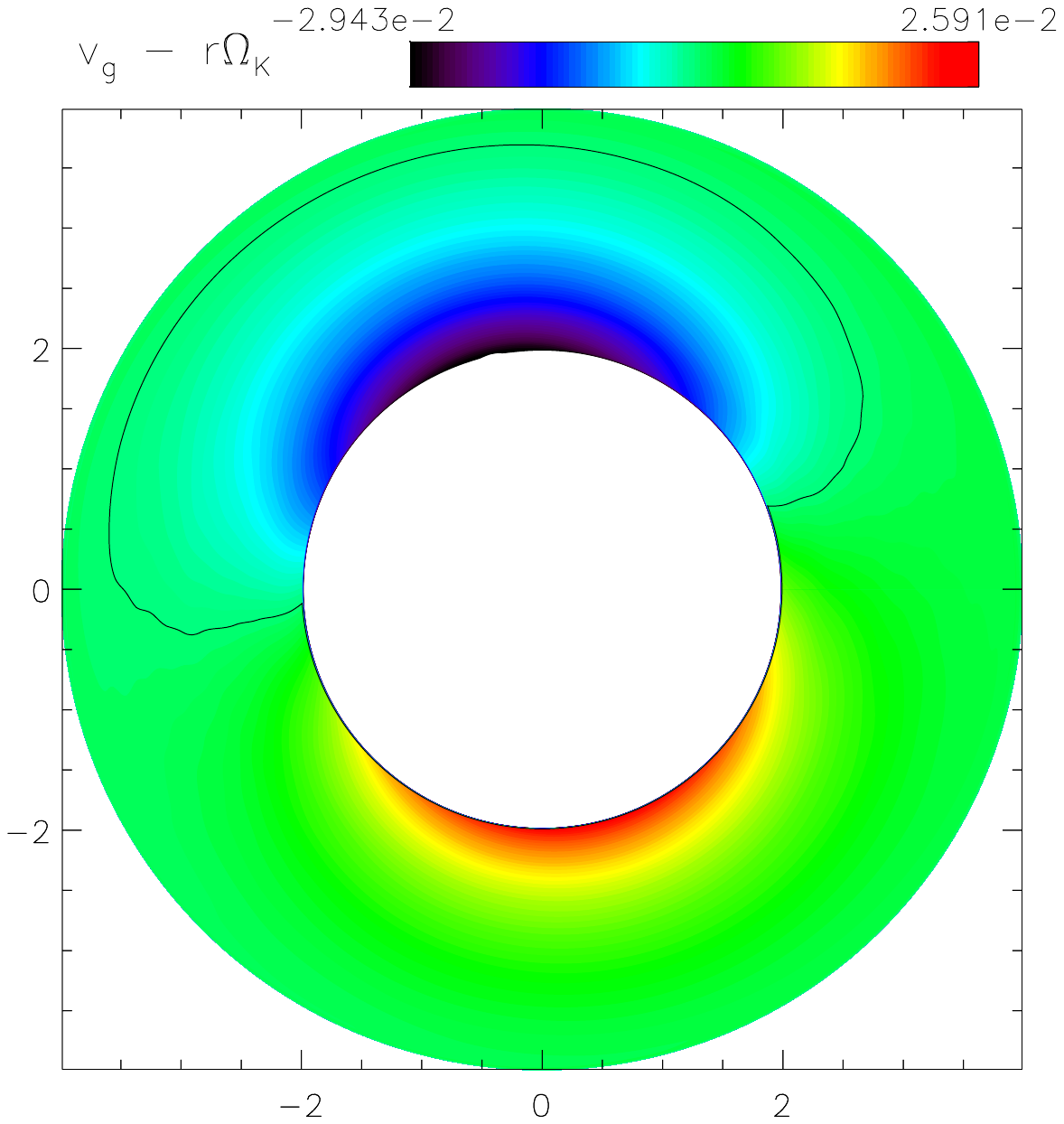}} &
\scalebox{0.5}{\includegraphics[bb=0.72in 5in 7.75in 10in]{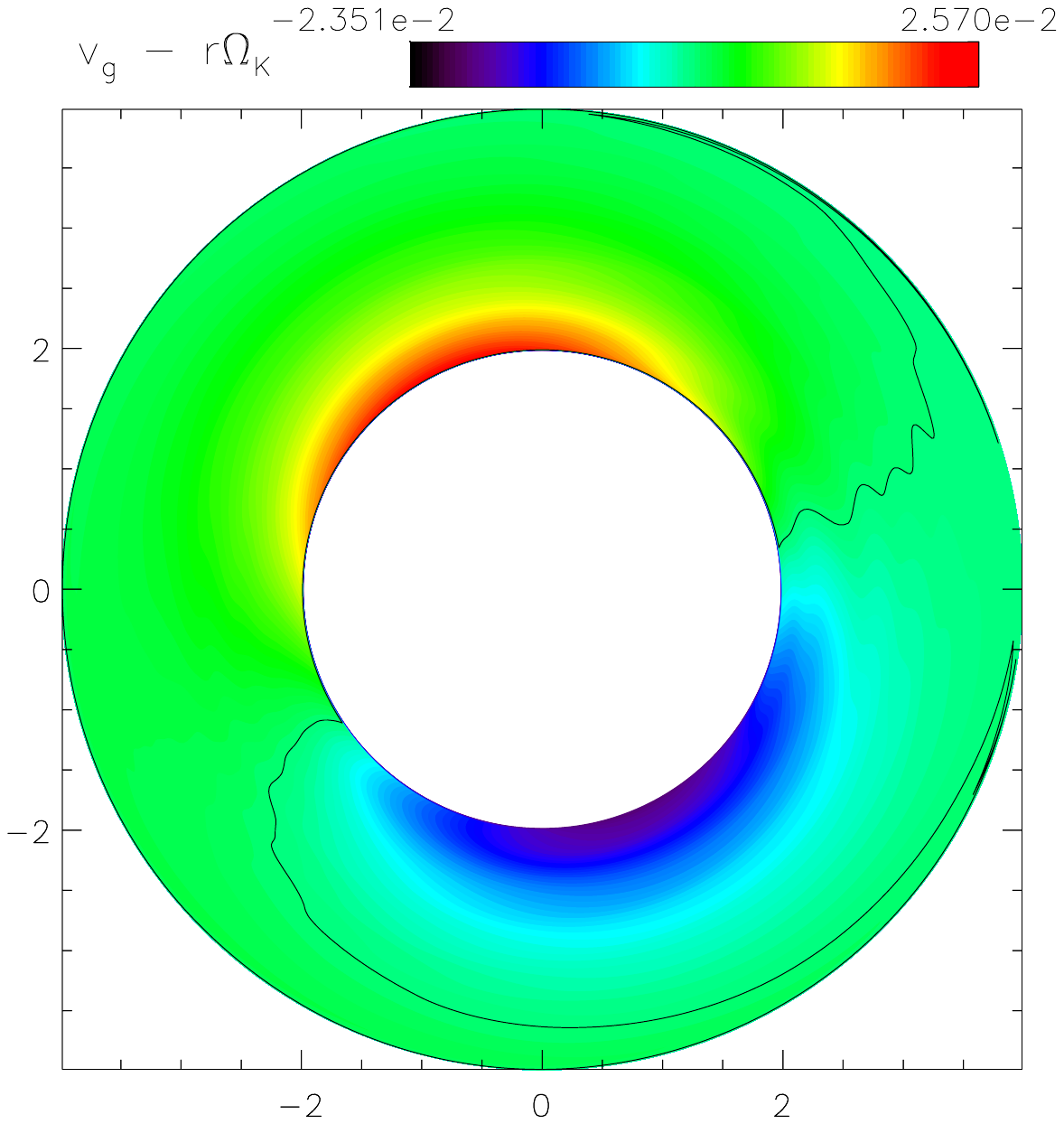}}
\end{tabular}
 \caption{Gas velocity fields in the exterior disk associated with the density map
 shown in the time averaged bottom panels of
Figure~\ref{fig:density_g}. The left panels show the radial gas velocity $u_g
\approx u_{g,0}'$ (top) and the azimuthal gas velocity departure from the Keplerian
circular velocity $v_g'=v_g-r\Omega_K\approx v_{g,0}'$ (bottom) for the $e_p=0$
case. The right panels display $u_g$ and $v_g'$ for the $e_p=0.1$ case. The zero
values are denoted by the black curves.} \label{fig:v_gas}
\end{figure}
\begin{figure}
\begin{tabular}{cc}
  \scalebox{0.42}{\includegraphics[bb=0.72in 5in 7.75in 10in]{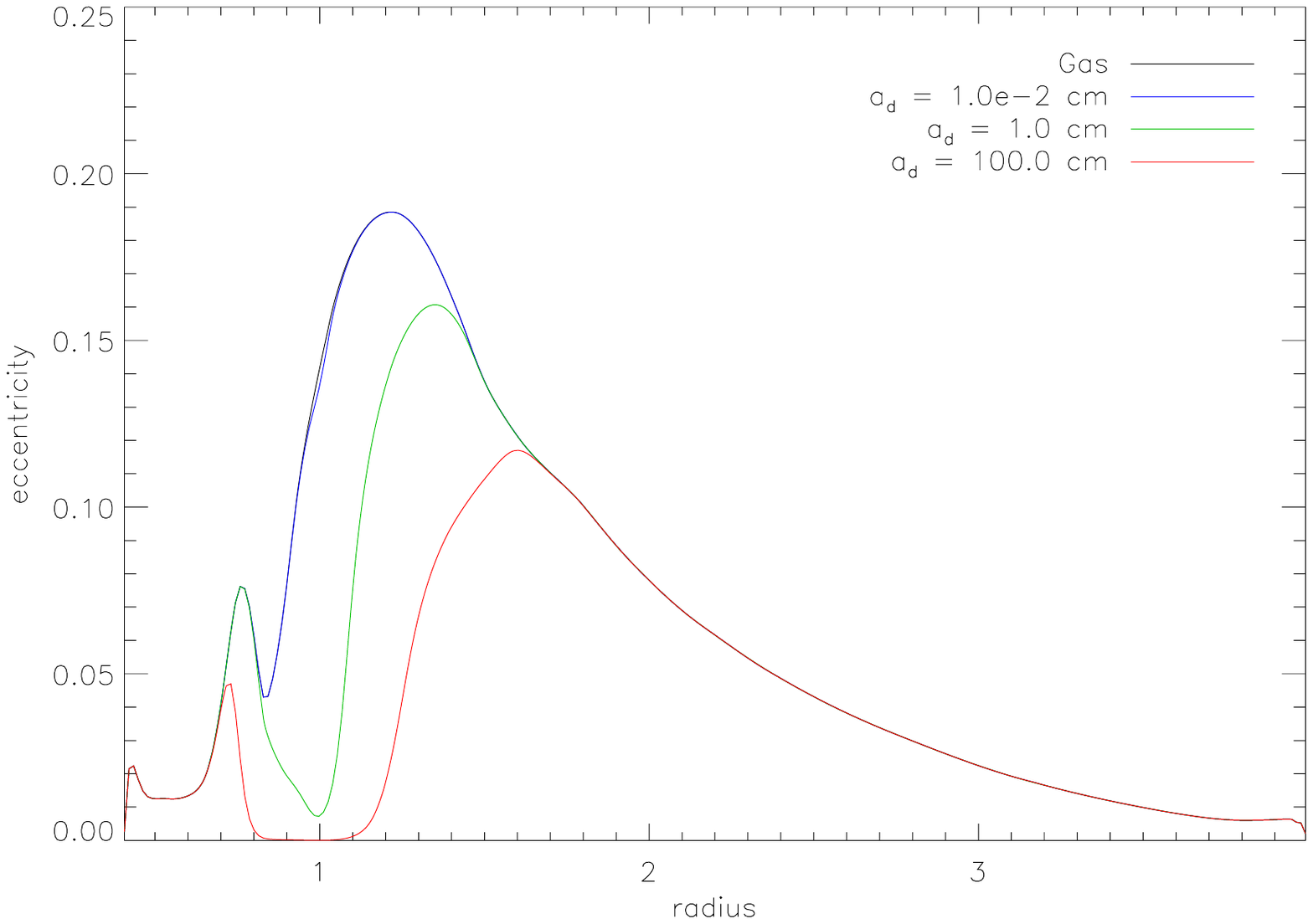}} &
 \scalebox{0.42}{\includegraphics[bb=0.72in 5in 7.75in 10in]{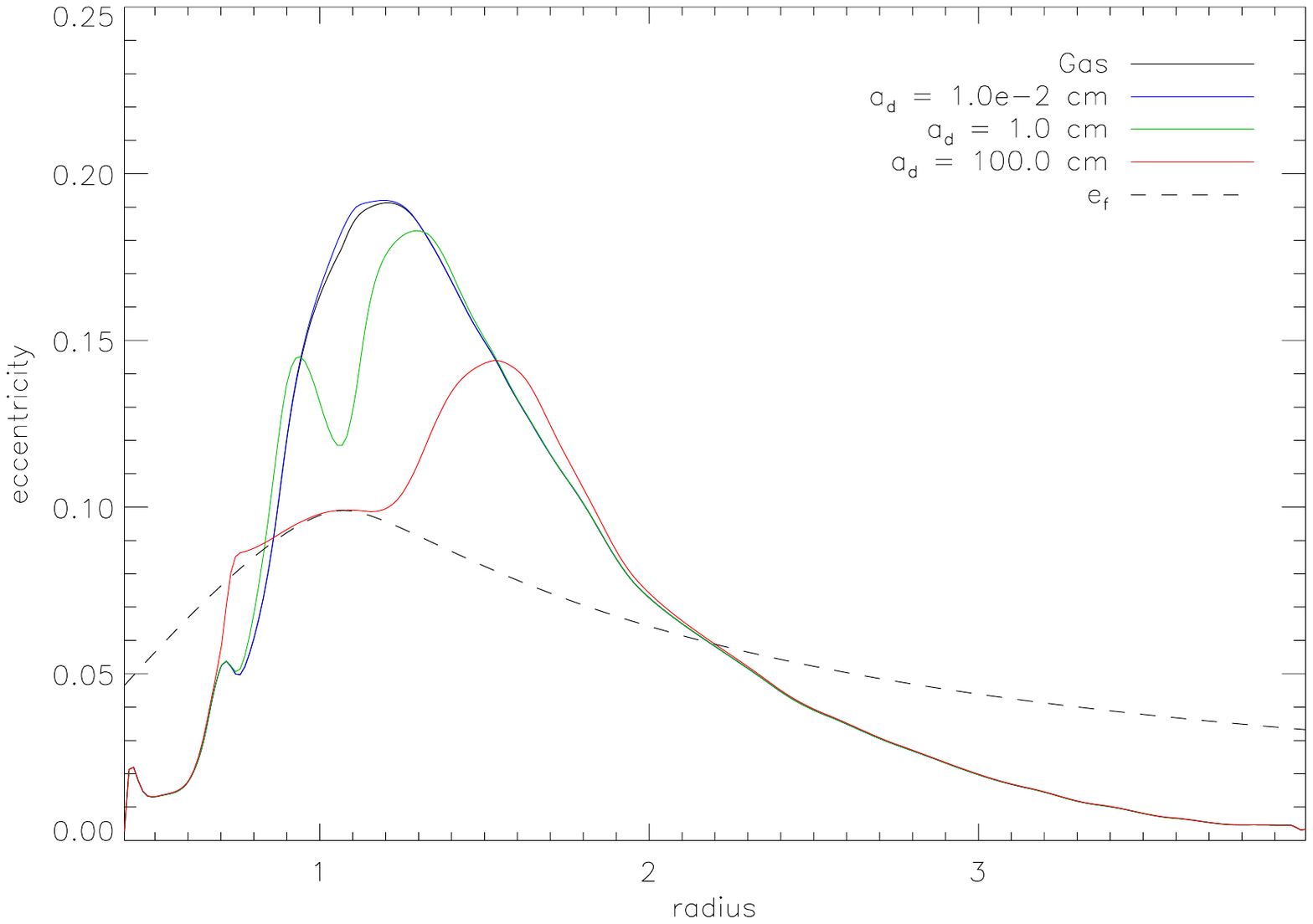}} \\
  \end{tabular}
\caption{Azimuthally averaged eccentricities of gas and dust particles of various sizes in
the cases of $e_p=0$ (left panel) and $e_p=0.1$ (right panel). The forced
eccentricity $e_f=|E_f|$ is also plotted for the $e_p=0.1$ case.}
\label{fig:ecc_ep=0}
\end{figure}
\begin{figure}
\centering
\begin{tabular}{cc}
\scalebox{0.5}{\includegraphics[bb=0.72in 5in 7.75in 10in]{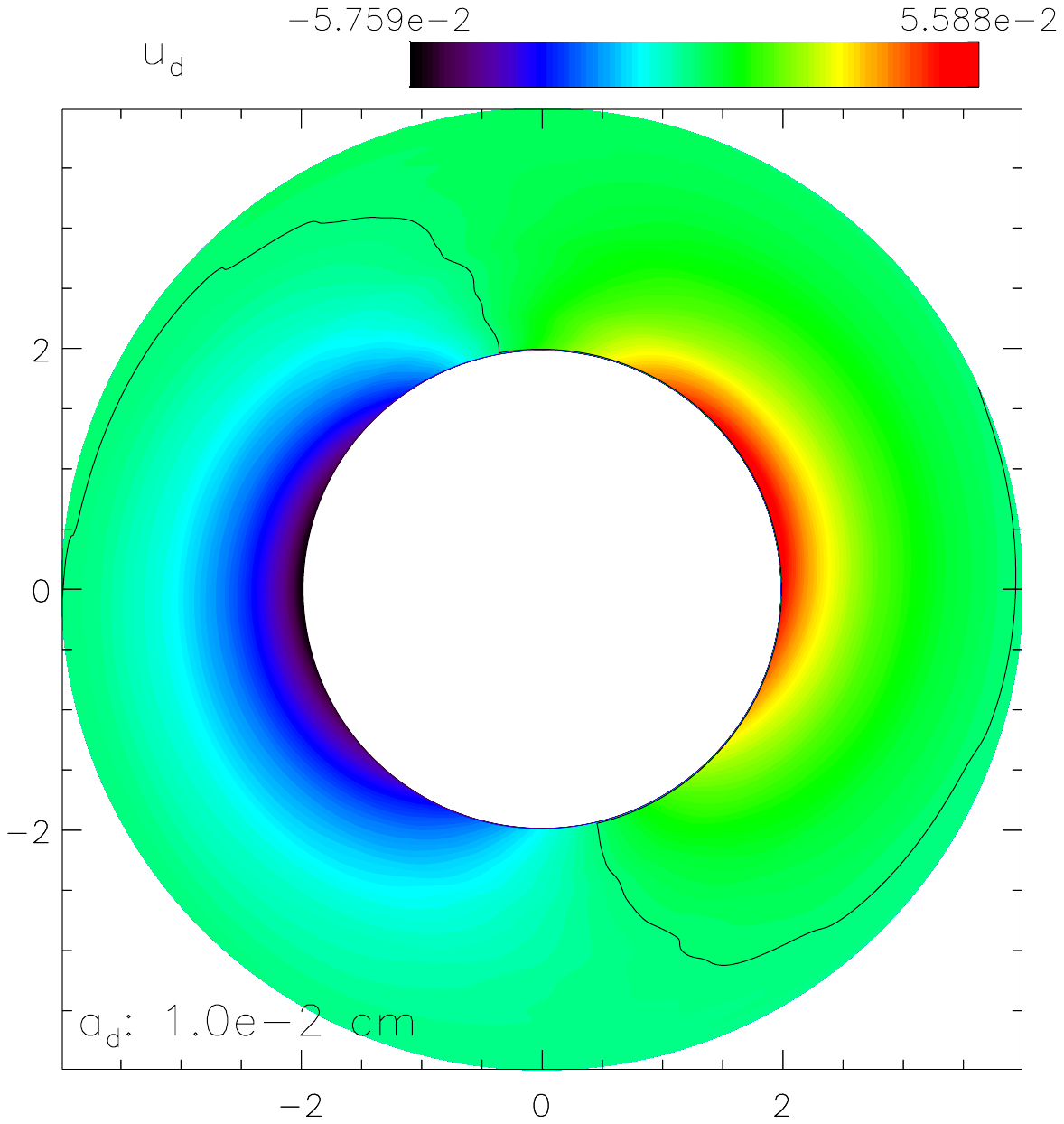}}
&
\scalebox{0.5}{\includegraphics[bb=0.72in 5in 7.75in 10in]{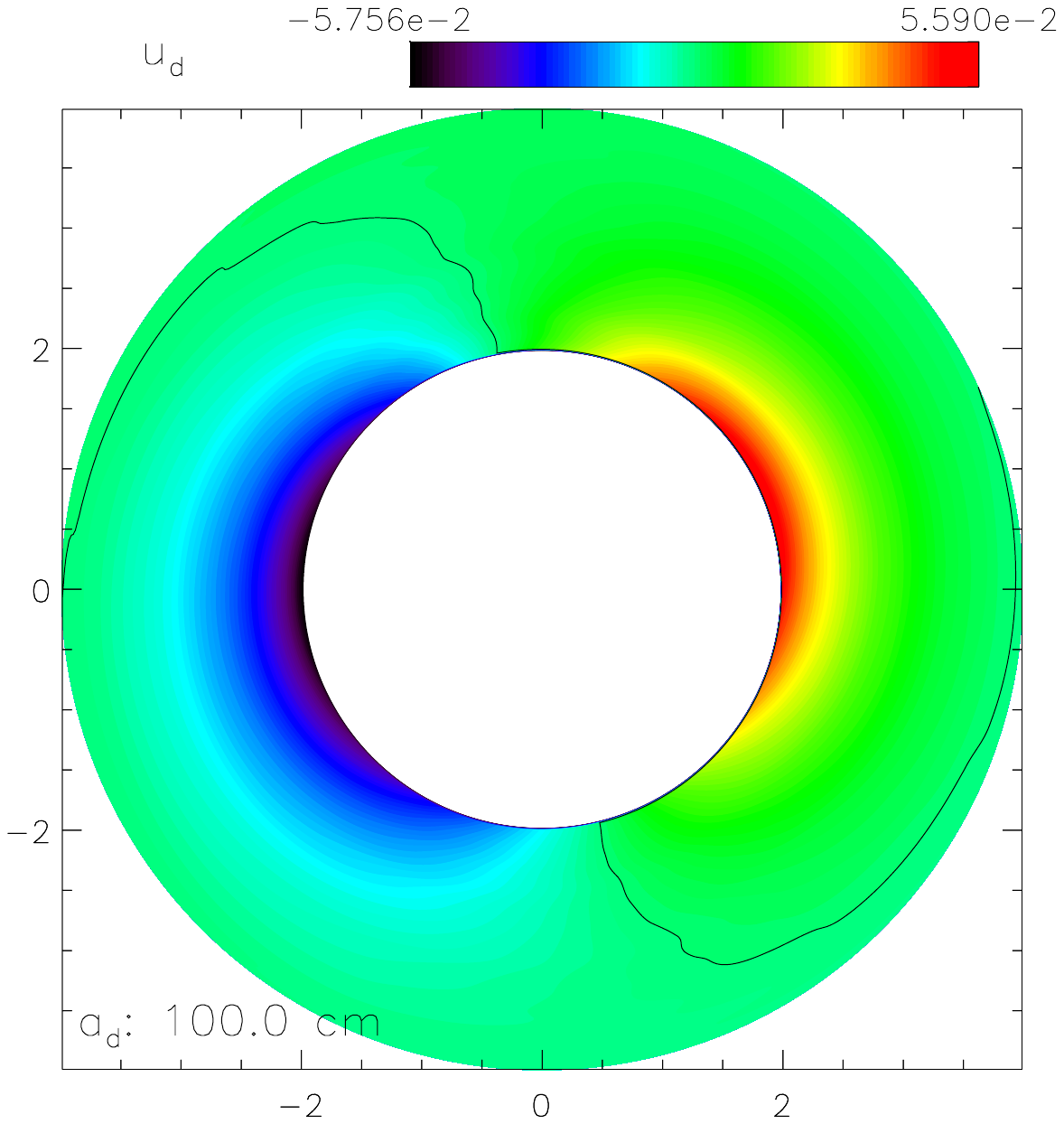}}\\
\scalebox{0.5}{\includegraphics[bb=0.72in 5in 7.75in 10in]{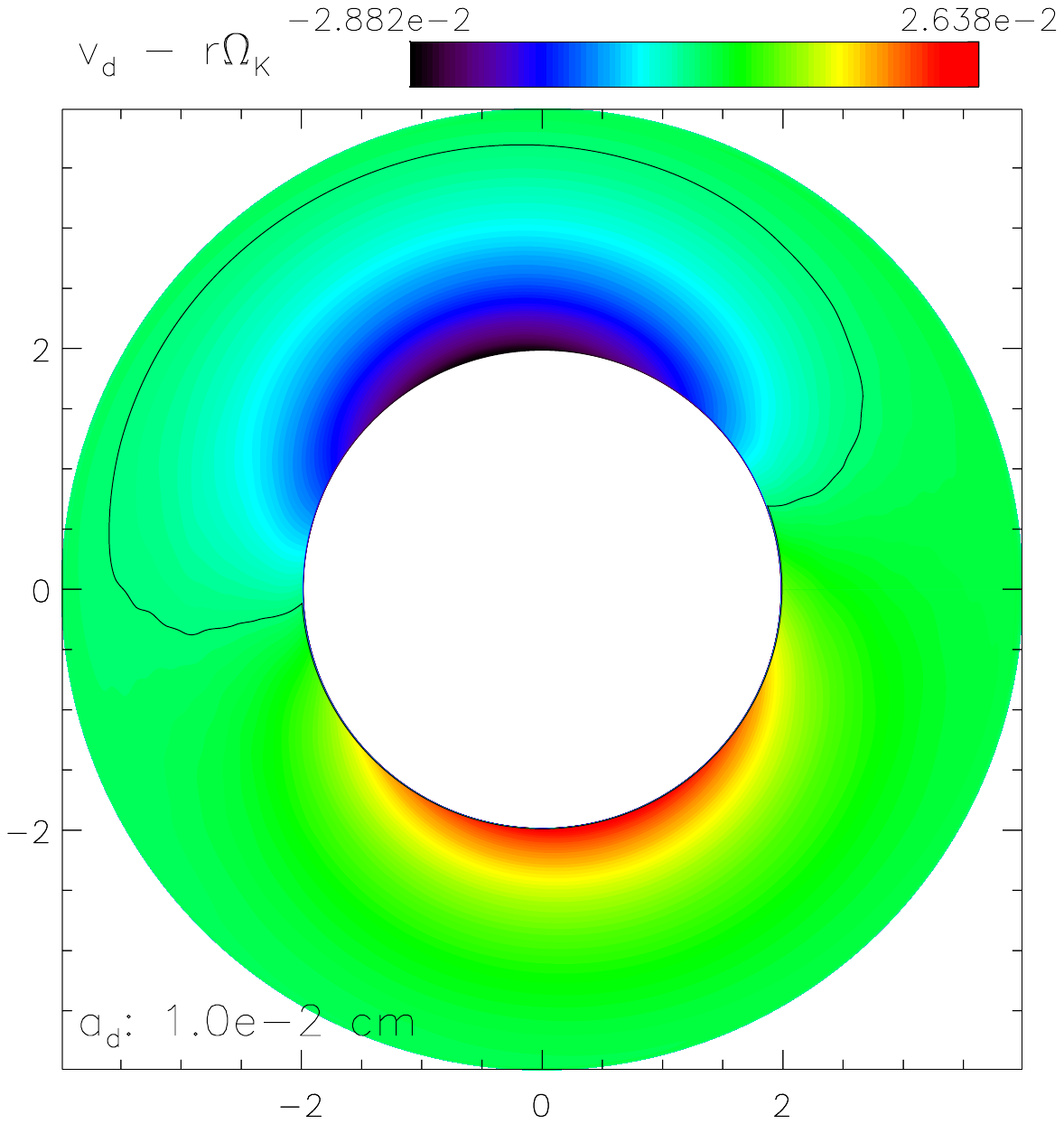}}&
\scalebox{0.5}{\includegraphics[bb=0.72in 5in 7.75in 10in]{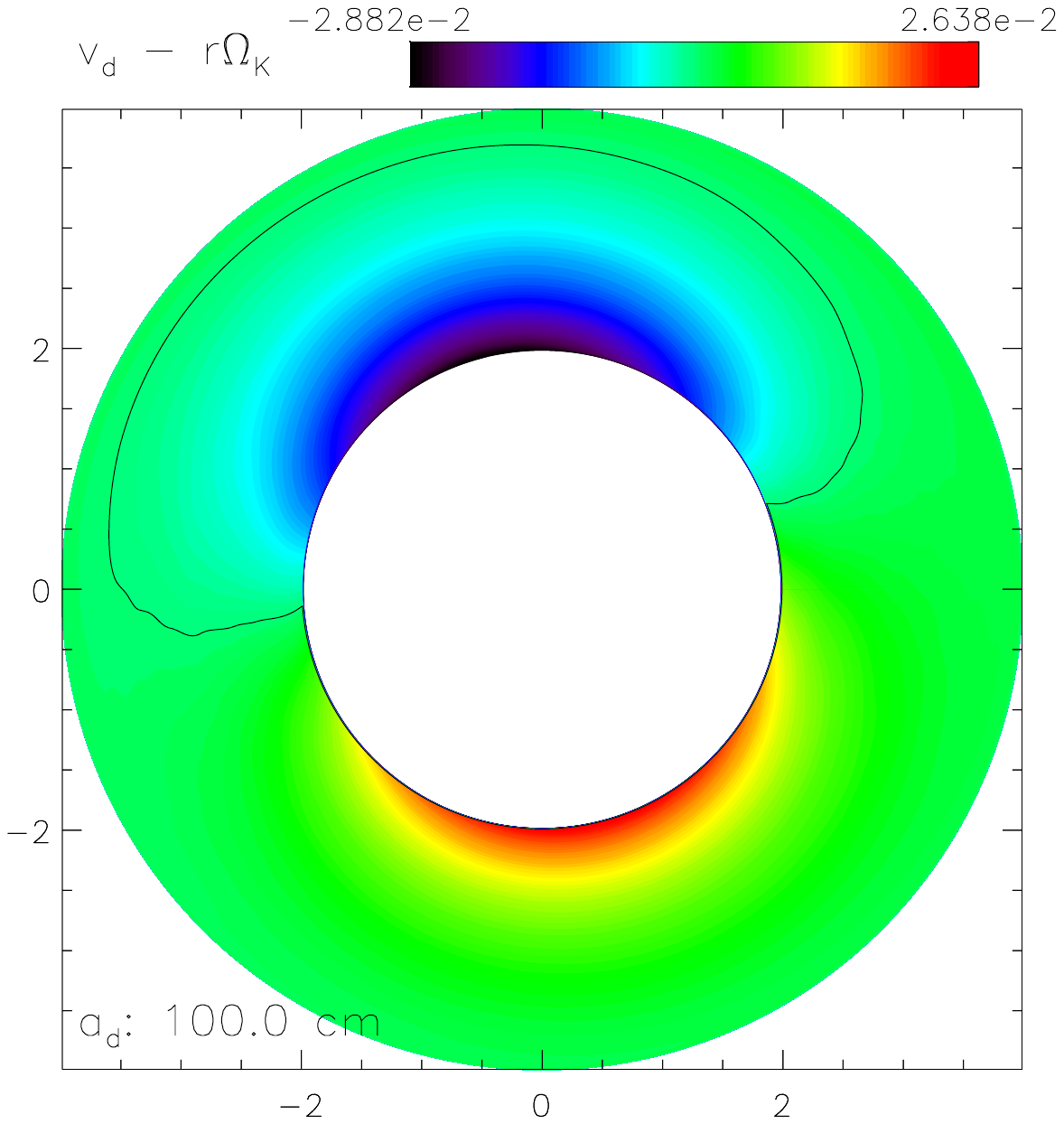}}
\end{tabular}
  \caption{Radial (top panels) and azimuthal (bottom panels) dust velocities departure from the Keplerian circular
  velocity for the $e_p=0$ case. The left panels show the results for size of $1\times 10^{-2}$ cm and the right panels
  show the results for $1$ m.
  The black lines are the zero contours.
  The unit of the velocity is about $1.33\times10^6$ cm s$^{-1}$.}
  \label{vrd_8000_5}
\end{figure}
\begin{figure}
\centering
\begin{tabular}{cc}
\scalebox{0.5}{\includegraphics[bb=0.72in 5in 7.75in 10in]{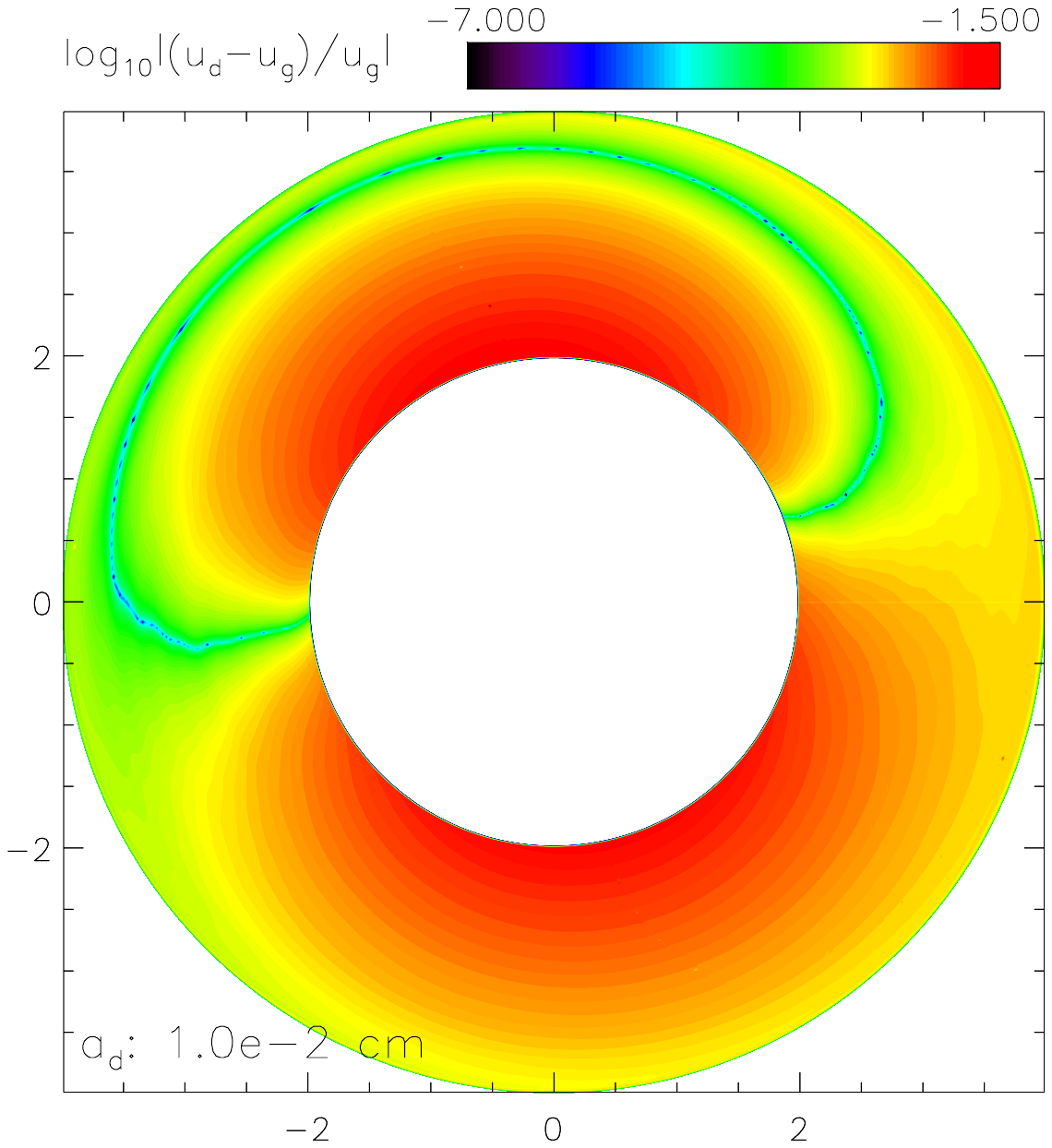}} &
\scalebox{0.5}{\includegraphics[bb=0.72in 5in 7.75in 10in]{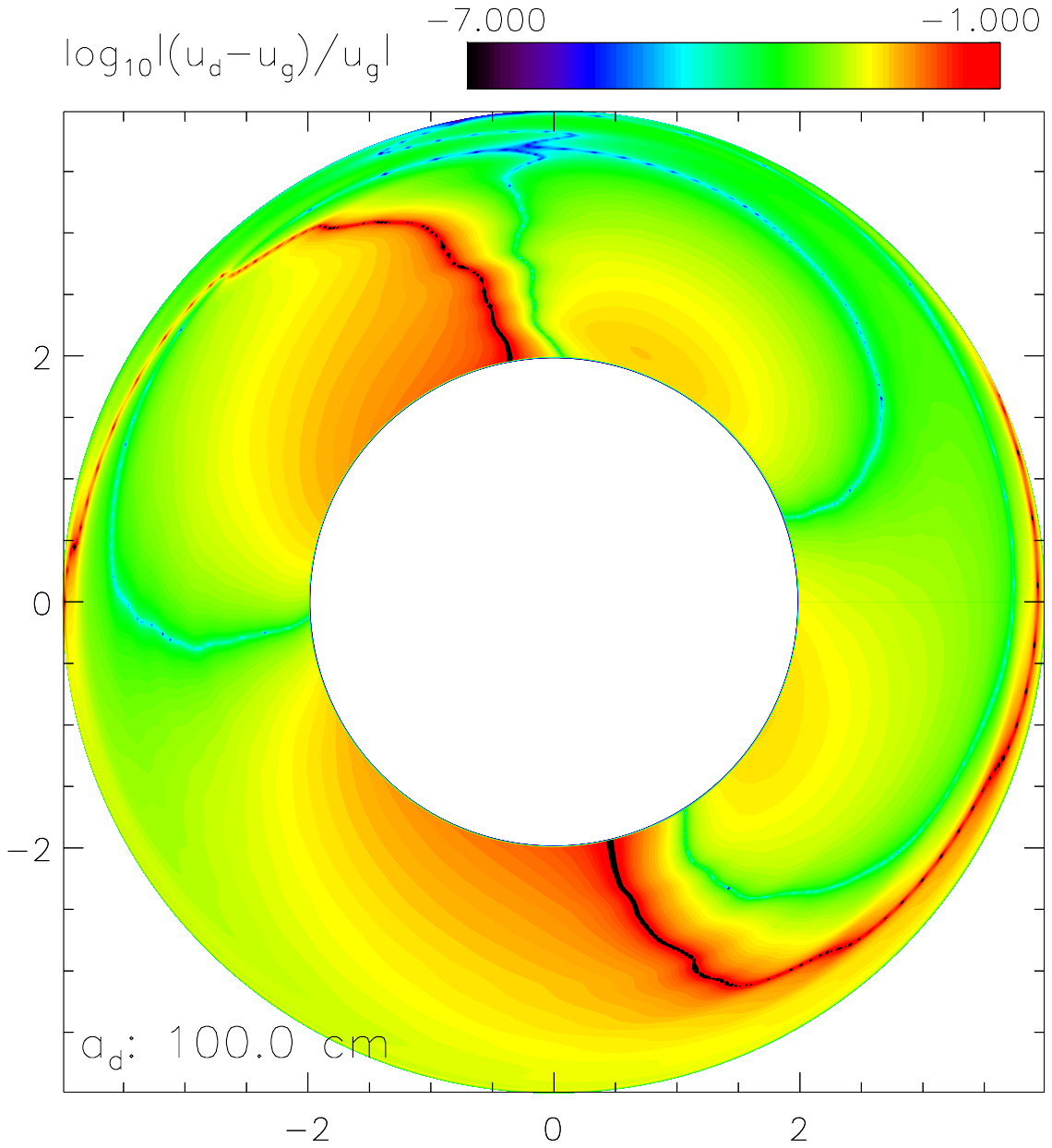}}\\
\scalebox{0.5}{\includegraphics[bb=0.72in 5in 7.75in 10in]{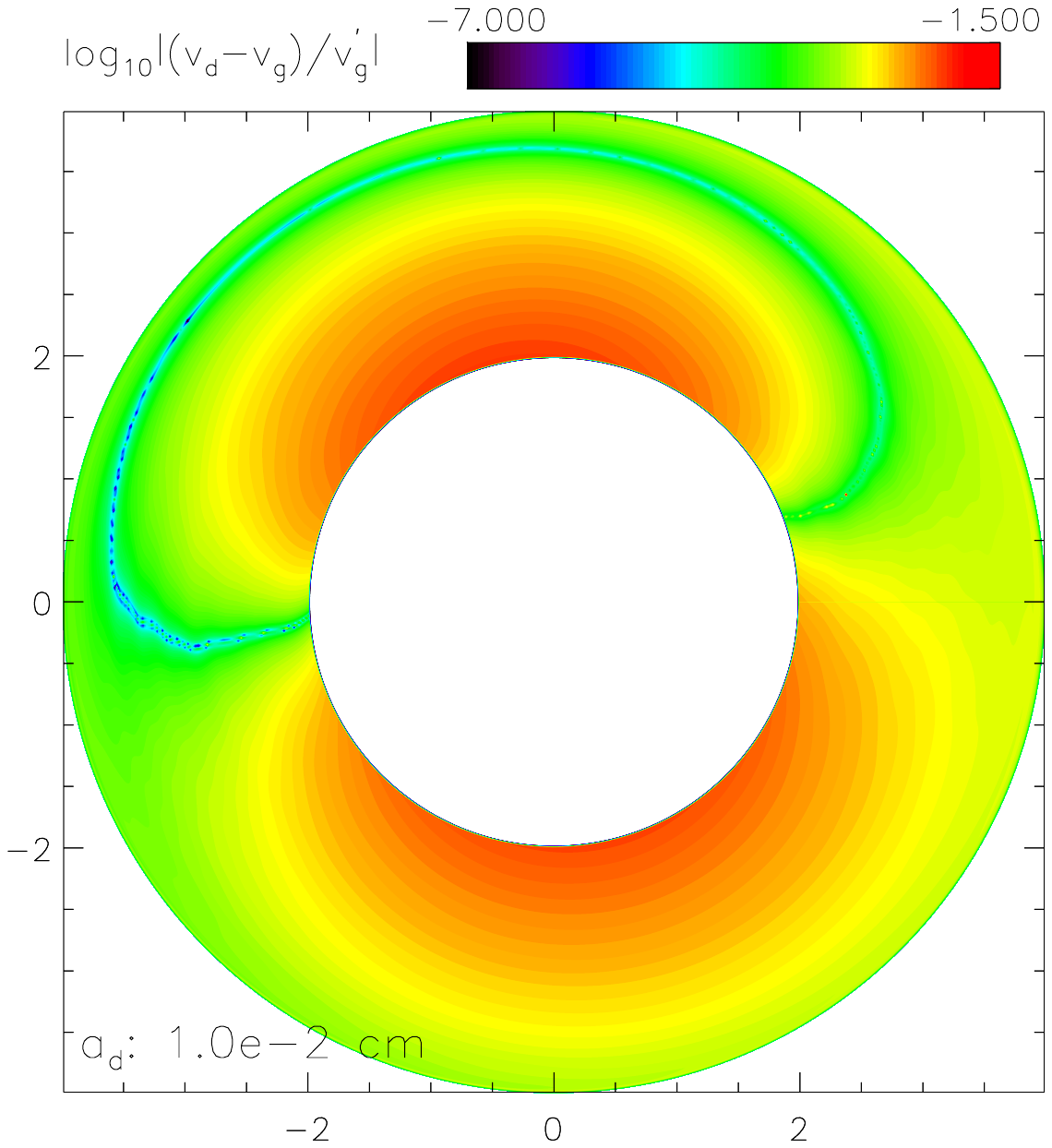}}&
\scalebox{0.5}{\includegraphics[bb=0.72in 5in 7.75in 10in]{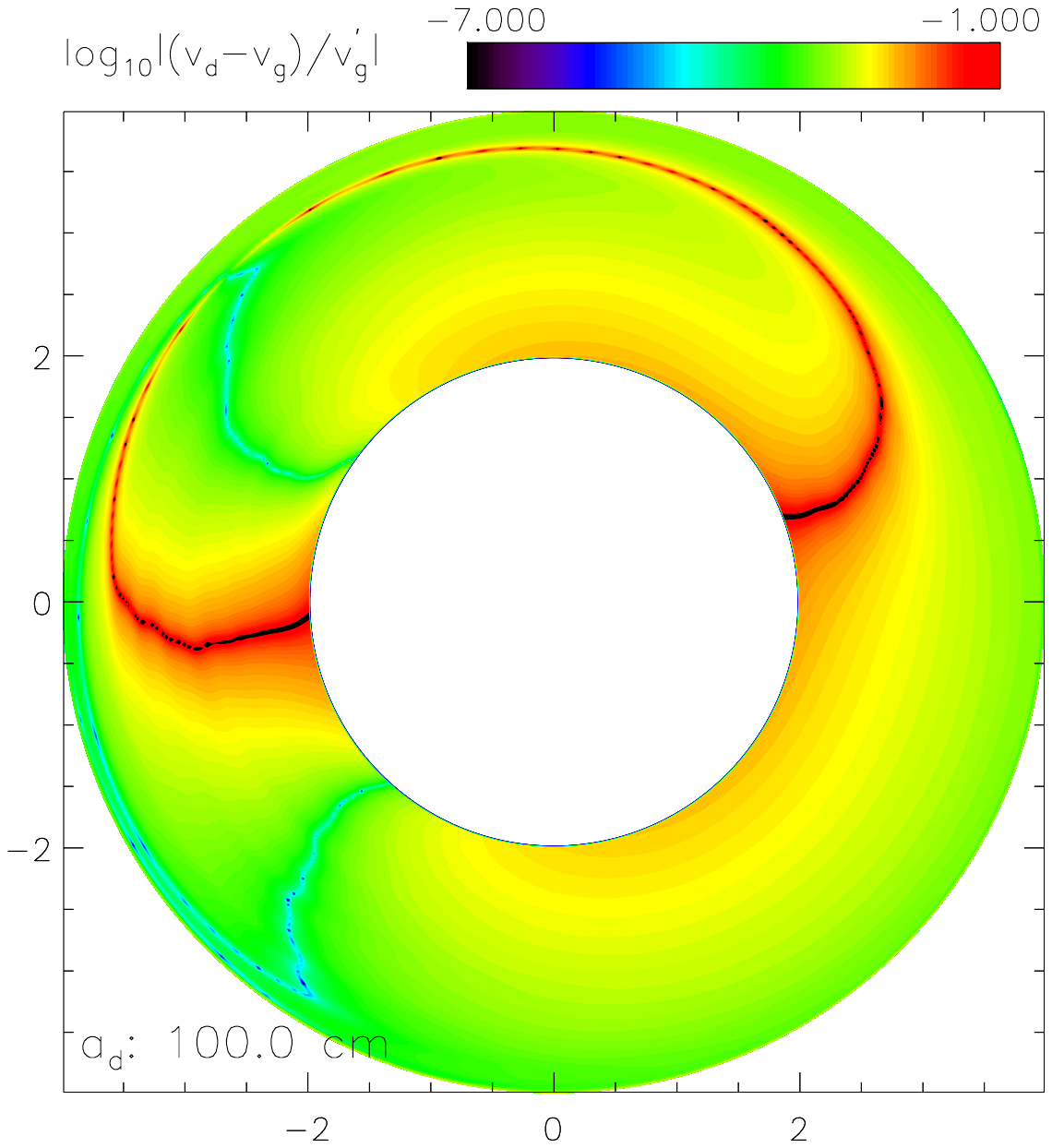}}
\end{tabular}
  \caption{Absolute values of the fractional differences between the gas velocities shown in the left panels of Figure
  \ref{fig:v_gas}
  and the dust velocities shown in Figure \ref{vrd_8000_5} for the $e_p=0$ case.
  The radial velocity difference is plotted in the top panels
  and azimuthal velocity difference is illustrated in the bottom panels.
  The left panels show the results for size of $1\times 10^{-2}$ cm and the right
panels
  show the results for $1$ m. The fractional differences can become large ($\gtrsim 1$) in the regions
  where $u_g$ or $v'_g$ is close to zero and thus we limit the top scale to show
  the value on the entire disk clearly.
  In the right panels for meter-sized particles,
  the narrow black curves indicate the regions where
  the fractional differences are too large and thus out of scale.}
  \label{fig:vd-vg}
\end{figure}
\begin{figure}
  \centering
  \begin{tabular}{cc}
  \scalebox{0.5}{\includegraphics[bb=0.72in 5in 7.75in 10in]{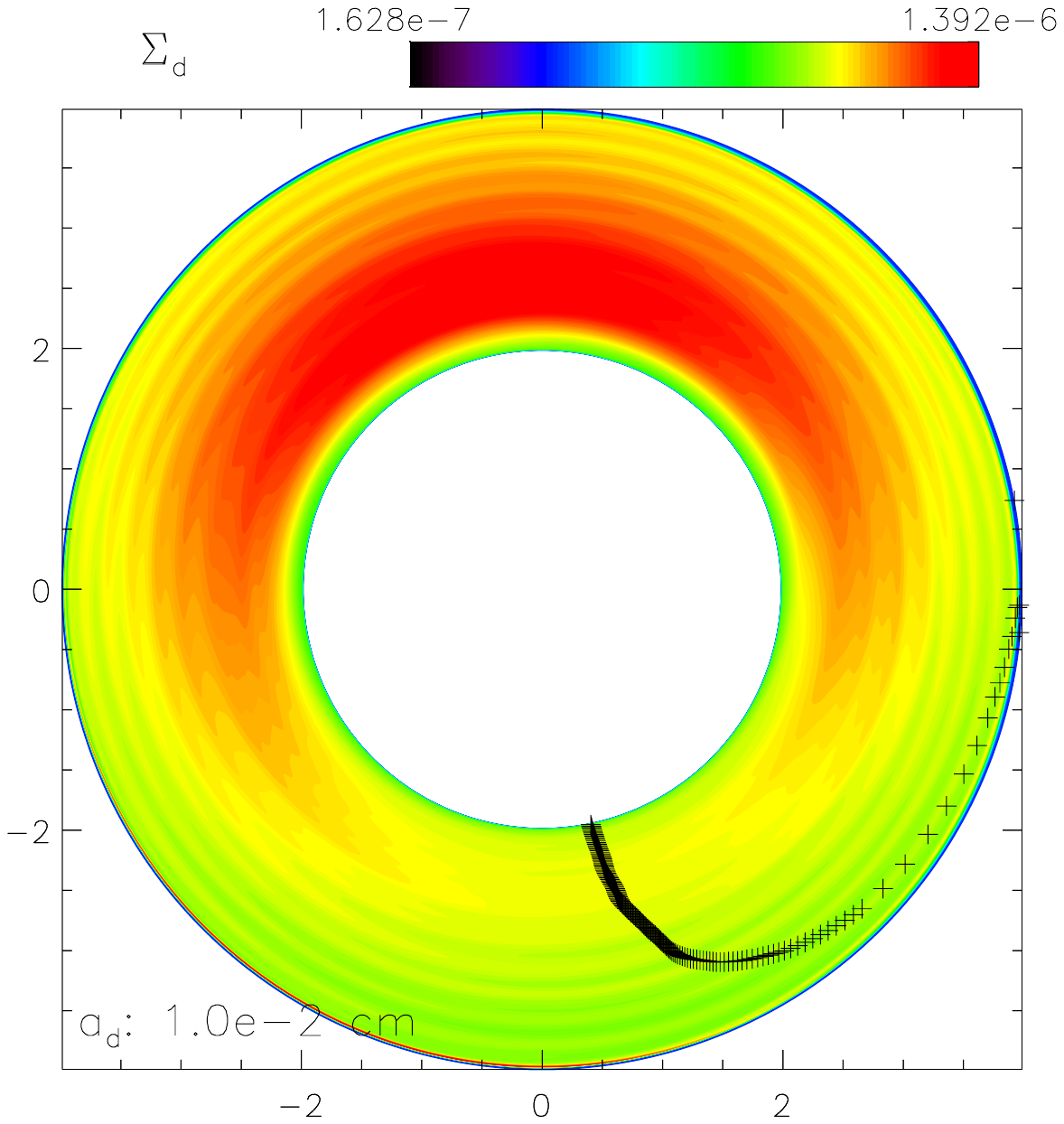}}&
  \scalebox{0.5}{\includegraphics[bb=0.72in 5in 7.75in 10in]{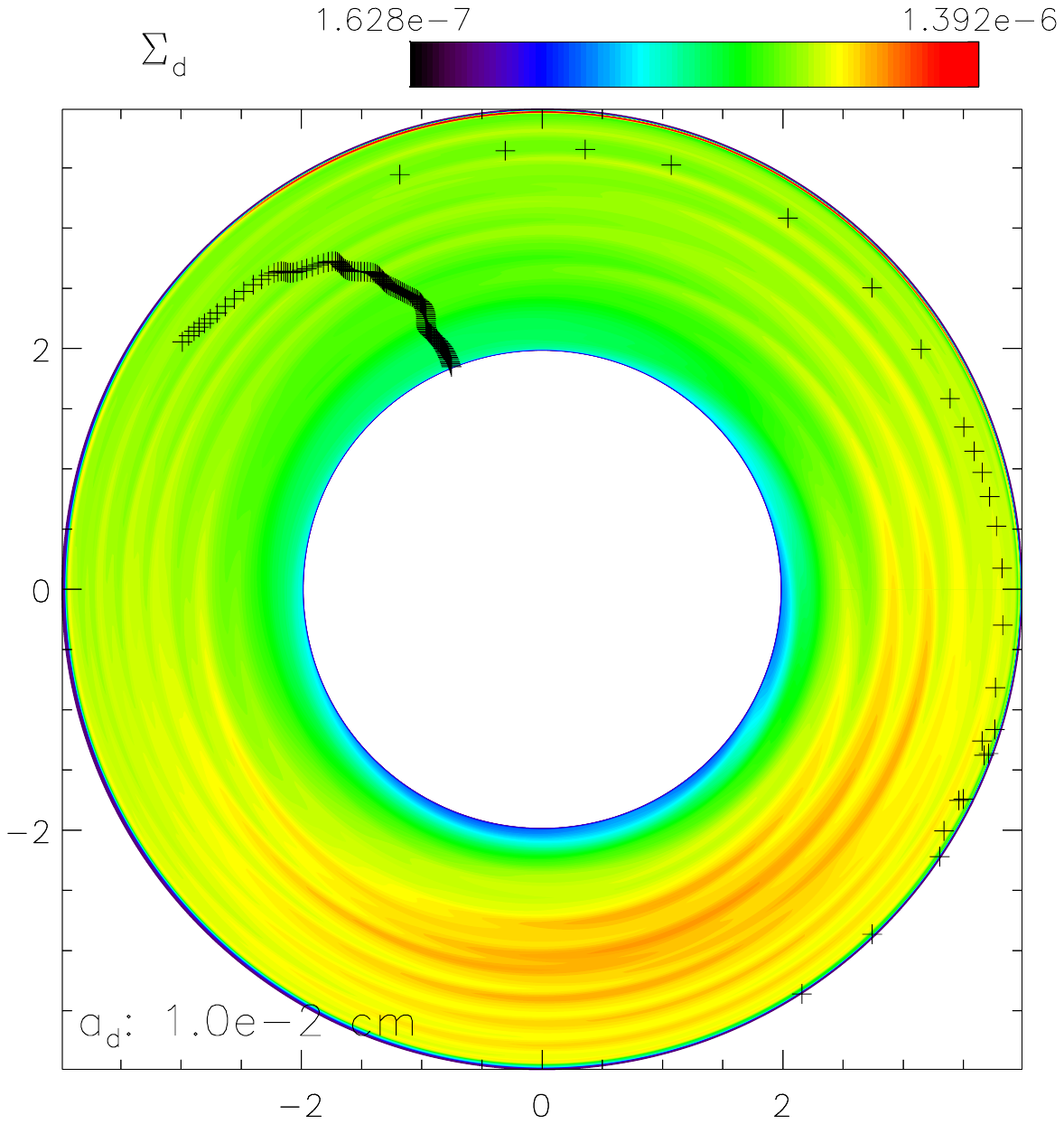}}\\
\scalebox{0.5}{\includegraphics[bb=0.72in 5in 7.75in 10in]{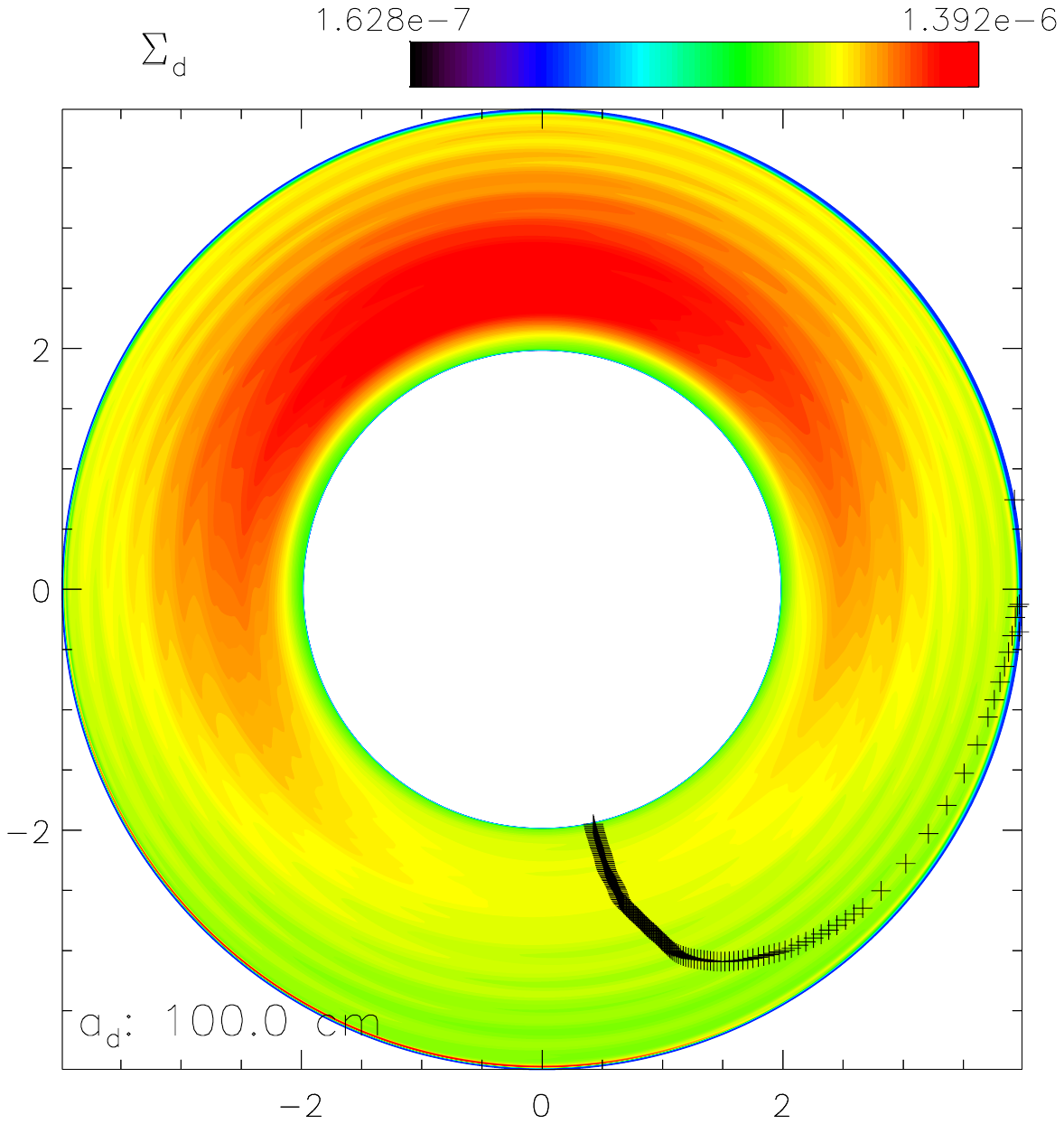}}&
\scalebox{0.5}{\includegraphics[bb=0.72in 5in 7.75in 10in]{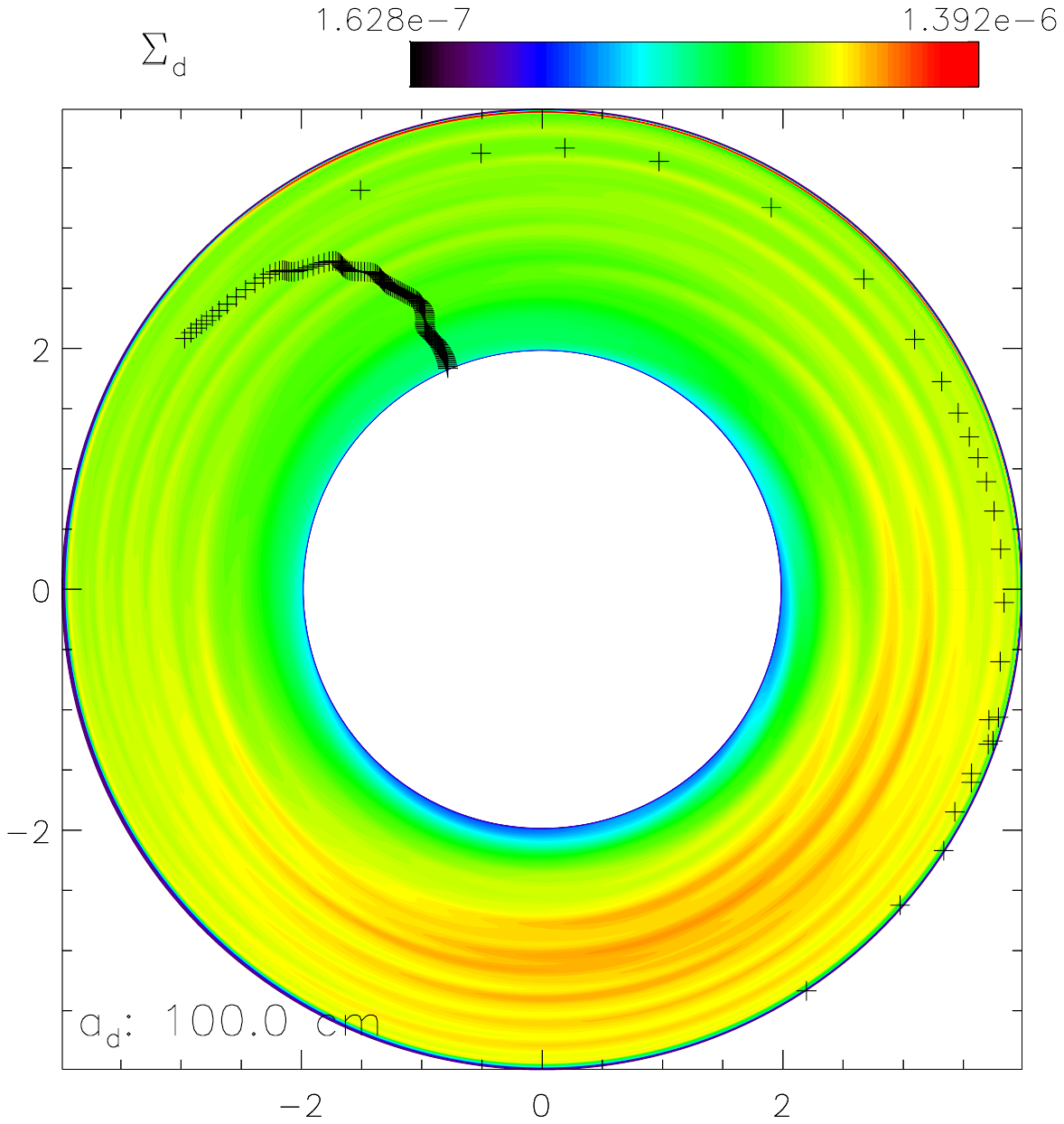}}
 \end{tabular}
  \caption{The spatial distribution of the dust surface density of various grain sizes for the $e_p=0$
  (left panels) and $e_p=0.1$ cases (right panels). The length unit is $5$ AU and the mass unit is 1 $M_\odot$. Hence,
   the unit of the surface density is about $3.56\times10^{-1}$ g cm$^{-2}$.
            The plus signs in the plots mark the azimuthally averaged longitude of pericenter for dust $\varpi_{d,ave}$.}
  \label{den_8000_5}
\end{figure}
\begin{figure}
\begin{tabular}{cc}
  \scalebox{0.42}{\includegraphics[bb=0.72in 5in 7.75in 10in]{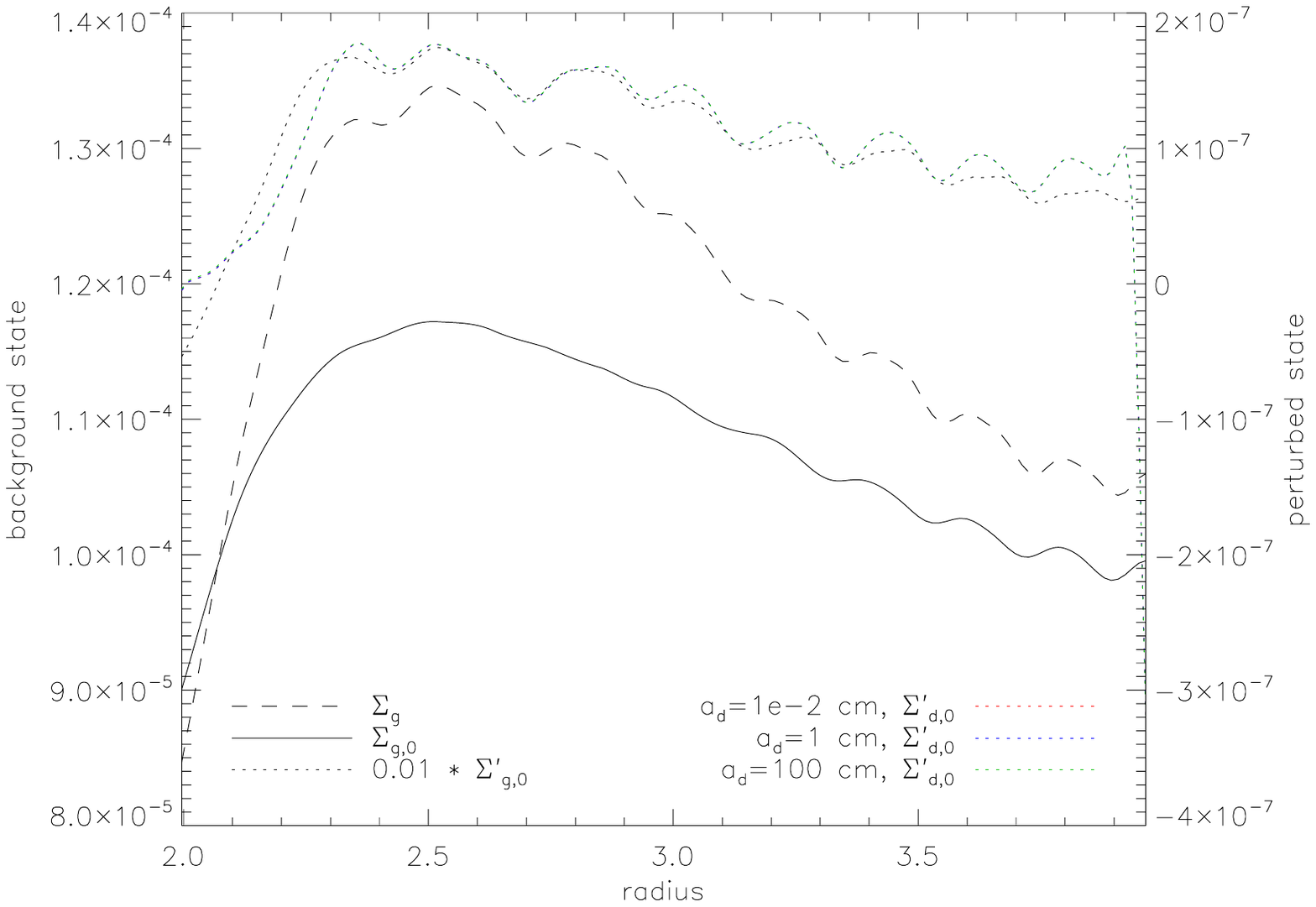}} &
 \scalebox{0.42}{\includegraphics[bb=0.72in 5in 7.75in 10in]{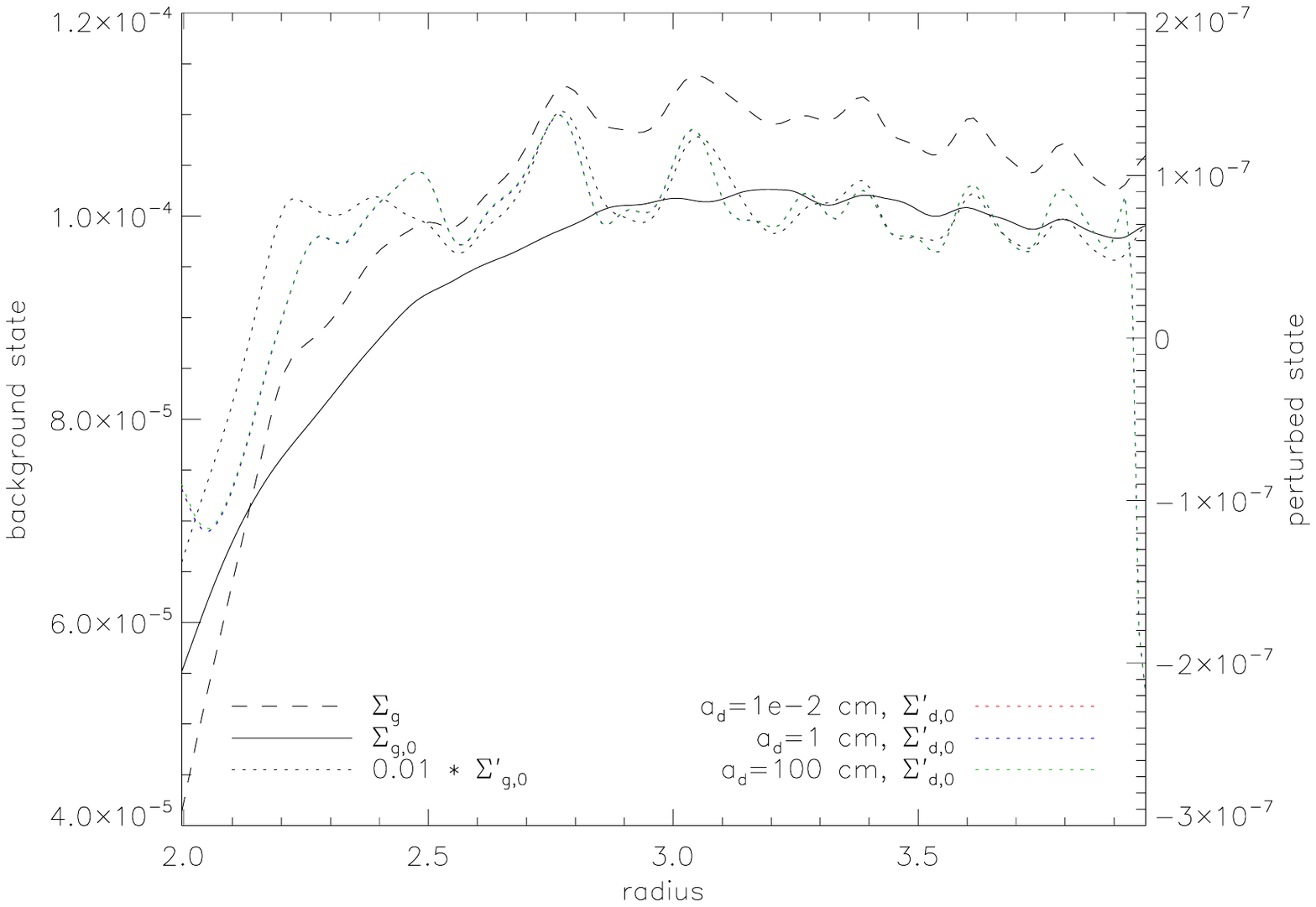}} \\
  \end{tabular}
\caption{The density radial profiles at $\phi=110^\circ$ for the $e_p=0$ case (left
panel) and at $\phi=300^\circ$ for the $e_p=0.1$ case (right panel). $\Sigma_g$ is
the gas density after time averaging. $\Sigma_{g,0}$ is the azimuthal average of
$\Sigma_g$ and therefore is the background gas density. $\Sigma_{g,0}'$ and
$\Sigma_{d,0}'$ are the perturbed gas and dust densities, respectively. The scales
for the background and perturbed values are shown on the left and right sides of
each panel, respectively. Note that the wave-like features on the perturbed density
profiles are the residuals of the short-period structures after time averaging.
Nevertheless, they are small enough for estimating the order of magnitude of the
secular density value $\Sigma_{d,0}'$.} \label{fig:Den_1D}
\end{figure}
\begin{figure}
\centering
\begin{tabular}{cc}
\scalebox{0.5}{\includegraphics[bb=0.72in 5in 7.75in 10in]{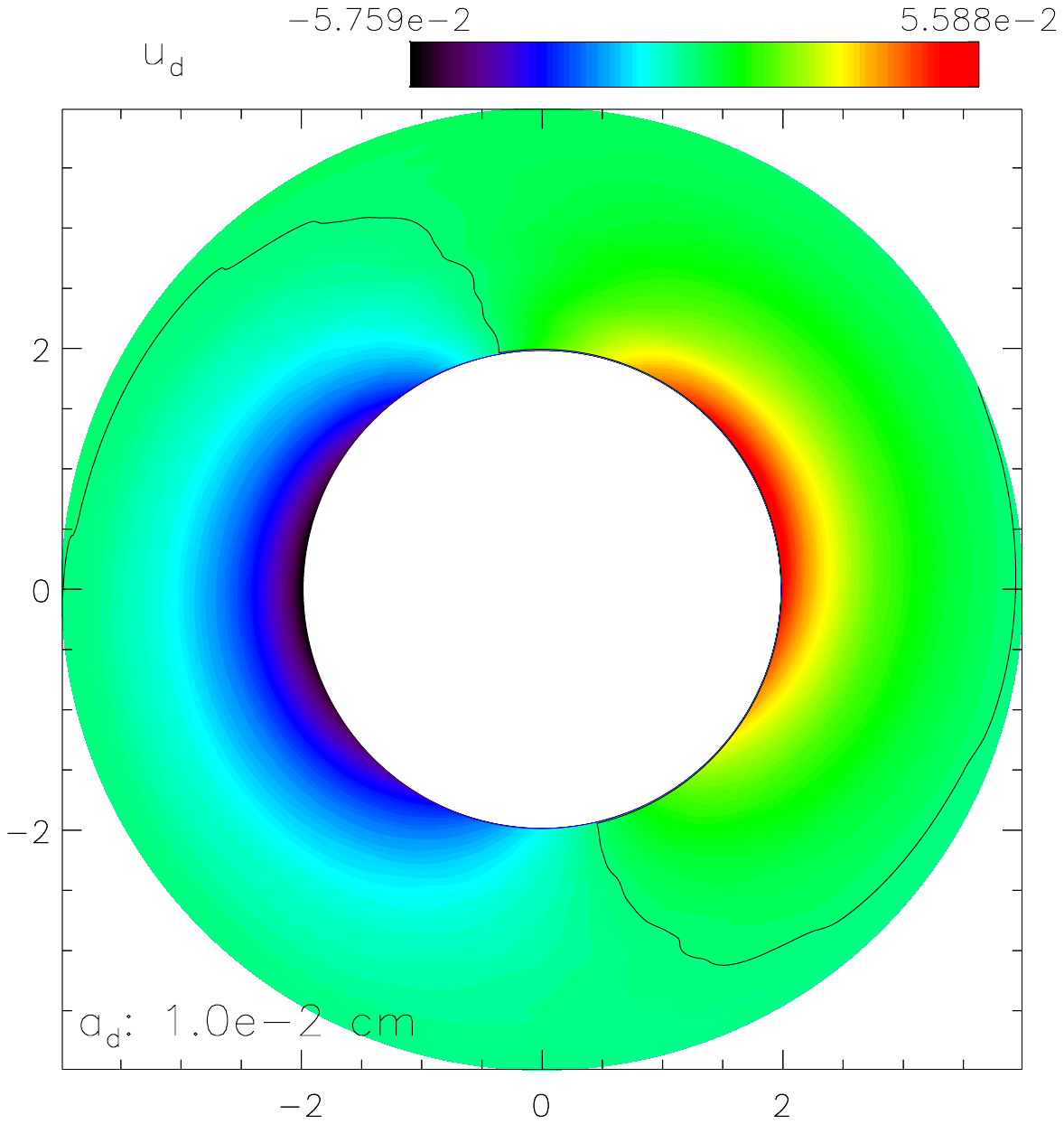}}
&
\scalebox{0.5}{\includegraphics[bb=0.72in 5in 7.75in 10in]{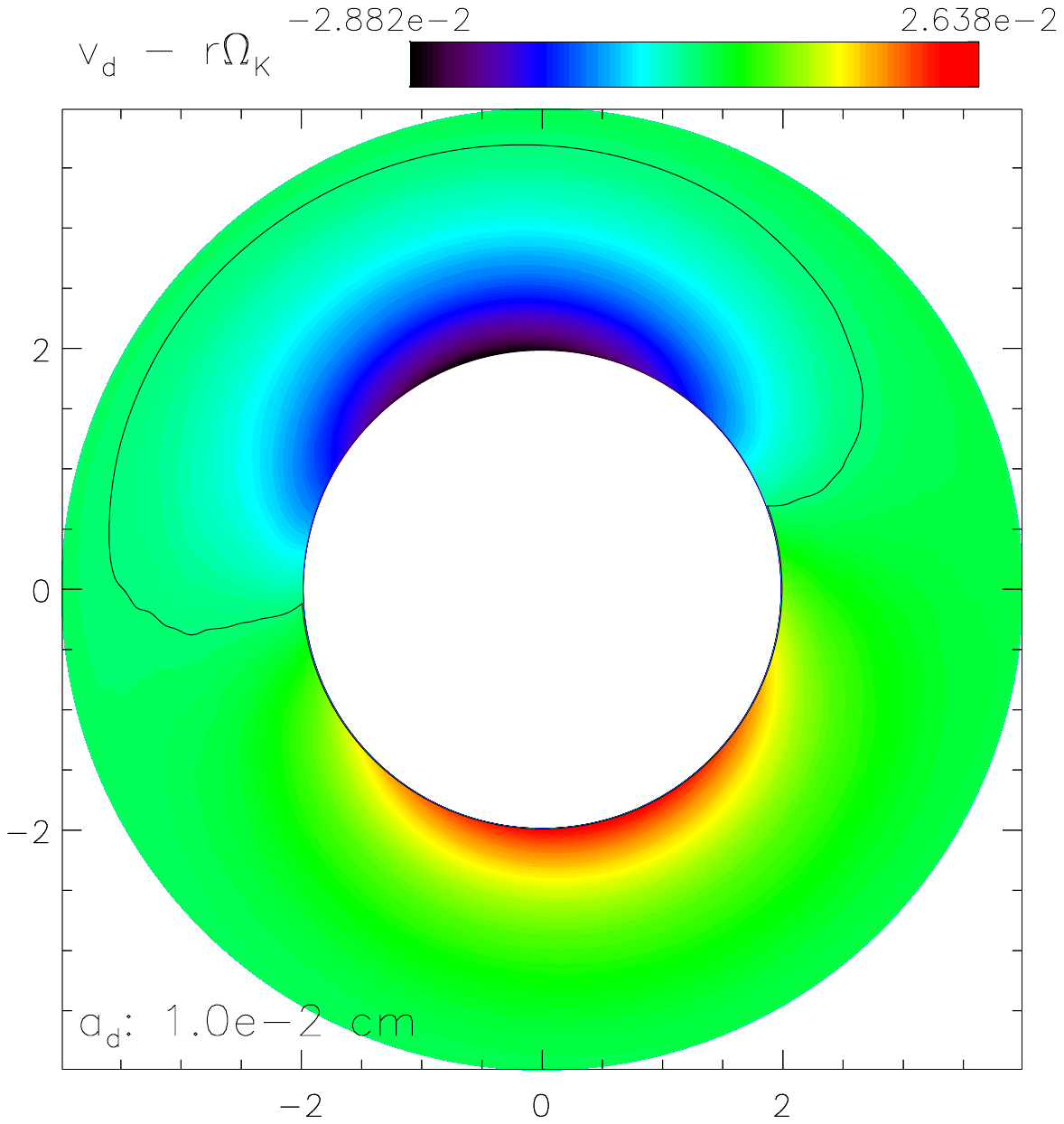}}\\
\scalebox{0.5}{\includegraphics[bb=0.72in 5in 7.75in 10in]{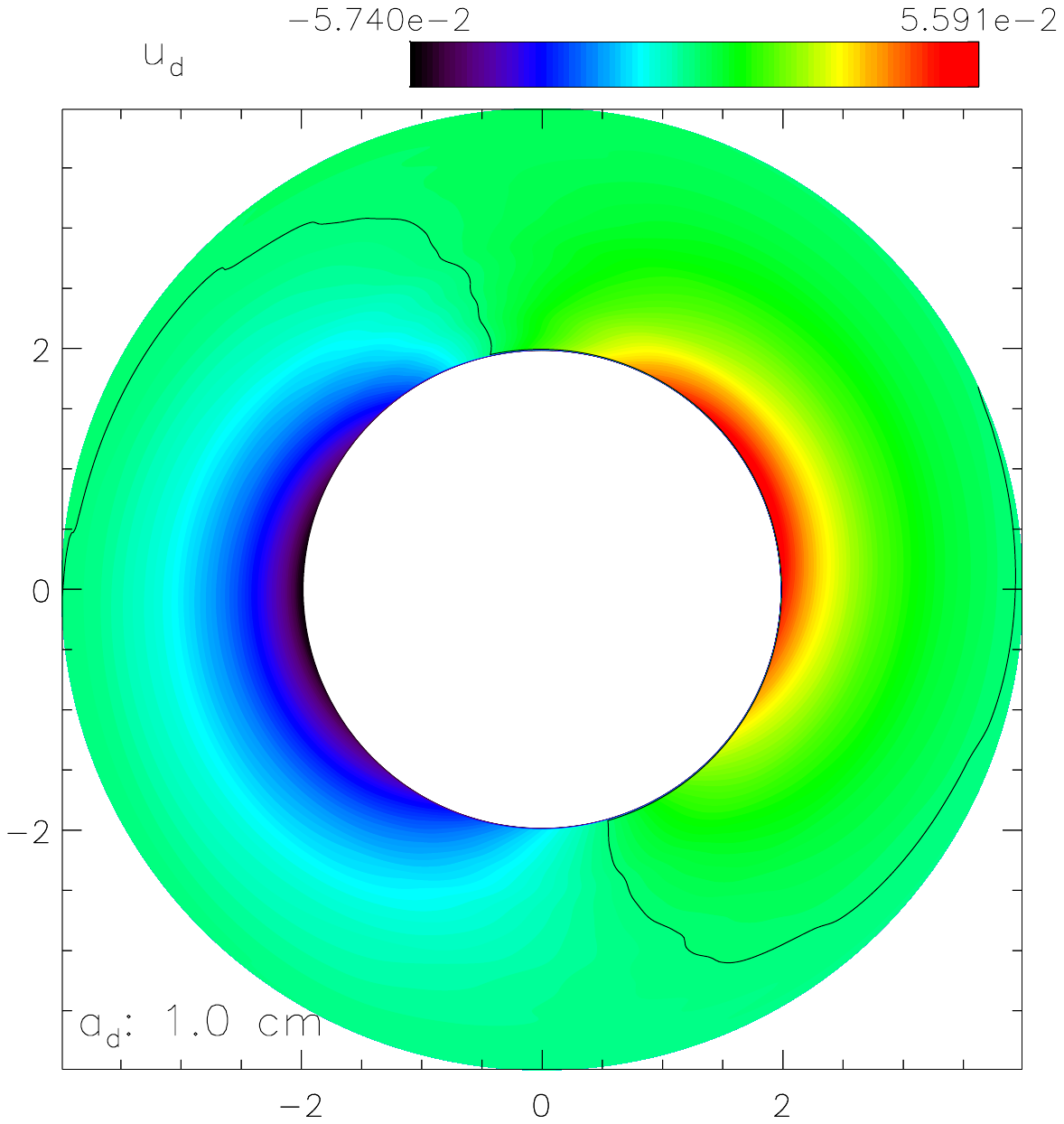}}
&
\scalebox{0.5}{\includegraphics[bb=0.72in 5in 7.75in 10in]{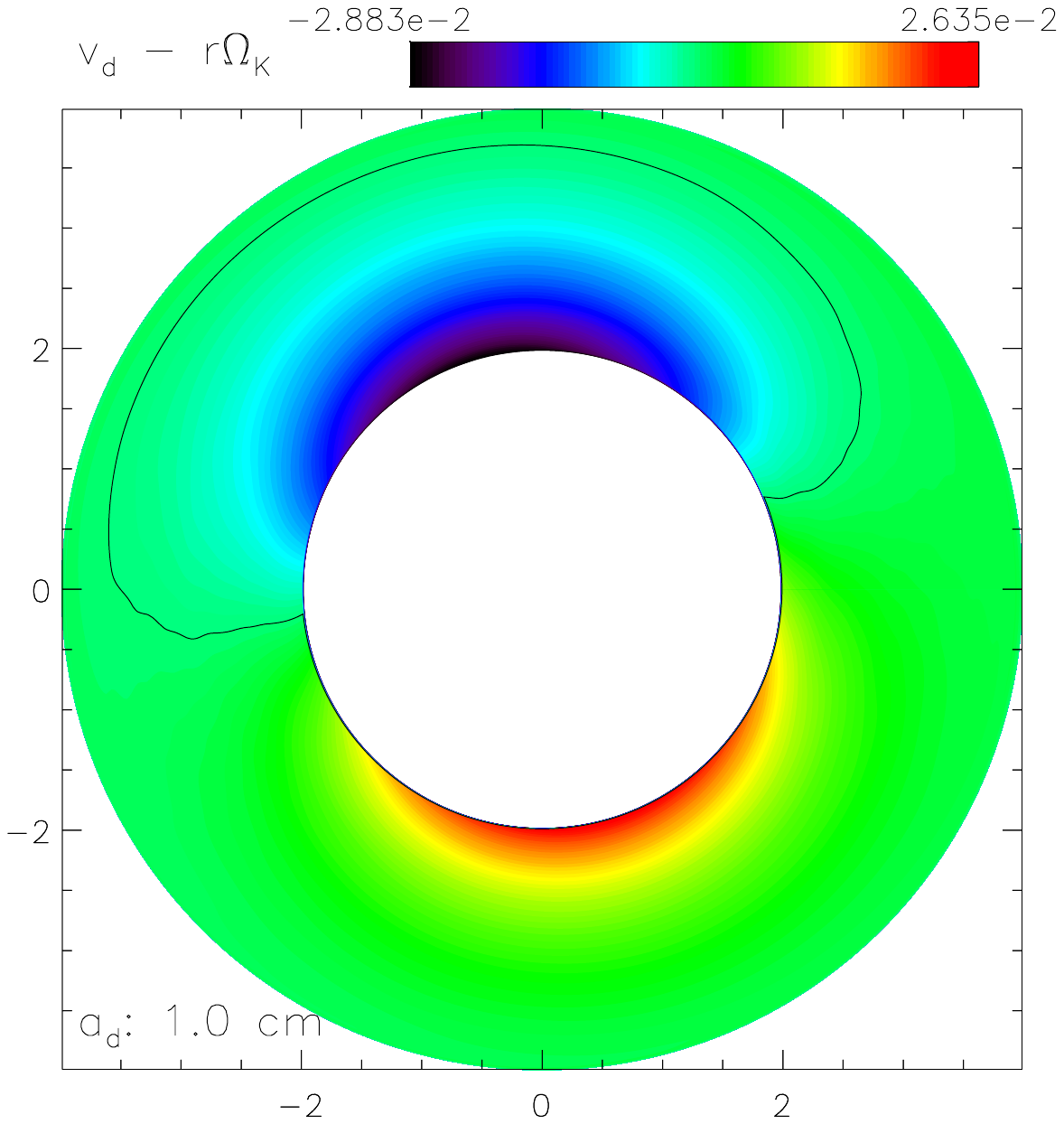}}\\
\scalebox{0.5}{\includegraphics[bb=0.72in 5in 7.75in 10in]{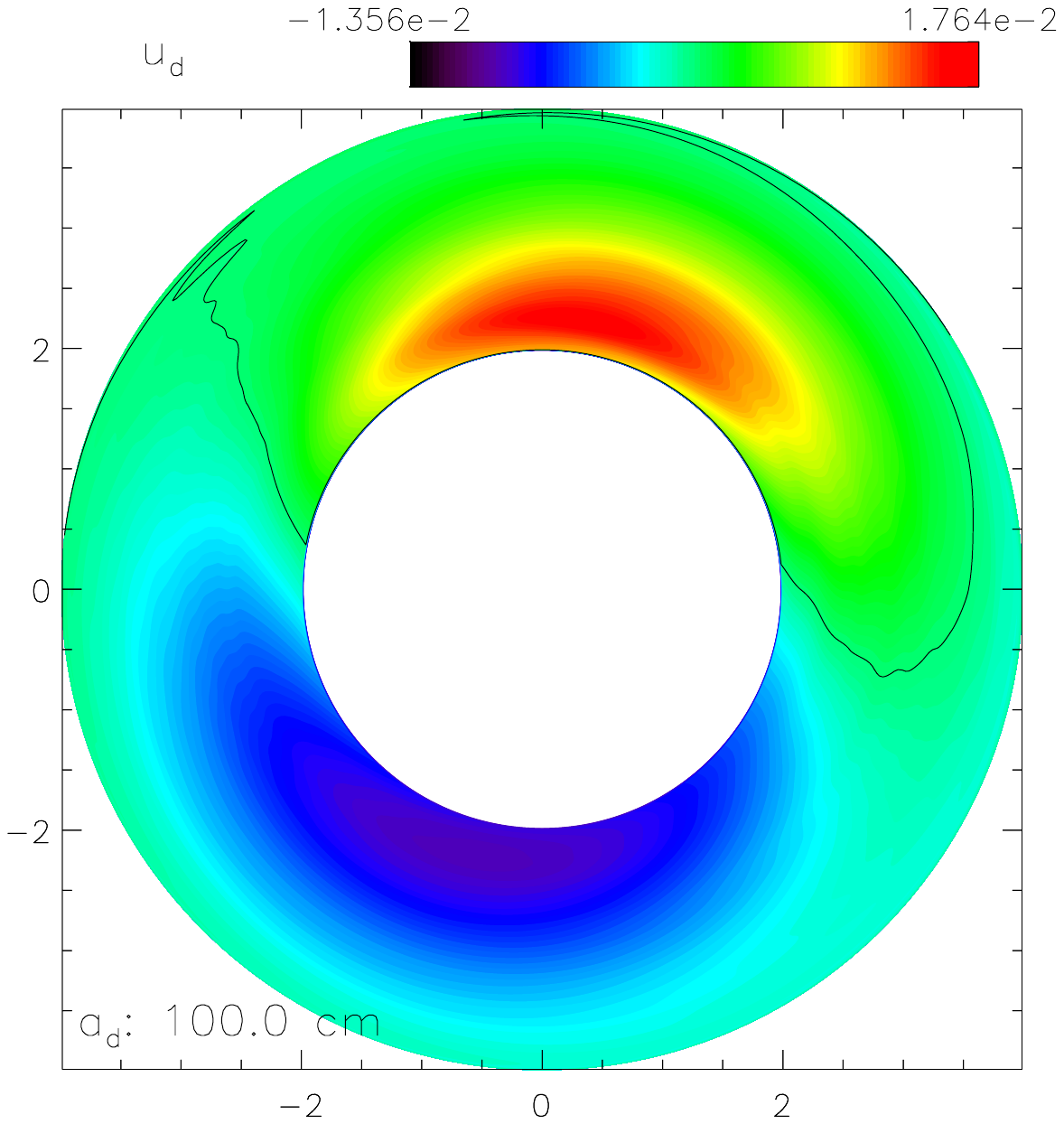}}
&
\scalebox{0.5}{\includegraphics[bb=0.72in 5in 7.75in 10in]{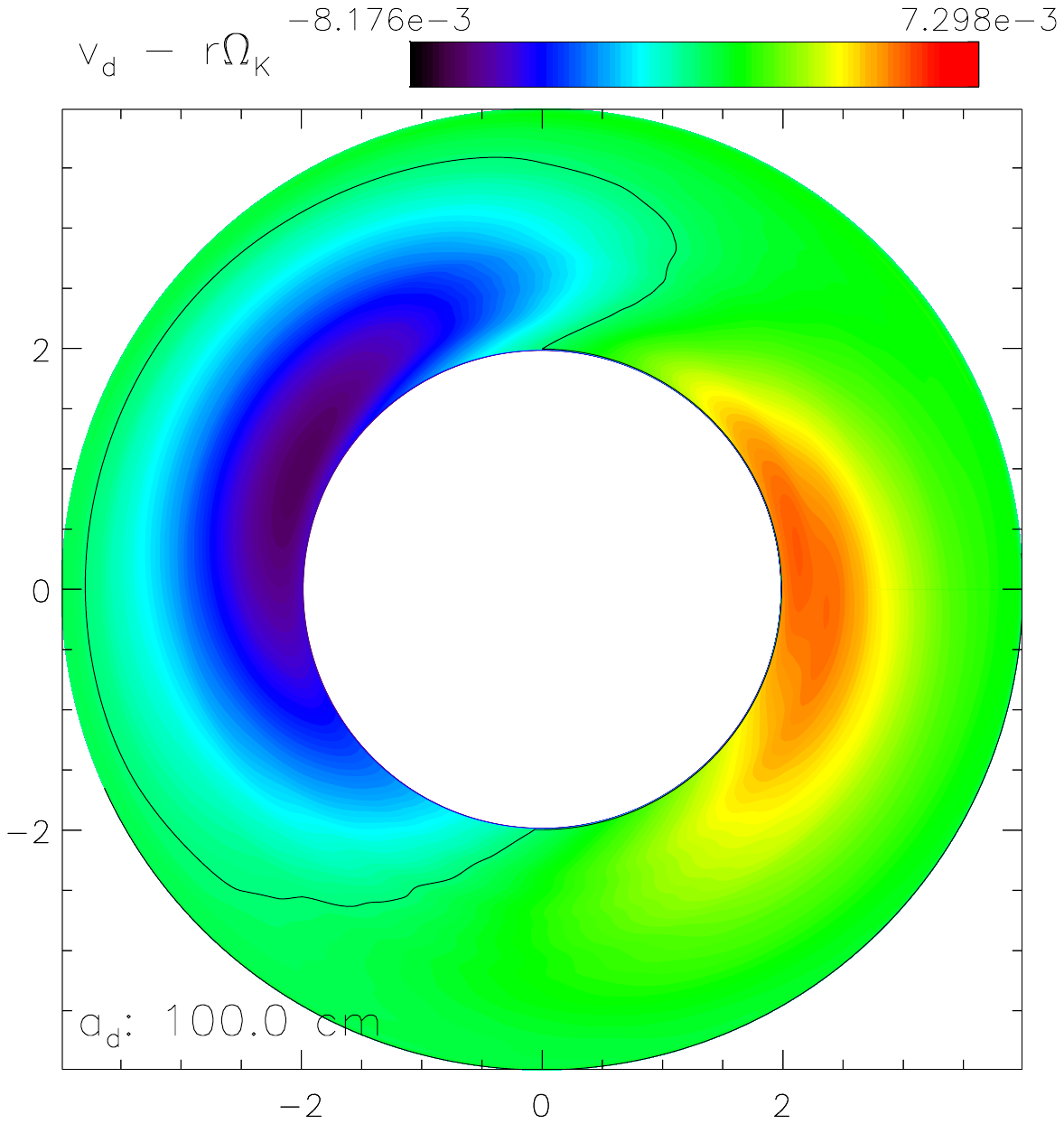}}
\end{tabular}
  \caption{Dust velocity fields (left - radial; right - azimuthal) of the transition disk for the $e_p=0$
  case for three particle sizes.
   The unit of the velocity is 2.98$\times 10^5$ cm/s.}
  \label{fig:vrd_8000_100_ep=0}
\end{figure}
\begin{figure}
\centering
\begin{tabular}{cc}
\scalebox{0.5}{\includegraphics[bb=0.72in 5in 7.75in 10in]{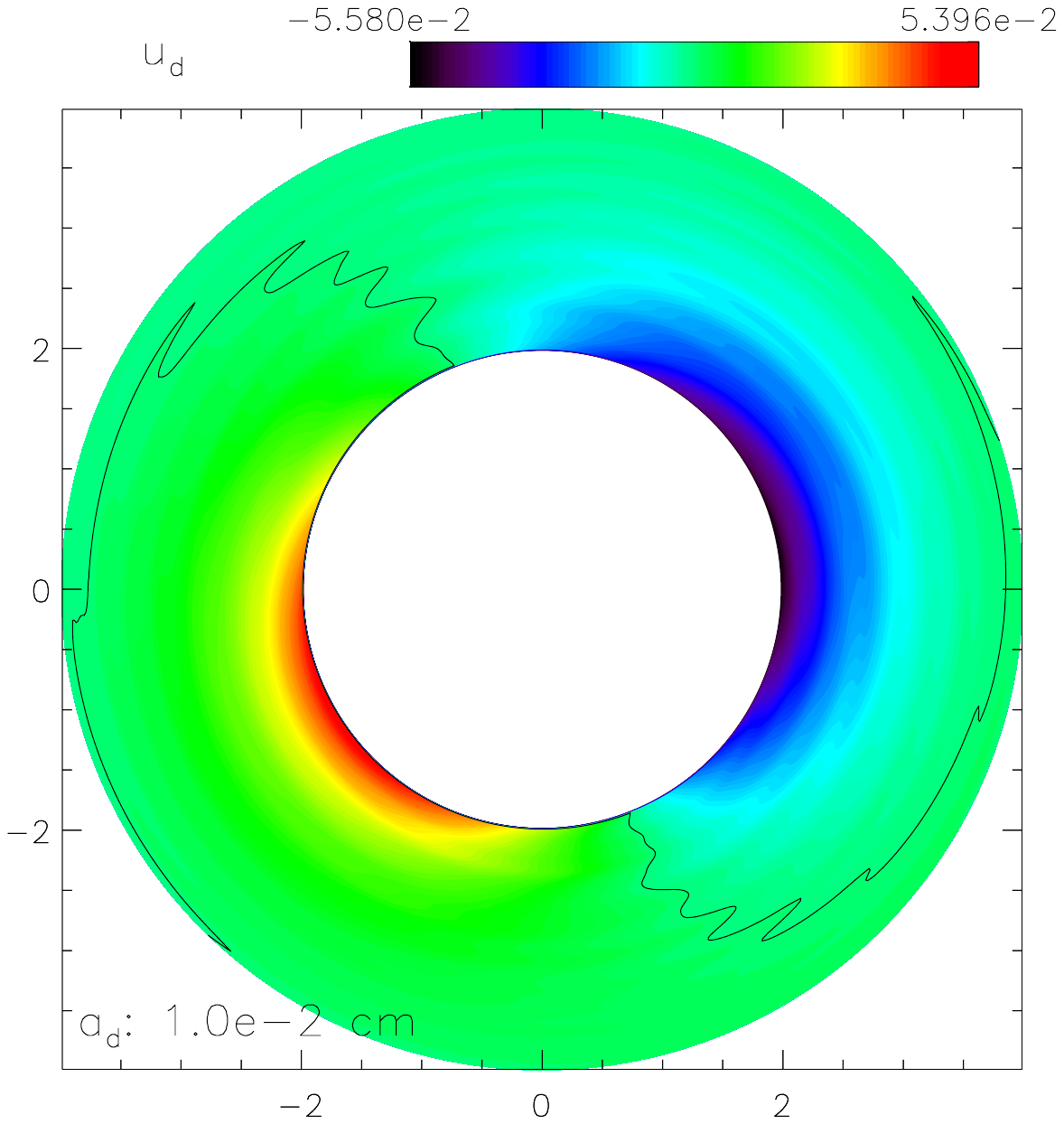}}
&
\scalebox{0.5}{\includegraphics[bb=0.72in 5in 7.75in 10in]{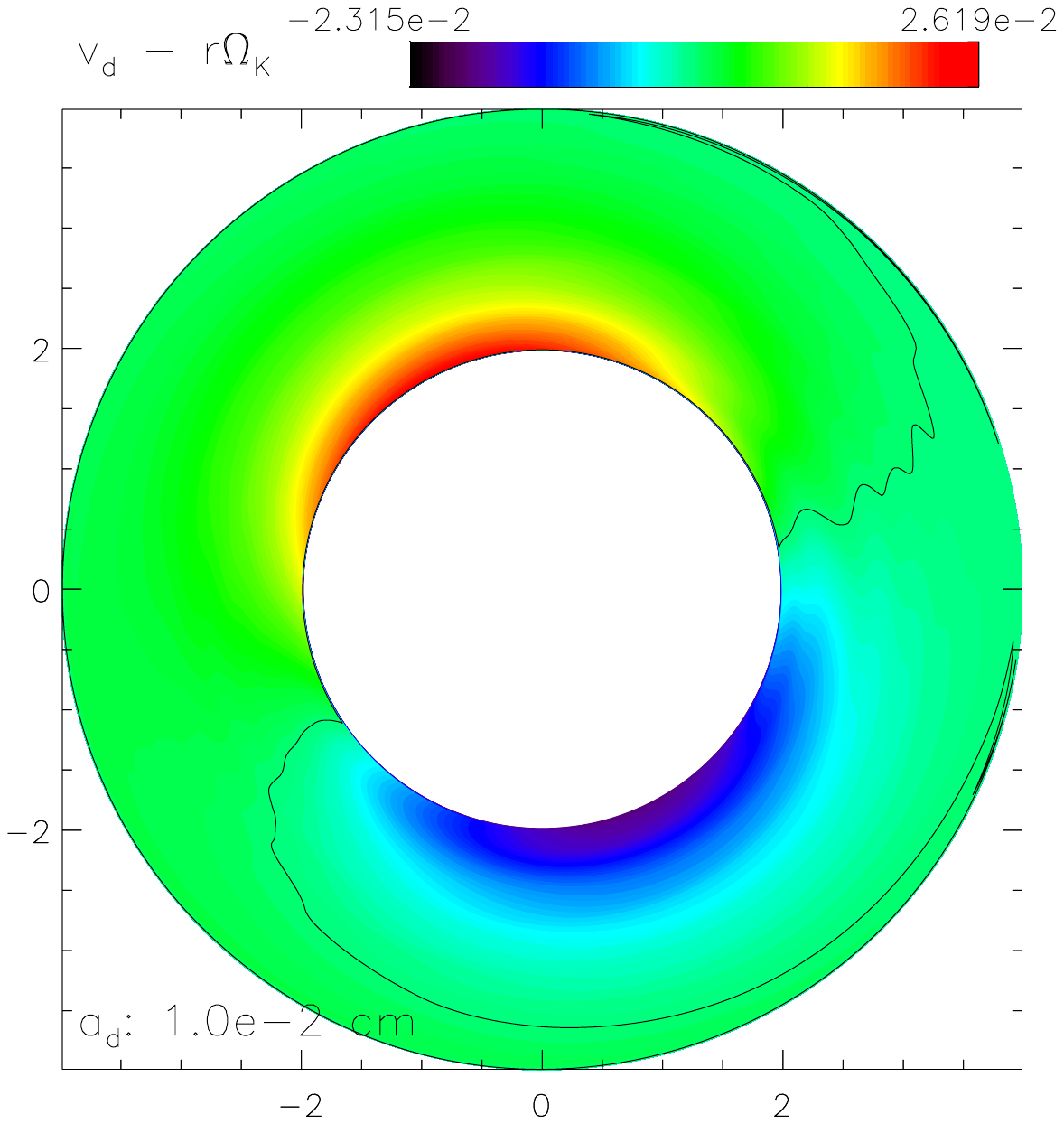}}\\
\scalebox{0.5}{\includegraphics[bb=0.72in 5in 7.75in 10in]{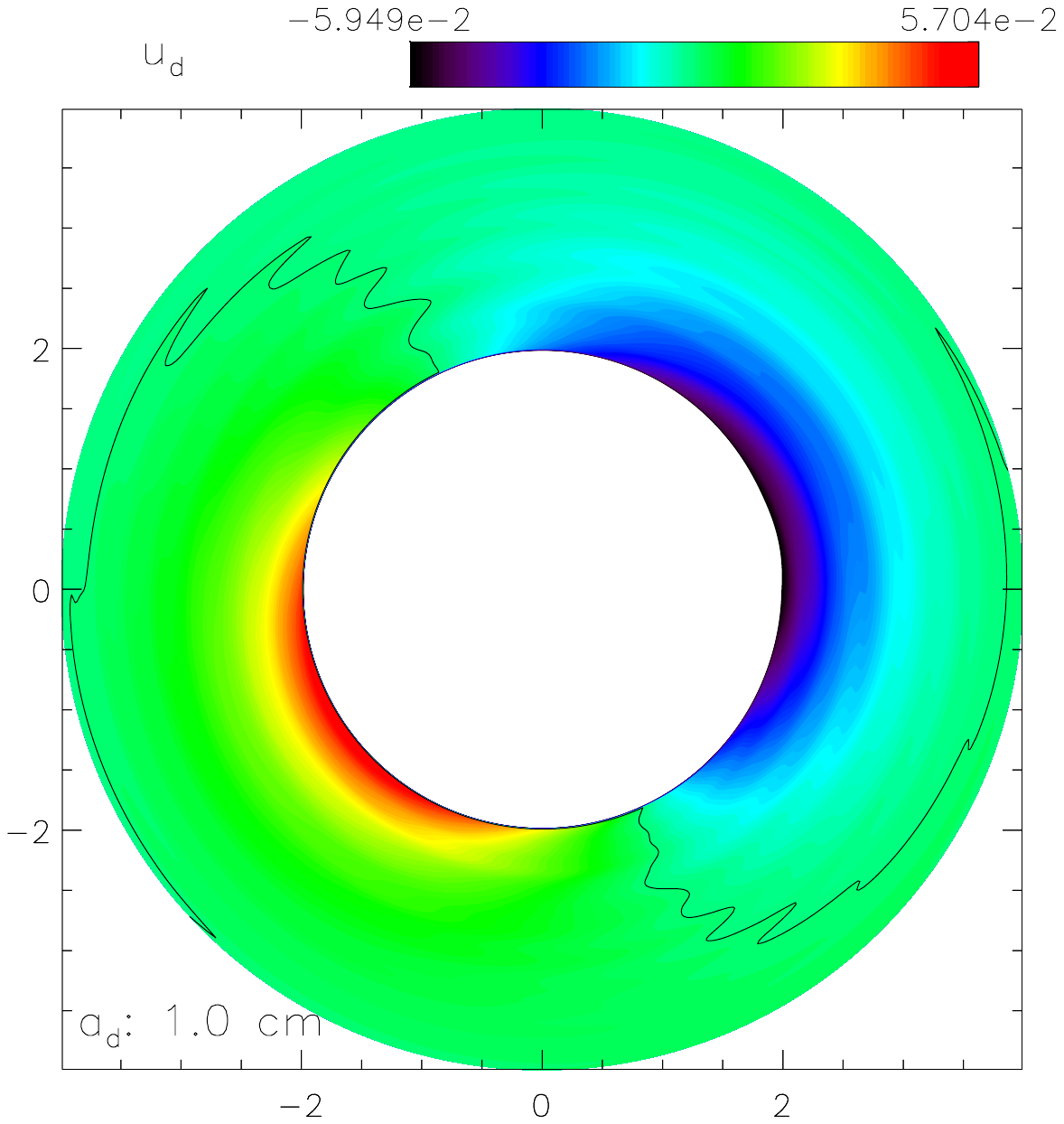}}
&
\scalebox{0.5}{\includegraphics[bb=0.72in 5in 7.75in 10in]{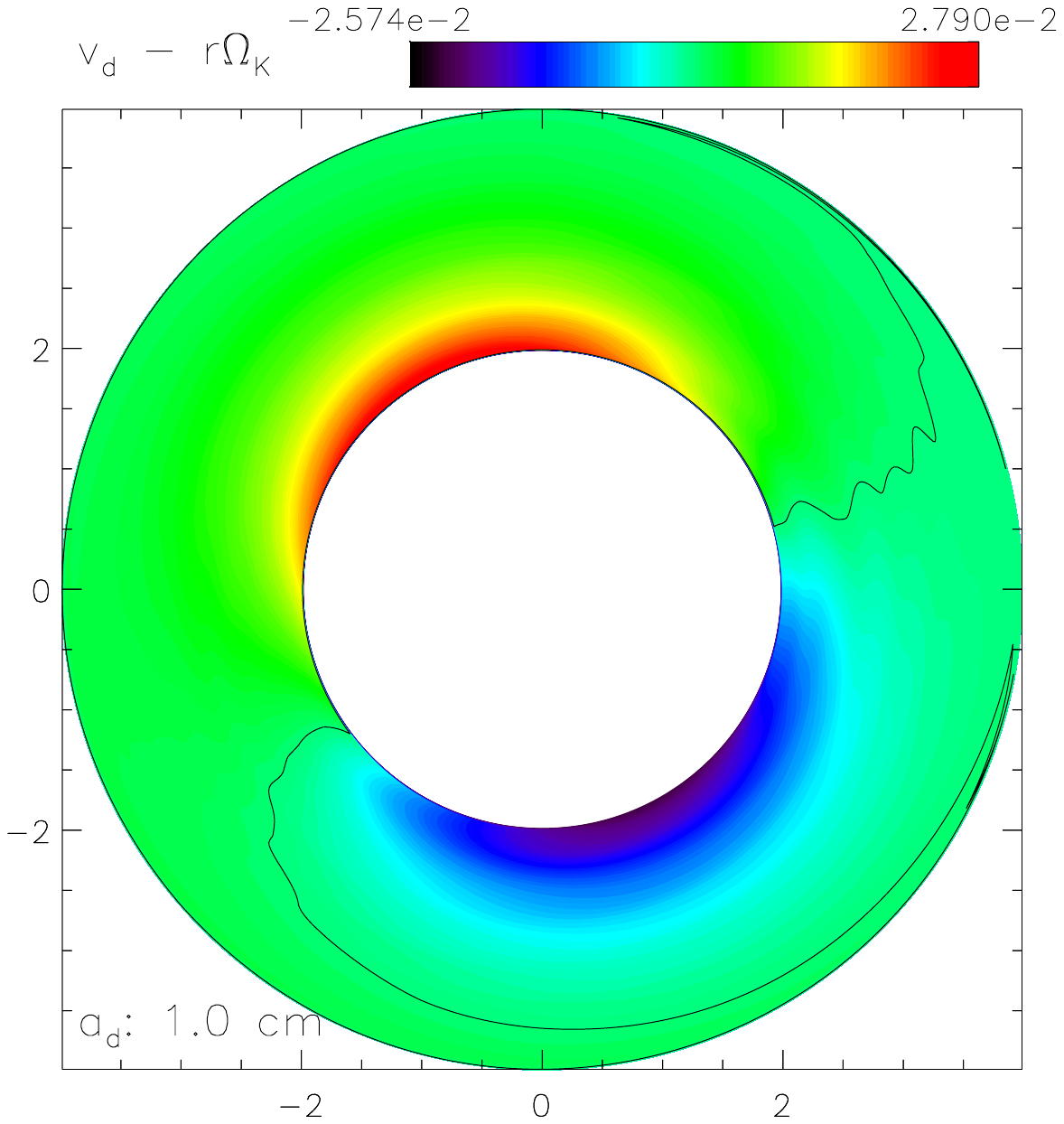}}\\
\scalebox{0.5}{\includegraphics[bb=0.72in 5in 7.75in 10in]{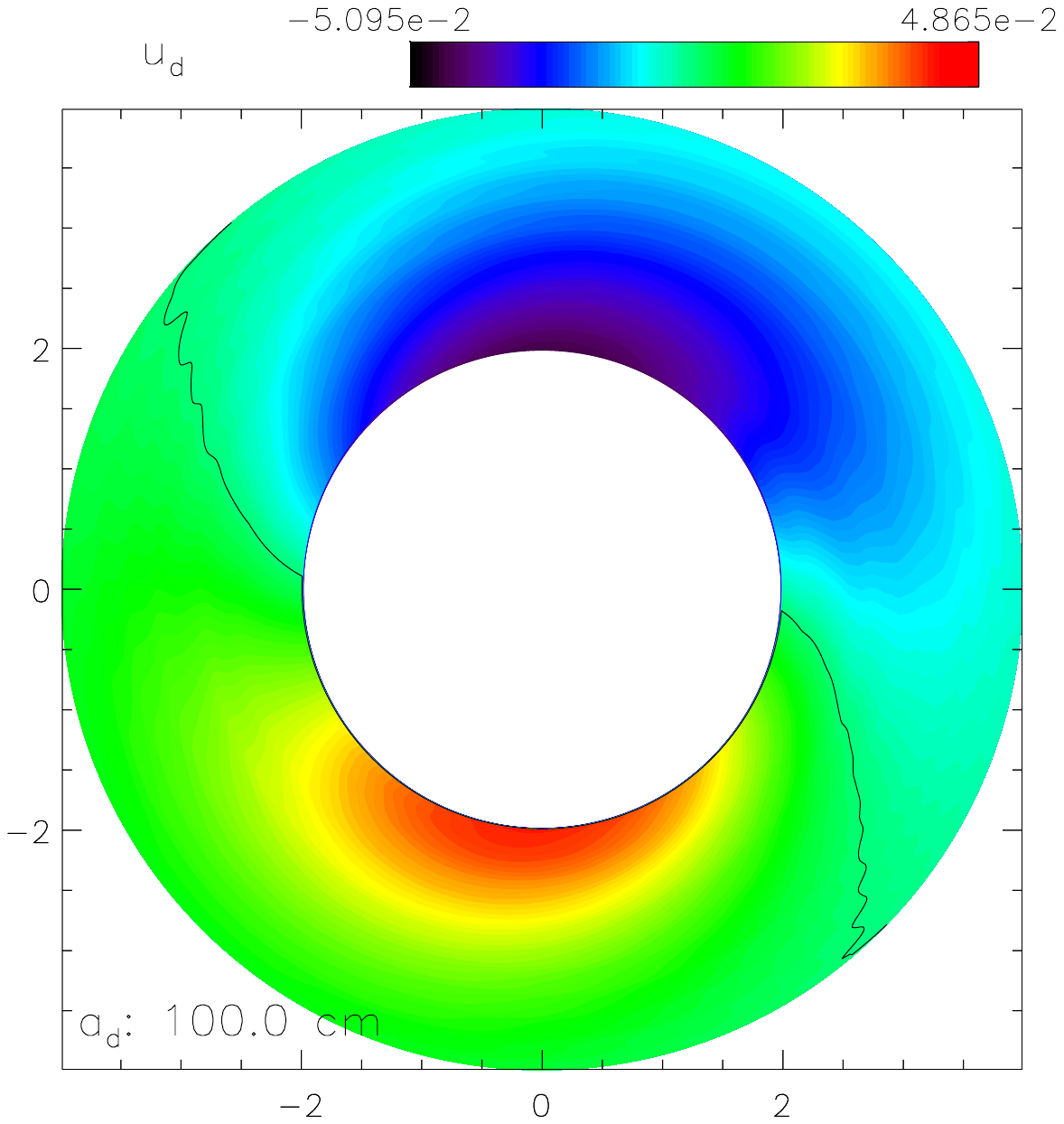}}
&
\scalebox{0.5}{\includegraphics[bb=0.72in 5in 7.75in 10in]{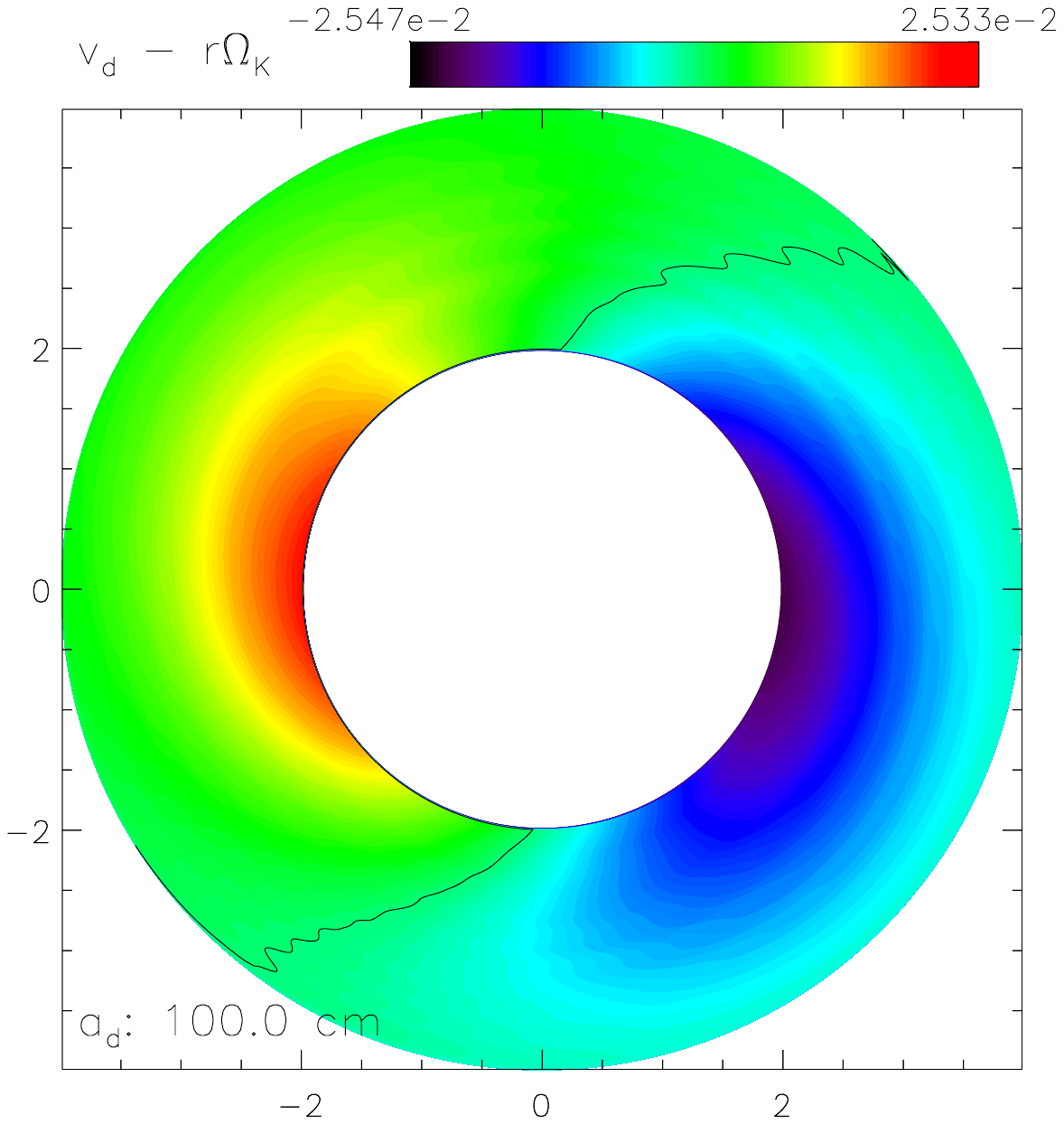}}
\end{tabular}
  \caption{Same as Figure \ref{fig:vrd_8000_100_ep=0} but for the $e_p=0.1$ case.}
  \label{fig:vrd_4000_100_ep=0.1}
\end{figure}
\begin{figure}
  \centering
  \begin{tabular}{cc}
   \scalebox{0.5}{\includegraphics[bb=0.72in 5in 7.75in 10in]{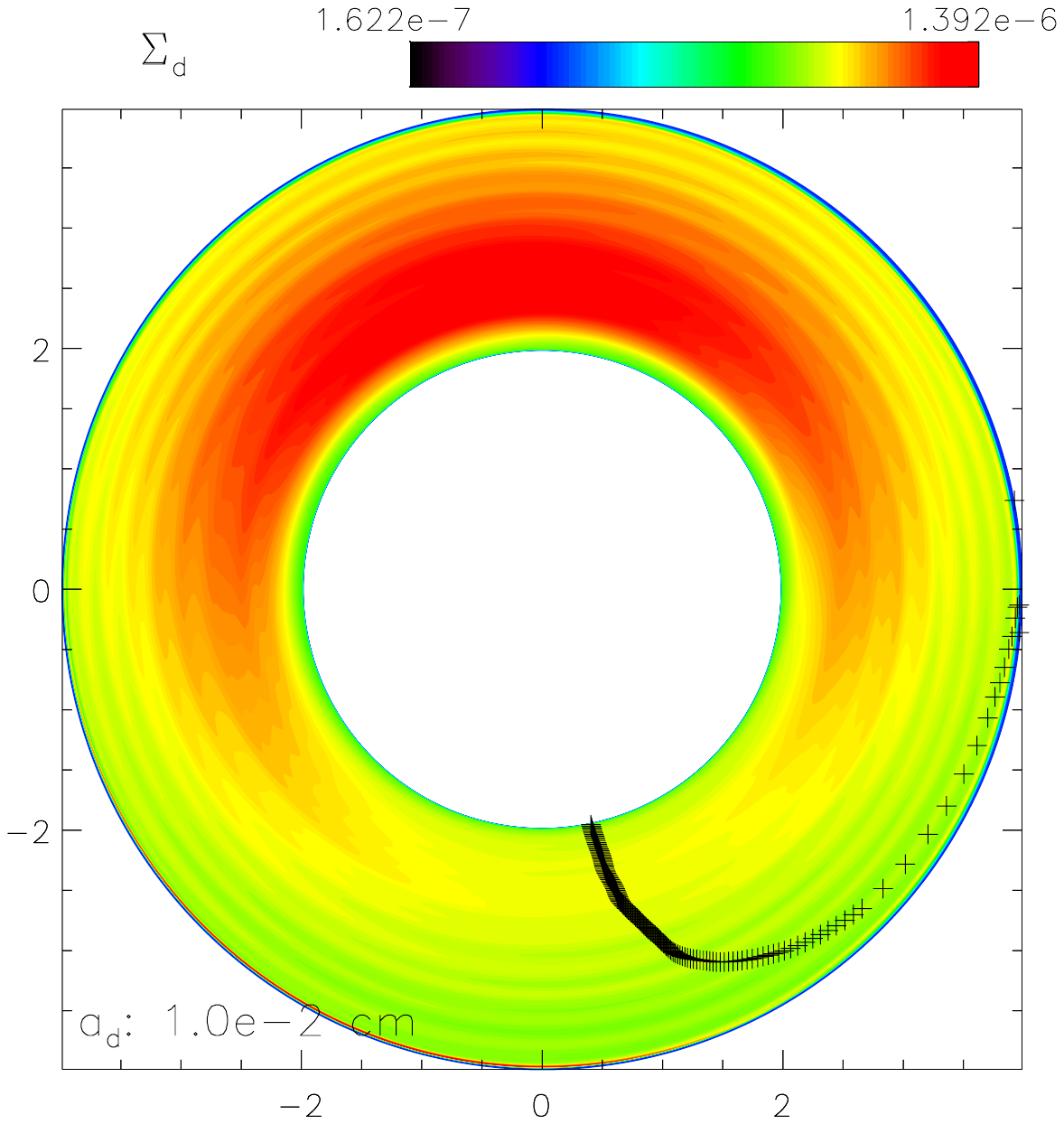}}&
\scalebox{0.5}{\includegraphics[bb=0.72in 5in 7.75in 10in]{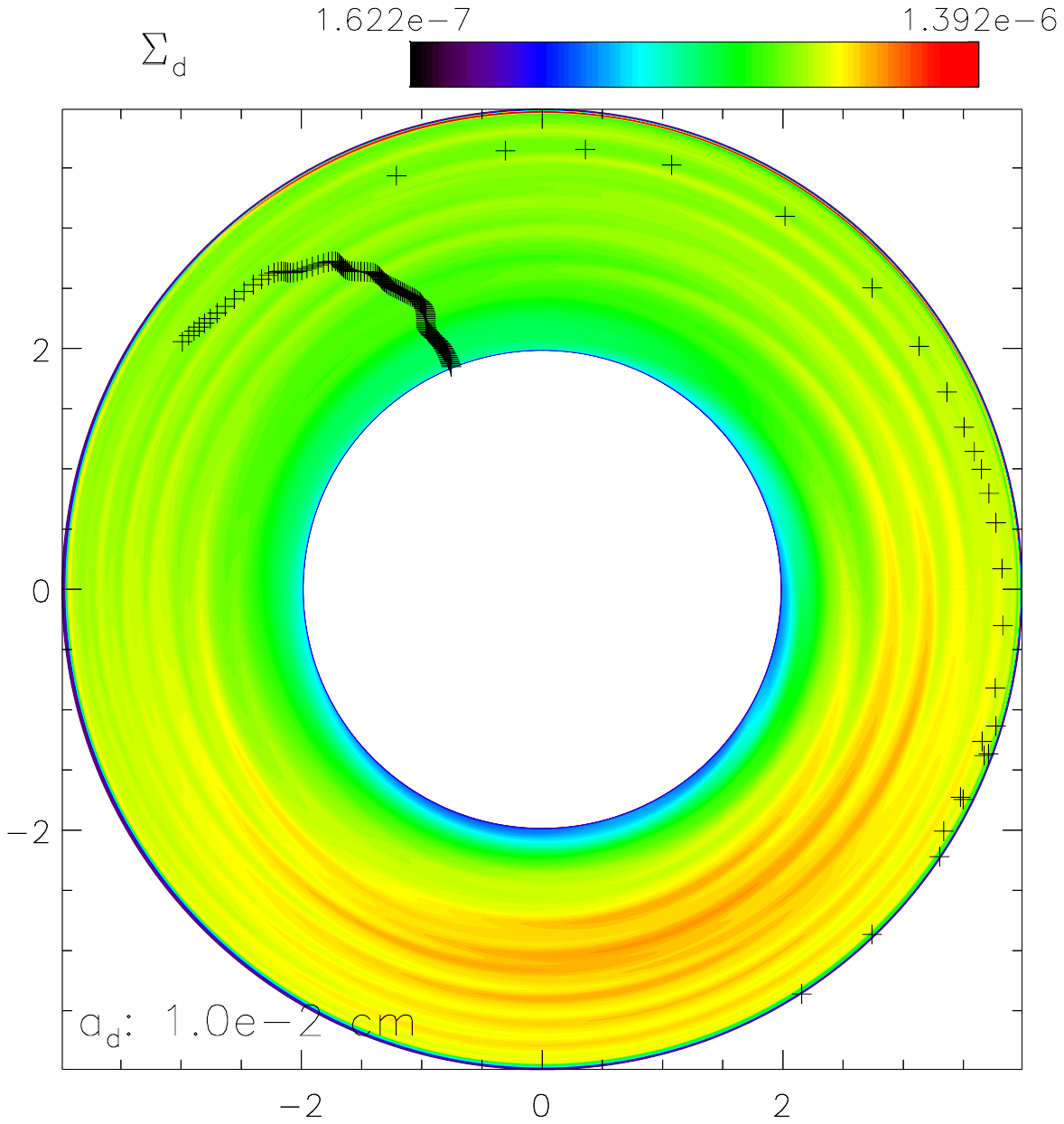}}\\
  \scalebox{0.5}{\includegraphics[bb=0.72in 5in 7.75in 10in]{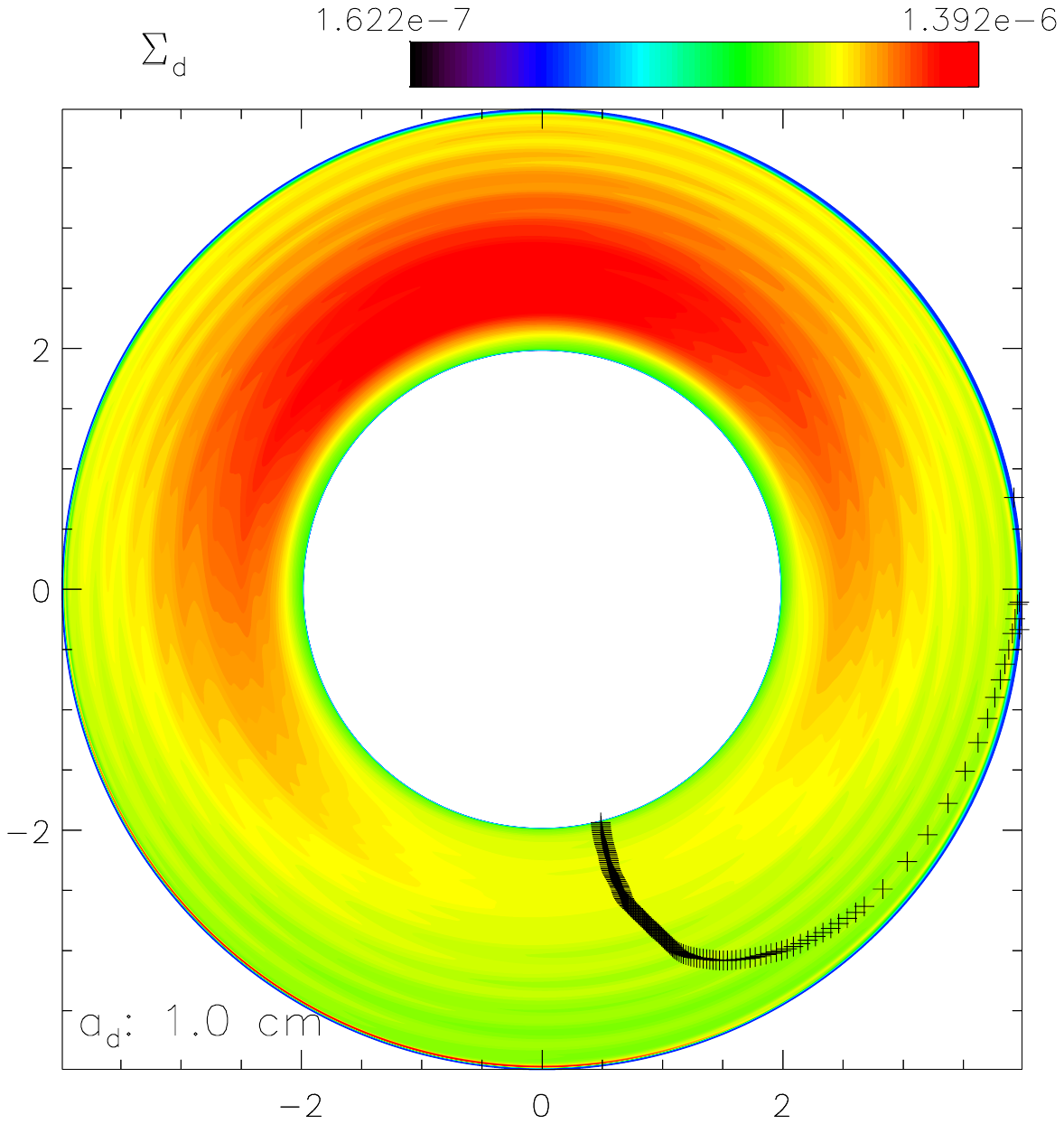}}
  &
  \scalebox{0.5}{\includegraphics[bb=0.72in 5in 7.75in 10in]{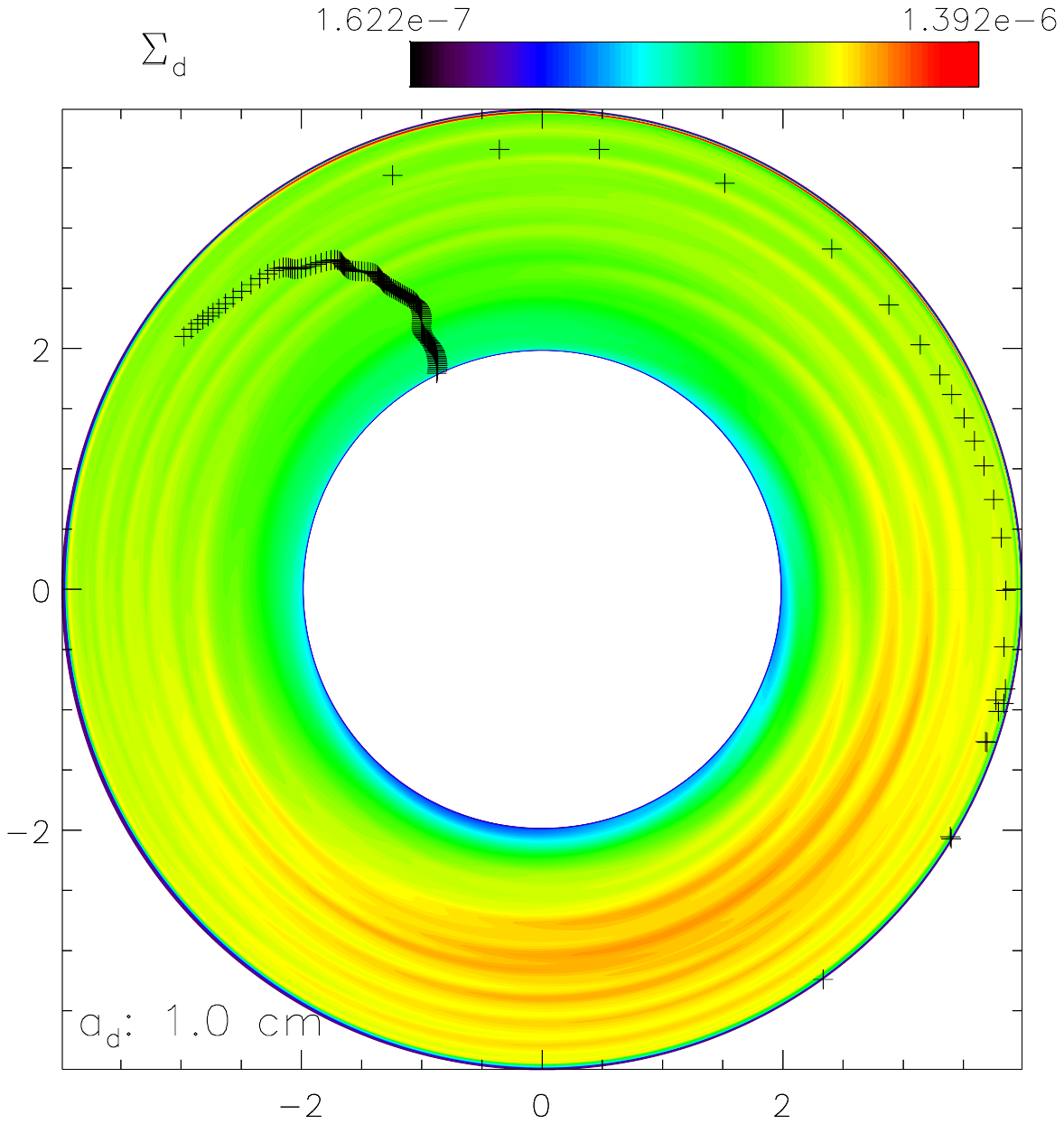}}\\
  \scalebox{0.5}{\includegraphics[bb=0.72in 5in 7.75in 10in]{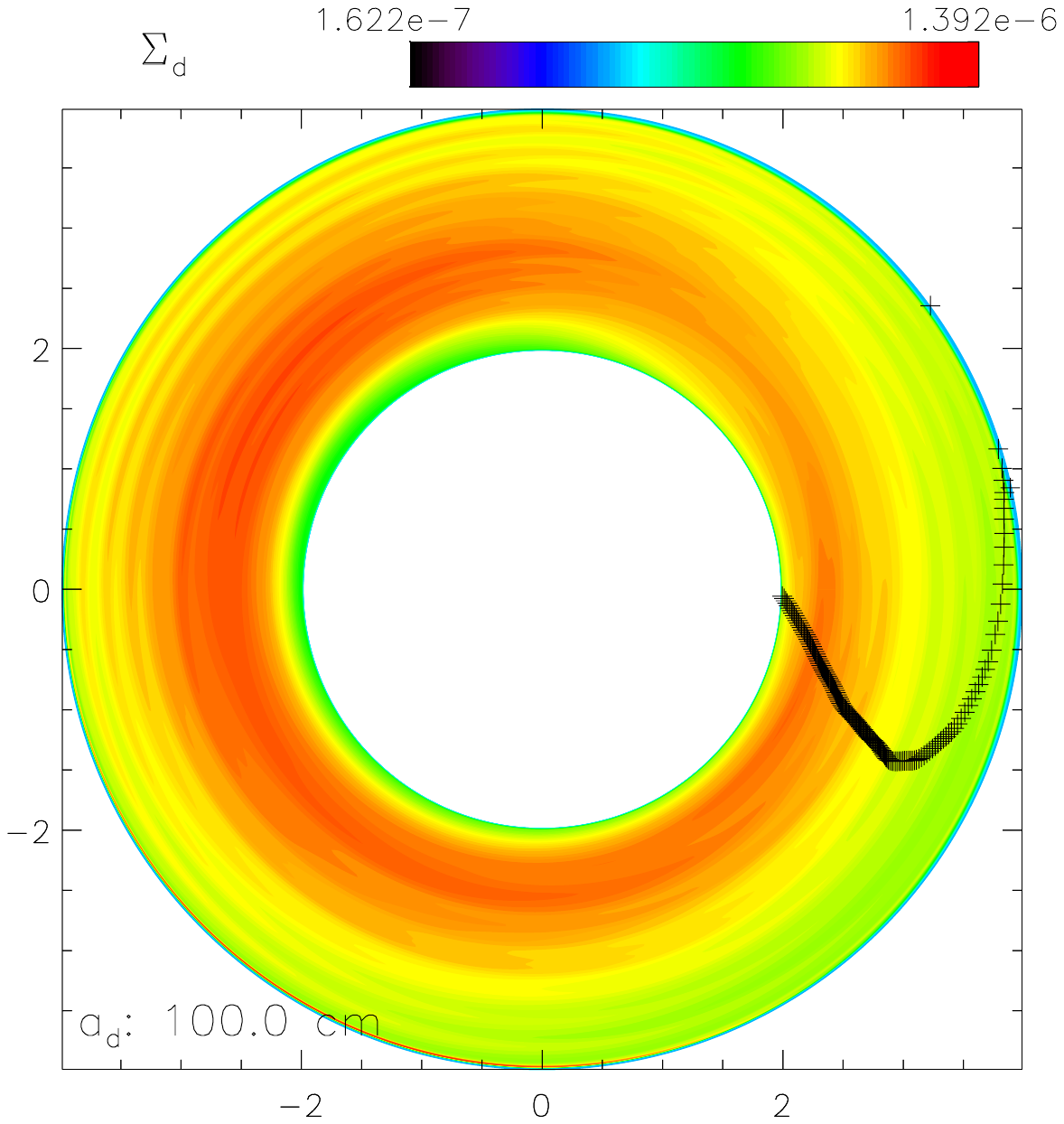}}&
\scalebox{0.5}{\includegraphics[bb=0.72in 5in 7.75in 10in]{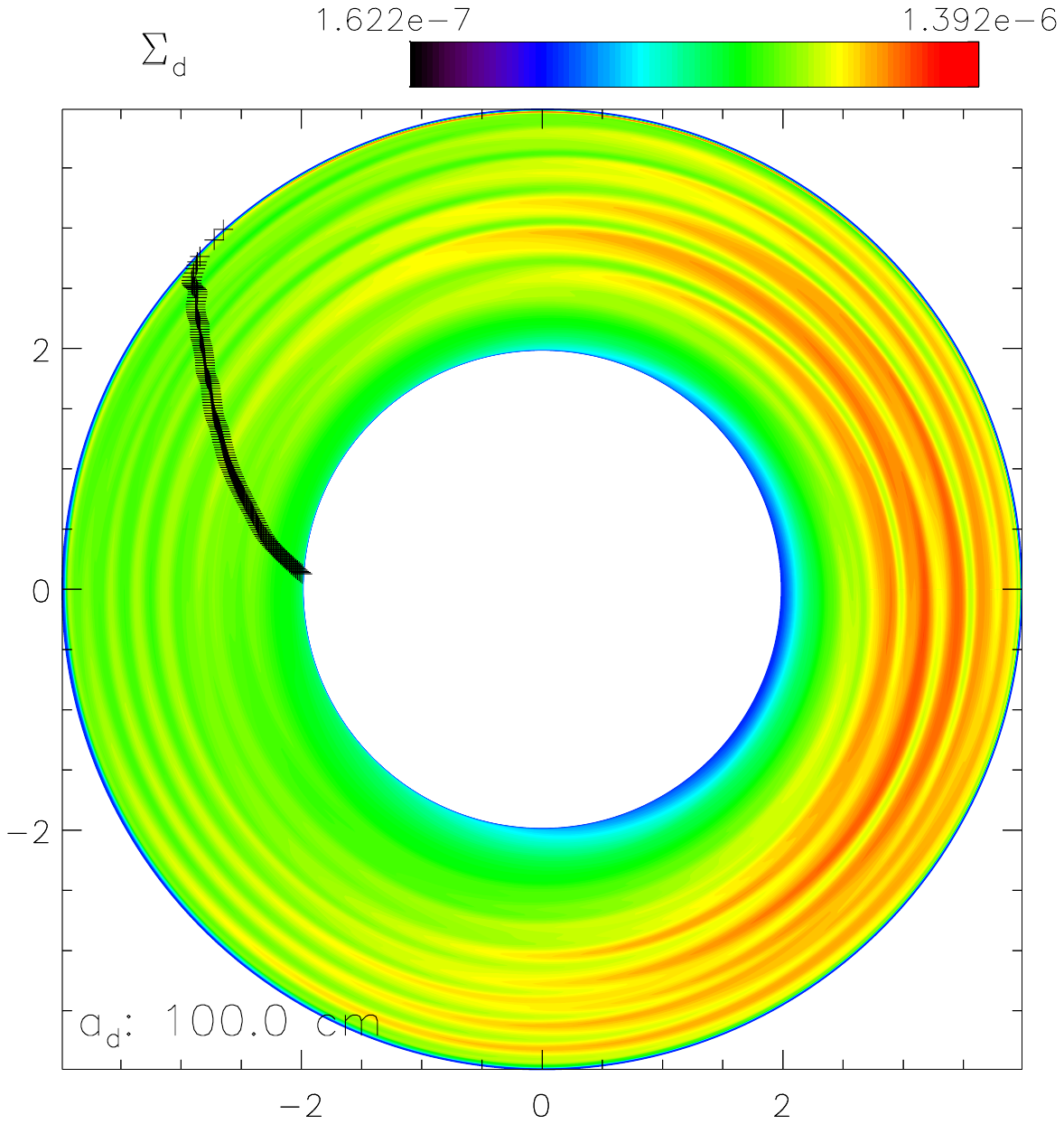}}
\end{tabular}
  \caption{The dust surface density of various sizes in the transition disk
  for the $e_p=0$ (left panels) and $e_p=0.1$ (right panels) cases.
            The length unit is $a_p=100$ AU.
            The unit of the surface density is $8.89\times10^{-4}$ g cm$^{-2}$.}
  \label{fig:den_100}
\end{figure}
\begin{figure}
\begin{tabular}{cc}
  \scalebox{0.42}{\includegraphics[bb=0.72in 5in 7.75in 10in]{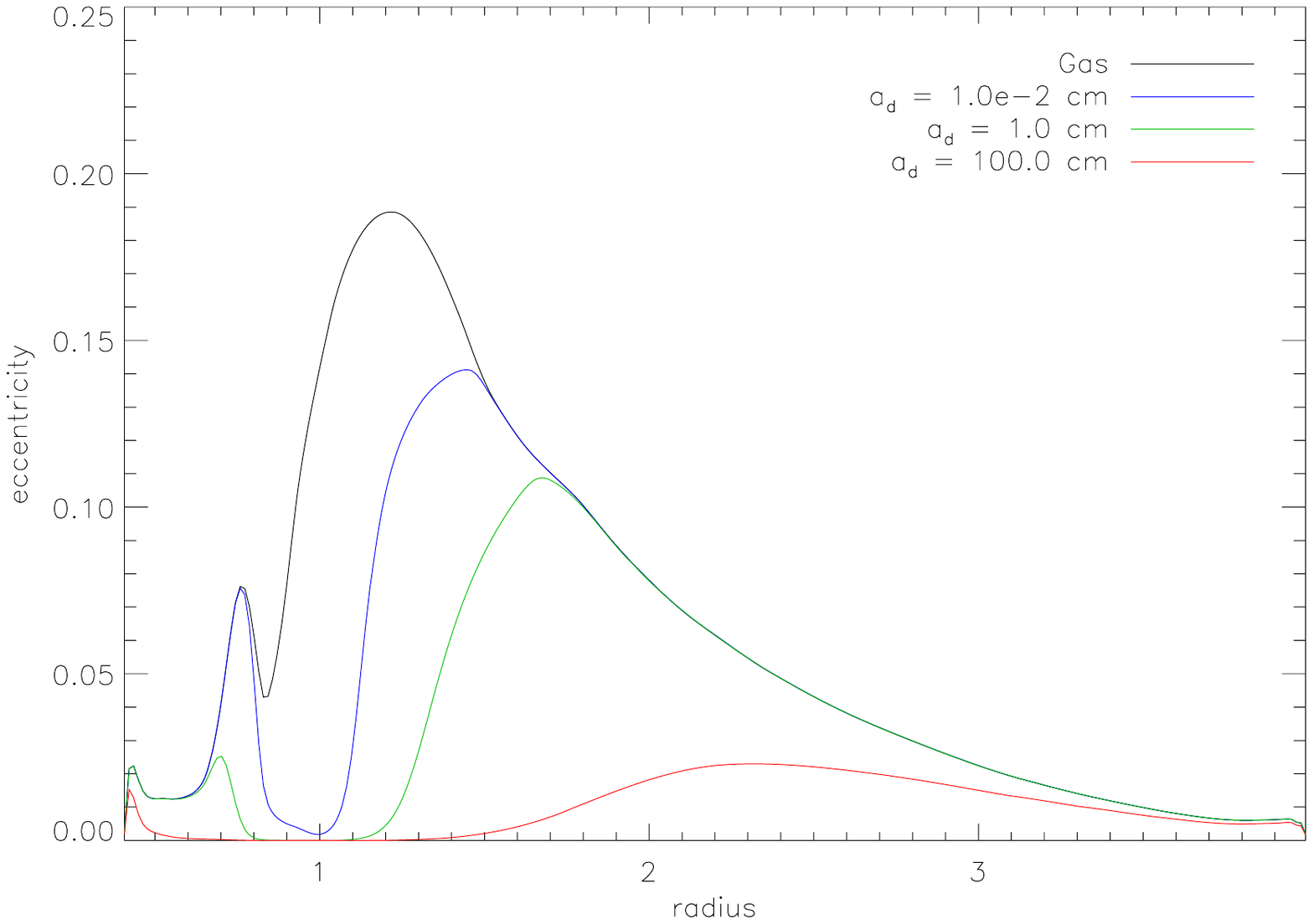}} &
 \scalebox{0.42}{\includegraphics[bb=0.72in 5in 7.75in 10in]{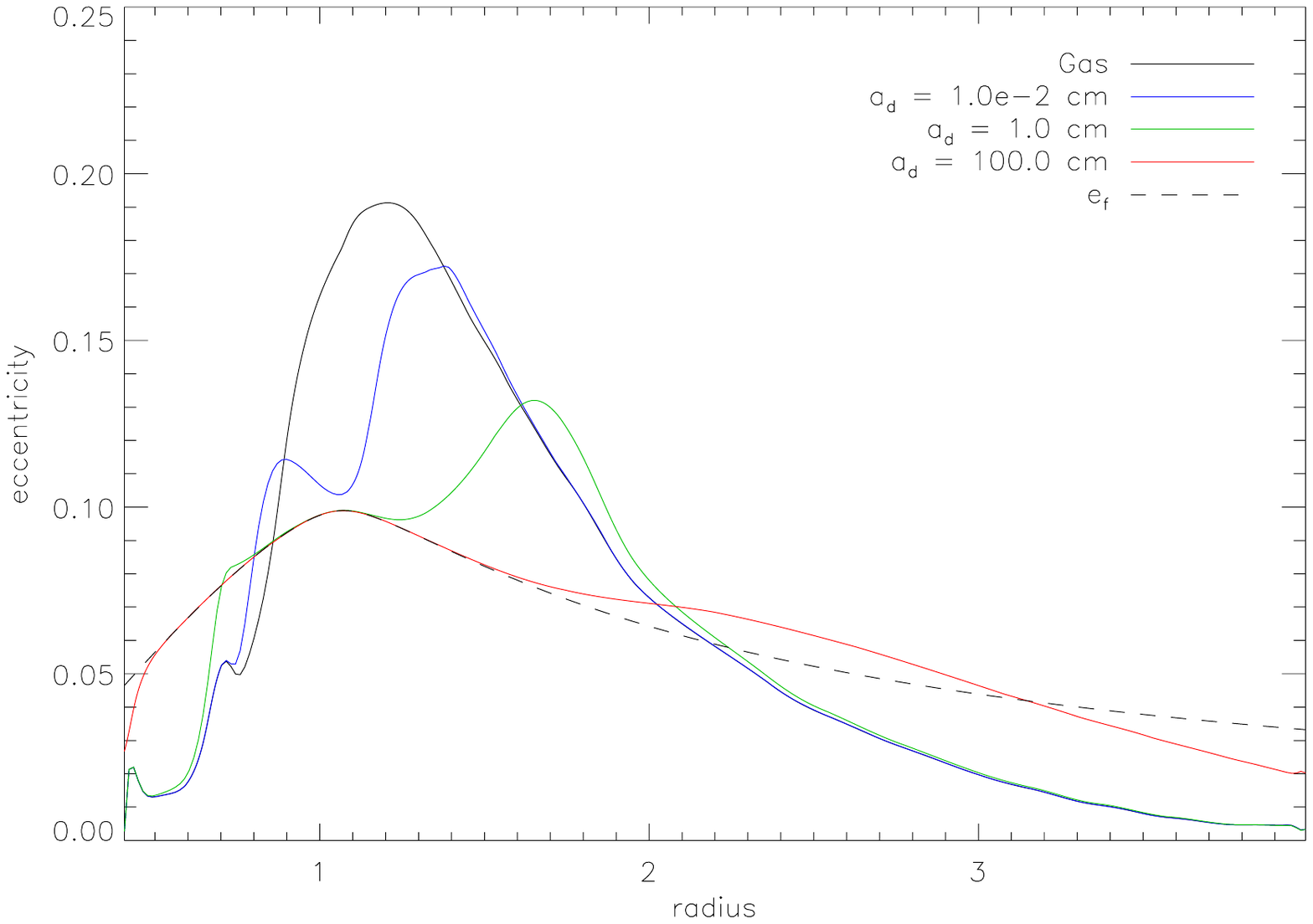}} \\
  \end{tabular}
  \caption{The azimuthal average of the dust and gas eccentricities for the transition disk. The left panel shows the
  case for $e_p=0$ and the right panel shows the cases for $e_p=0.1$. The forced eccentricity $e_f=|E_f|$ is also plotted for the $e_p=0.1$ case.}
  \label{ecc_8000_100}
\end{figure}

The gap of the gaseous disk for $e_p=0.1$ is slightly shallower but wider than that
for $e_p=0$ in our model for $q\approx 5\times 10^{-3}$. This outcome seems to be
consistent with the parameter study by \citet{Hosseinbor07} for smaller planetary
masses $q\leq 10^{-3}$ on fixed orbits. Since we are concerned with dust dynamics
exterior to the gap, the gap shape does not introduce significant differences in the
results between the $e_p=0$ and $e_p=0.1$ cases.
Nonetheless, it has been known from theories that in the presence of eccentricity,
the planet-disk interaction does not depend only on the 3:1 eccentric Lindblad
resonance but may rely on other resonances \citep{GT80}. Using the same code with
the standard disk parameters, \citet{Hosseinbor07} suggested that the planetary
eccentricity enhances the
torques on the gas via eccentric corotation resonances against
the
opposite torques from Lindblad resonances and thus results in more gas in the gap.
The account does not explain why the gap opened up by an eccentric planet is wider,
which occurs beyond the disk radius $r\approx 1.2$-1.3 as shown in their Figure 4
and in our Figure \ref{fig:DenAve_vs_ep}. Note that at $r\approx 1.587$, the 2:1
eccentric corotation resonance overlaps with the 4:2 eccentric Lindblad resonance
as well as the 2:1 ordinary Lindblad resonance. Unfortunately, one is unable to
deduce sufficient information from resonant torques \citep{GT80} to determine the
evolution of the system when both the disk and planet possess eccentricity
\citep{O07}. Investigation following the method provided by \citet{O07} is
desirable for understanding the gap opening due to mean-motion resonances in an
eccentric disk.

In addition, \citet{Hosseinbor07} has cautioned that it is not self-consistent to
assume a fixed planetary orbit in their study. Our work, based on the same
assumption, is certainly plagued with the same problem. The eccentricity evolutions
of the disk and the planet should be coupled \citep{dAngelo06}, especially when the
disk-to-planet mass ratio is not negligibly small \citep{Papa01,Riceet08}. In the
linear secular theory solving for eccentric modes of a coupled planet-disk system,
the pericenter of the planet's orbit should not stand still as is assumed in our
model but rather precess as well. The precession rate is the eigenvalue of the
coupled planet-disk problem \citep{GO06} and therefore should be the same as that of
the disk. A self-consistent linear calculation as well as numerical works should be
conducted in the context that we have presented here.

Needless to say, our fiducial model for protoplanetary and transition disks only
presents suggestive results for azimuthal asymmetry of dust distributions in an
eccentric disk. Parameter studies built on our fiducial model should be conducted
to provide a more complete picture that encompasses a large variety of physical
conditions for protoplanetary disks at different stages.



\acknowledgments We thank Xuening Bai, Cl\'ement Baruteau,
Shigehisa Takakuwa, and Ya-Wen Tang for useful discussions. This work is partly supported by the NSC
grant in Taiwan through NSC 100-2112-M-001-005-MY3.

\clearpage



\clearpage


\clearpage

\clearpage

\clearpage
\clearpage

\clearpage








\clearpage


\clearpage

\clearpage

 \clearpage

\clearpage

\end{document}